Charles University in Prague
Faculty of Mathematics and Physics

# DOCTORAL THESIS

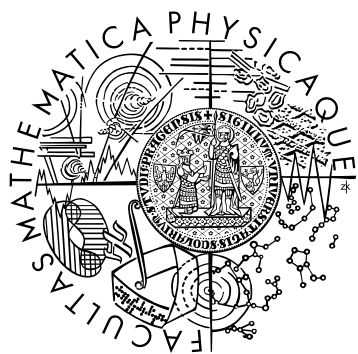

Jakub Řípa

## Statistical Analysis of the Observable Data
## of Gamma-Ray Bursts

Astronomical Institute of Charles University

Supervisor: **doc. RNDr. Attila Mészáros, DrSc.**

Study Programme: Physics
Study Branch: F1

Praha 2011

# Errata:

- Acknowledgements (next page, 3. paragraph, line 1): Mézáros → Mészáros

- Page 51, Chap. 4.1.3, lines 3−4: ... energy bands $(120 − 1500)\,\text{keV}$ / counts in the band $(25 − 120)\,\text{keV}$.

- Page 62, first paragraph, line 6: ... content of this section ...

- Page 64, 1. paragraph, line 10−12: ... In 286 cases the signal-to-noise ratio, in at least one of the two used channels, during $T_{90}$ was lower than 3.5 or ...

- Page 66, last paragraph, line 1: accurate → accurately

- Page 69, Tab. 5.13, line 4: emphasize → emphasized

- Page 76, Fig. 6.60., line 2: Dependence of $z_{\text{turn}}$ on $q$ for $\Omega_{\text{M}} = 0.27$, $\Omega_{\Lambda} = 0.73$ and $N = 1$.

May 11, 2011, Jakub Řípa

This study was supported by the Charles University Grant Agency grant No. 46307, by the Czech Science Foundation grants No. 205/08/H005, and P209/10/0734, by the project SVV 261301 of the Charles University in Prague, and by the Research Program MSM0021620860 of the Ministry of Education of the Czech Republic.

The data used in this thesis were mainly obtained by the Reuven Ramaty High-Energy Solar Spectroscopic Imager (RHESSI). A large acknowledgement is therefore to the RHESSI team and to the working group from The Paul Scherrer Institute involved in this satellite project. They provided and introduced me the relevant software. Other data, used in this thesis, were obtained by the BATSE, Swift, and Fermi instruments. The outcomes of the technical teams involved in these projects are also widely acknowledged.

At this place I would like thank my supervisor Doc. Attila Mézáros for his leadership, support and helpful advices. I very appreciate the help of my advisors, Dr. Claudia Wigger, Dr. Lajos G. Balázs, Dr. István Horváth, and Dr. René Hudec for discussions, for a fruitful collaboration, and for the assistance with the final corrections of this manuscript. I also thank Péter Veres, David Huja, and Dr. Zsolt Bagoly for collaboration and for useful discussions. I appreciate the support from Doc. Marek Wolf during the whole period and help by Maruška Hrudková with the LaTeX software.

I would like to express my warm gratefulness to whole my family and especially to my parents for the persistent support throughout all my studies.

Prohlašuji, že jsem tuto disertační práci vypracoval samostatně a výhradně s použitím citovaných pramenů, literatury a dalších odborných zdrojů.

Beru na vědomí, že se na moji práci vztahují práva a povinnosti vyplývající ze zákona č. 121/2000 Sb., autorského zákona v platném znění, zejména skutečnost, že Univerzita Karlova v Praze má právo na uzavření licenční smlouvy o užití této práce jako školního díla podle § 60 odst. 1 autorského zákona.

Fer. 14, 2011

_________________________            _________________________
V Praze                                      Jakub Řípa

*Název práce:* Statistické zpracování pozorovatelných dat gama záblesků
*Autor:* Jakub Řípa, ripa@sirrah.troja.mff.cuni.cz
*Ústav:* Astronomický ústav UK
*Vedoucí disertační práce:* doc. RNDr. Attila Mészáros, DrSc., meszaros@cesnet.cz

*Abstrakt:*
Gama záblesky stále ještě nejsou zcela pochopené jevy. Nicméně jejich prozkoumání může poskytnout užitečný nástroj pro lepší porozumění ranného vesmíru, protože patří mezi nejvzdálenější a nejintenzivnější objekty, které astronomové znají. Tato práce se snaží přinést více informací o tzv. skupině prostředních záblesků objevené již dříve různými autory v různých databázích. Nejdříve jsou statisticky studovány délky trvání a spektrální tvrdosti gama záblesků pozorovaných přístrojem "Reuven Ramaty High-Energy Solar Spectroscopic Imager". Získané výsledky naznačují, že tyto prostřední záblesky tvoří samostatnou skupinu. Za druhé, tyto záblesky jsou dále podrobně zkoumány s ohledem na jejich "spektrální lagy", špičkové toky, rudé posuvy, výskyt supernovy, atd. Za třetí, dlouhé gama záblesky se známými rudými posuvy a s odvozenými pseudo-rudými posuvy detekovanými přístrojem "The Burst and Transient Source Experiment" a dále záblesky se změřenými rudými posuvy z přístrojů Swift a Fermi jsou použity pro studium kosmologických efektů na rozdělení jejich pozorovaných toků záření.

*Klíčová slova:* gama záblesky, gamma astrofyzika, kosmologie

*Title:* Statistical analysis of observable data of gamma-ray bursts
*Author:* Jakub Řípa, ripa@sirrah.troja.mff.cuni.cz
*Department:* Astronomical Institute of Charles University
*Supervisor:* doc. RNDr. Attila Mészáros, DrSc., meszaros@cesnet.cz

*Abstract:*
Gamma-ray bursts are still not fully understood events. However, their exploration could provide a useful tool for a better understanding of the early Universe because they belong to the most distant and violent objects that astronomers know. This thesis tries to bring more information about a so-called group of intermediate-duration bursts claimed by different authors employing different data samples. Firstly, duration and spectral hardness properties of bursts from the Reuven Ramaty High-Energy Solar Spectroscopic Imager are statistically analysed. The obtained results bring a suspicion that these intermediate bursts gather into a separate group. Secondly, these bursts are investigated in more detail with respect to their spectral lags, peak count rates, redshifts, supernova observations, and so forth. Thirdly, long-duration bursts with known redshifts and with derived pseudo-redshifts detected by The Burst and Transient Source Experiment, Swift and Fermi bursts with known redshifts, are used to study the cosmological effects on the observed flux and fluence distributions.

# Contents

# Introduction

This thesis summarizes the results of my doctoral study on the topic of gamma-ray bursts. Gamma-ray bursts (GRBs) are observationally sudden and nonrepetitive violent flashes initially of gamma rays (prompt emission) followed by electromagnetic emission at lower wavebands (afterglow). Typical photon energies of the prompt emission are of the order of hundreds of keV, however, as known from the latest observations, they can reach up to 300 GeV. On the other side, afterglow detections can go down to radio wavebands. Their sources lie in the deep universe and they belong to the most distant objects that astronomers know. The fluxes of these gamma flashes often over-shine for a short moment the whole sky and even over-shine sources in our relative vicinity like our Sun or the Milky Way Galaxy. These observational facts lead to the astonishing finding that if the gamma emission were isotropic, the source would be able to transfer half of the solar mass into radiation energy during tens of seconds. Although GRBs are studied for about forty years there are still many puzzles about their exact sources and radiation mechanisms.

The content and the focus of this thesis is to bring more information about the so-called intermediate GRB group on the basis of the statistical tests of the observational data obtained mainly from the Reuven Ramaty High-Energy Solar Spectroscopic Imager (RHESSI). One part of this work also discusses the cosmological effects on the observed GRB fluxes.

This manuscript is organized as follows. A brief overview of the present knowledge about gamma-ray bursts is presented in Section 1. Section 2 introduces the RHESSI satellite whose data are widely employed in the following chapters. In Section 3 an insight into the classification of different GRB groups is presented. In Section 4 a statistical analysis of GRB durations and spectral hardnesses is performed in order to confirm an intermediate group of bursts claimed by different authors and using different data samples in the past. Section 5 investigates these intermediate bursts in more detail with respect to their spectral lags, peak count rates, redshifts, supernova observations, and so forth. Section 6 focuses on the cosmological effects on the observed flux and fluence distributions of GRBs. Hence, Sections $1-3$ give an overview of the topic and Sections $4-6$ contain only the results of the author's research. These results are summarized in Section 7. This thesis contains three Appendices A, B, and C of data samples used throughout this work, used statistical analyses, and a description of the uncertainties of the decimated RHESSI data. Finally, the list of the cited publications, the list of used abbreviations, the list of the author's publications, and the list of the independent citations of the author's publications is enclosed.

# 1   Overview of Gamma-Ray Bursts

This chapter provides a brief overview of the present knowledge of gamma-ray bursts and draws form the following survey publications, in which a curious reader can find more complex and detailed information: Band (1999); Daigne (2004); Djorgovski et al. (2002); Fishman & Meegan (1995); Gehrels et al. (2009); Hudec (1998); Hurley et al. (2006); Klose (2000); Mészáros (2001, 2002, 2006); Piran (1999, 2004); Ramana Murthy & Wolfendale (1993); Řípa (2006); Schönfelder (2001); Štoček & Mészáros (2000); Topinka (2002, 2009); Vedrenne & Atteia (2009); Zhang & Mészáros (2004).

## 1.1   Discovery

The discovery of gamma-ray bursts came up fortuitously in the year 1967 (Fig. 1.1) when the U.S. military satellites Vela, designed to patrol a possible violation of the nuclear test ban treaty signed by nuclear powers in 1963, detected unknown gamma flashes coming from the outer space. After the declassifying of the secret data, astronomers began to study these flashes. The first article about GRBs was published by Klebesadel et al. (1973) and described 16 proved detections in the period from July 1969 to July 1972 (Fig. 1.2) by the Vela 5A, 5B, 6A and 6B satellites with CsI scintillation detectors onboard arranged in a certain way to achieve the isotropic sensitivity. The energy sensitivity was in the range $0.2 - 1.0$ MeV for the Vela 5 and $0.3 - 1.5$ MeV for the Vela 6 satellites. GRB observations were soon after confirmed by the Soviet Cosmos$-461$ satellite (Mazets et al. 1974), which observed gamma flashes at the range $0.05 - 0.3$ MeV.

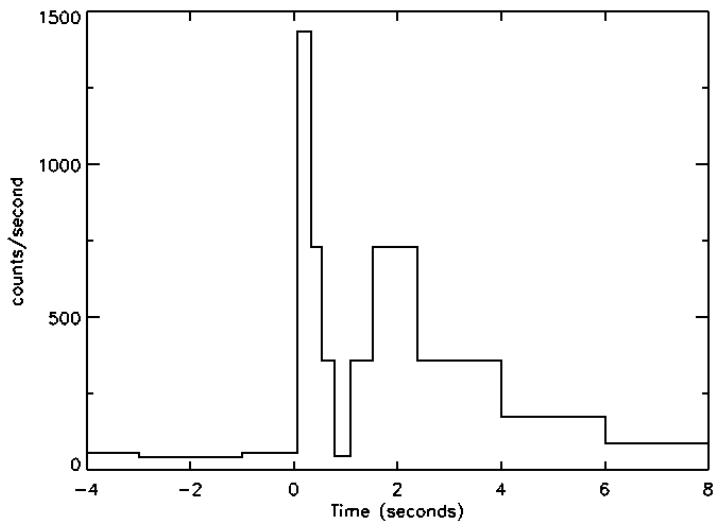

**Fig. 1.1.** The first detected gamma-ray burst recorded on July 2, 1967 by the Vela 4A satellite and also triggered by the Vela 3 satellites (Credit: Bonnell 1995).

Many subsequent observations were performed, e.g., with IMP 6 (Cline et al. 1973), OSO$-7$ (Wheaton et al. 1973), Helios 2 (Cline et al. 1979), Venera (Mazets & Golenetskii 1981a; Mazets et al. 1983), Prognoz (Barat et al. 1981; Boer et al. 1986), HEAO$-1$ (Knight et al. 1981), ISEE$-3$ (Klebesadel et al. 1981; Teegarden & Cline 1981), PVO (Klebesadel et al. 1981; Evans et al. 1981), SMM (Forrest et al. 1980), Ginga (Murakami et al. 1989), Granat (Terekhov et al. 1994), Ulysses (Hurley et al. 1992), EURECA (Brandt et al. 1994), Mars Observer (Laros et al. 1994) satellites which extended our knowledge of these new cosmic phenomena.

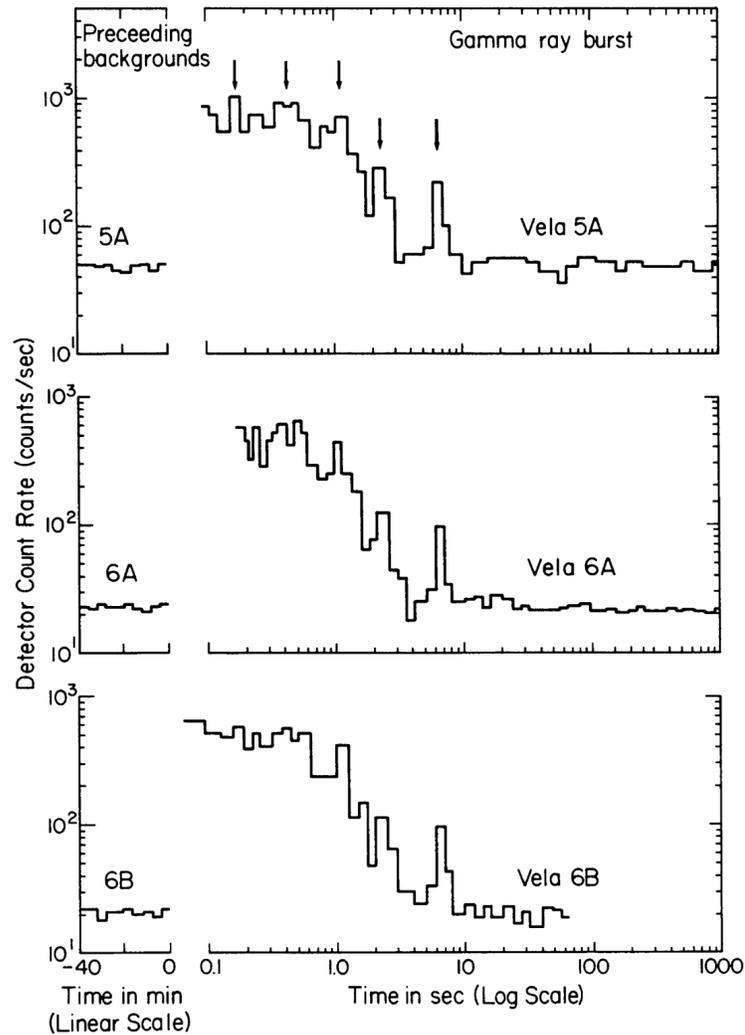

**Fig. 1.2.** Count light curve of GRB 700822 detected by three Vela satellites. The arrows show structures that are the same at all three records (Credit: Klebesadel et al. 1973).

## 1.2 Following missions

### 1.2.1 CGRO

A big step forward in the observations was made with Compton Gamma-Ray Observatory[1] (CGRO) (Gehrels et al. 1993) launched in April 1991 and ended its activity in June 2000. The satellite had four gamma instruments on-board providing measurements over a wide energy range from 30 keV to 30 GeV.

---

The Burst and Transient Source Experiment (BATSE[1]) was an all-sky system made of eight pairs of NaI detectors with a sensitivity 5 to 10 times better than the previous GRB detectors. The energy range was $30\,\mathrm{keV} - 1.9\,\mathrm{MeV}$, the time resolution $100\,\mu s$ and the directional accuracy a few degrees. This detector had a key importance in the history of gamma-ray astronomy, because during its whole lifetime it had captured 2704 GRBs (Paciesas et al. 1999), 185 Soft Gamma-ray Repeaters (SGRs) (Lazzati et al. 2005) and other events such as solar flares or Terrestrial Gamma Flashes (TGFs) (Fishman et al. 1994a).

Further instruments aboard CGRO were The Oriented Scintillation Spectrometer Experiment (OSSE), Imaging Compton Telescope (COMPTEL), and Energetic Gamma Ray Experiment Telescope (EGRET). The sensitivity of EGRET began at $20\,\mathrm{MeV}$ and extended observations up to $30\,\mathrm{GeV}$ (Dingus 1995).

### 1.2.2  Wind-Konus

The next instrument providing a large sample of GRB observations is a satellite Wind[2]. The main mission of this probe is studying solar wind and it was launched in November 1994 (still active). Its Russian Konus experiment (Aptekar et al. 1995) on-board with omnidirectional detectors has provided approximately 3000 GRB observations[3,4] so far in the energy range $10\,\mathrm{keV} - 10\,\mathrm{MeV}$ (Cline et al. 2003).

### 1.2.3  HETE-2

The High Energy Transient Explorer[5] (Ricker et al. 2003) was launched in October 2000. It had two X-ray instruments SXC and WXM and gamma telescope FREGATE (French Gamma-Ray Telescope) on-board. FREGATE is made of four NaI (Tl) detectors with the sensitivity from $6\,\mathrm{keV}$ to $400\,\mathrm{keV}$, and with the field of view approximately $3\,\mathrm{srad}$. HETE-2 detected more that 200 GRBs[6] (Vanderspek et al. 2004).

### 1.2.4  IPN

The precise localization of gamma-ray bursts is not an easy task. The majority of the first GRB detectors was not able to localize bursts. Although the BATSE experiment had the directional accuracy $\approx 1°$, it was not precise enough to identify an optical GRB counterpart. The Inter-Planetary Network[7] (IPN) is a group of spacecraft equipped with gamma-ray burst detectors. By timing the arrival of the gamma rays from a burst at several spacecraft, the precise location can be found. It is also called a triangulation method (Fig. 1.3). So far, 27 spacecraft have participated in the network. The accuracy of the IPN localization can go down to the order of arcmin, but unfortunately the time needed for gathering up the data from particular spacecraft is too long to allow fast follow up with, e.g., ground-based optical telescopes.

---

[1]http://gammaray.nsstc.nasa.gov/batse/
[2]http://heasarc.gsfc.nasa.gov/docs/heasarc/missions/wind.html
[3]http://lheawww.gsfc.nasa.gov/docs/gamcosray/legr/konus/konus_2003.html#tc1
[4]http://www.ioffe.ru/LEA/shortGRBs/Catalog/
[5]http://heasarc.nasa.gov/docs/hete2/hete2.html
[6]http://space.mit.edu/HETE/Bursts/Data/
[7]http://www.ssl.berkeley.edu/ipn3/

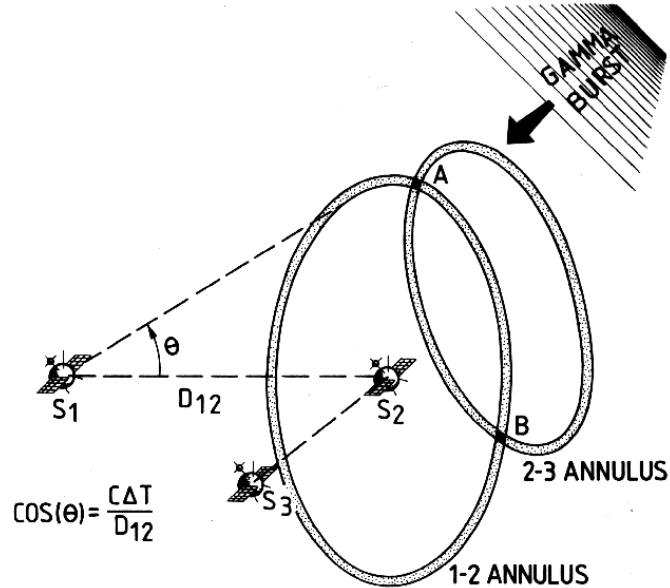

**Fig. 1.3.** The illustration of the triangulation method used by IPN. If the difference $\delta T$ between the times in which gamma rays from a GRB reach two spacecraft 1 and 2 in the mutual distance $D$ is known, the angle of the burst with respect to the spacecraft can be easily calculated ($C$ is the speed of light). Measurements from two spacecraft provide one annulus of the possible positions of the source. The sine of the width of the annulus ($\sin \delta\theta$) is proportional to the accuracy of the time difference measurements and inversely proportional to the distance $D$. By employing measurements of three spacecraft the three annuli, which intersect in one error-box, are obtained (Credit: IPN webpage[1]).

For the sake of completeness, a curious reader can find actual and past GRB observation notices, e.g., in the GRB Coordinates Network (GCN[1]) service, in the Kevin Hurley's Master Burst List[2], Jochen Greiner's webpage[3], Stephen Holland's webpage[4], GRBlog webpage[5], or GRBOX webpage[6].

## 1.3 Prompt emission properties

### 1.3.1 Fluxes

Typical energies of prompt photons are $\approx 30-1000$ keV with the peak energy around 180 keV. The observed peak fluxes are usually $\approx 0.3-50$ ph.cm$^{-2}$.s$^{-1}$, while the lower limit is given by the sensitivity of the instrument, and fluences (overall time integrated flux) $\approx 10^{-8}-10^{-4}$ erg.cm$^{-2}$.

---

[1]http://gcn.gsfc.nasa.gov/
[2]http://www.ssl.berkeley.edu/ipn3/masterli.html
[3]http://www.mpe.mpg.de/~jcg/grbgen.html
[4]http://lheawww.gsfc.nasa.gov/users/sholland/grb/index.html
[5]http://grblog.org/grblog.php
[6]http://lyra.berkeley.edu/grbox/grbox.php

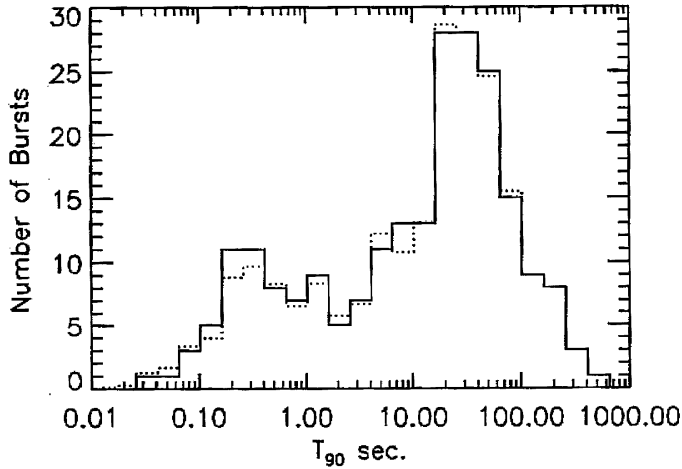

**Fig. 1.4.** Distribution of $T_{90}$ for the BATSE GRBs (Credit: Kouveliotou et al. 1993a).

### 1.3.2  Durations

The durations range from very short $\approx 0.01$ s up to long as $\approx 1000$ s. Fig. 1.4 shows the distribution of the duration $T_{90}$ from the first BATSE catalog. $T_{90}$ is defined as the time during witch the cumulative counts increase from $5\%$ to $95\%$ above the background, thus encompassing $90\%$ of the total GRB counts (Kouveliotou et al. 1993a). It is immediately seen that there are at least two groups of so-called short bursts with $T_{90} \lesssim 2$ s and long bursts with $T_{90} \gtrsim 2$ s.

However, a third group of GRBs with an intermediate duration has also been proposed (Horváth 1998; Mukherjee et al. 1998). The possible existence of such a third GRB group and its properties is widely discussed in Sections 4 and 5 of this thesis.

Figure 1.5 shows the burst hardness ratio as a function of the duration $T_{90}$. The hardness ratio is the ratio of the count number in a high energy band over the count number in a low energy band and thus gives information about the spectral hardness. Its logarithm is an analogy to a color index, e.g. B-V, known from classical astronomy. There is a clear tendency of the short group being harder and the long group being softer.

### 1.3.3  Time profiles

A very interesting feature of GRBs is the large variety of their time profiles, i.e., the variations of the observed intensities in course of time. It can be said that each burst has a unique profile. However, some common patterns can be found. There are: I. single pulse or spike events, II. smooth, either single or multiple, well-defined peaks (BATSE trigger 1076 or 1974 in Fig. 1.6), III. distinct, well-separated episodes of emission (BATSE trig. 143), IV. very erratic, chaotic and spiky bursts (BATSE trig. 408) as divided by Fishman & Meegan (1995). There are also bursts with a small precursor before the main peak (BATSE trig. 219) (Koshut et al. 1995). A precursor activity is still not fully understood. Some bursts have fast rise and exponential decay of its profile (FRED type) (BATSE trig. 563). Some bursts have atypical time profiles (BATSE trig. 1453). Short burst can be also multiple-peaked (BATSE trig. 551). Norris et al. (1996) suggested that time profiles of GRBs could be composed of the sum of a few individual pulses (Fig. 1.7).

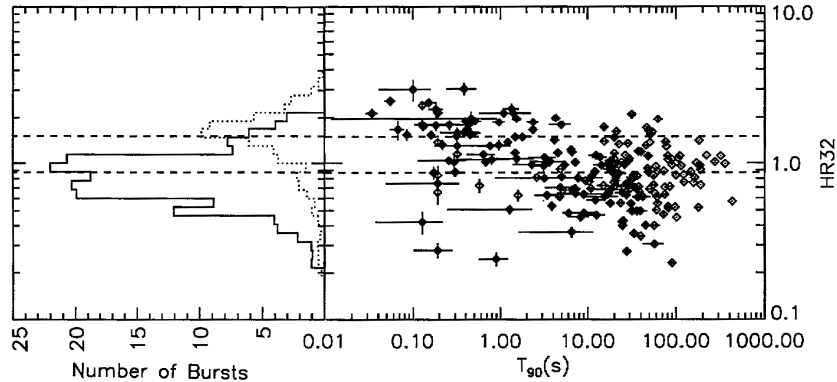

**Fig. 1.5.** *Left panel:* Hardness ratio histograms of short bursts, i.e., with $T_{90} < 2\,\mathrm{s}$ (dotted line) and long bursts, i.e., with $T_{90} > 2\,\mathrm{s}$ (solid line). *Right panel:* Hardness ratio $HR_{32}$ vs. $T_{90}$ scatter plot of the BATSE data. The dashed lines on both plots correspond to the mean hardness ratio of the two duration groups (Credit: Kouveliotou et al. 1993a).

The shortest observed time variability in GRB light curves was $\Delta t \approx 1\,\mathrm{ms}$. The fastest variations measured in an astrophysical source constrain its size $R$, because all the fluctuations shorter than the light-crossing time of the source will be smeared out by propagation delays within the source. The variation $\Delta t \approx 1\,\mathrm{ms}$ suggests the source to be a compact object of the size $R \lesssim c\Delta t = 300\,\mathrm{km}$ for a stationary source, or $R \lesssim \Gamma^2 c\Delta t = 3 \times 10^6\,\mathrm{km}$ for a relativistic flow with the Lorentz factor $\Gamma \approx 100$ (see Section 1.5.1).

---

[1]http://www.batse.msfc.nasa.gov/batse/grb/lightcurve/

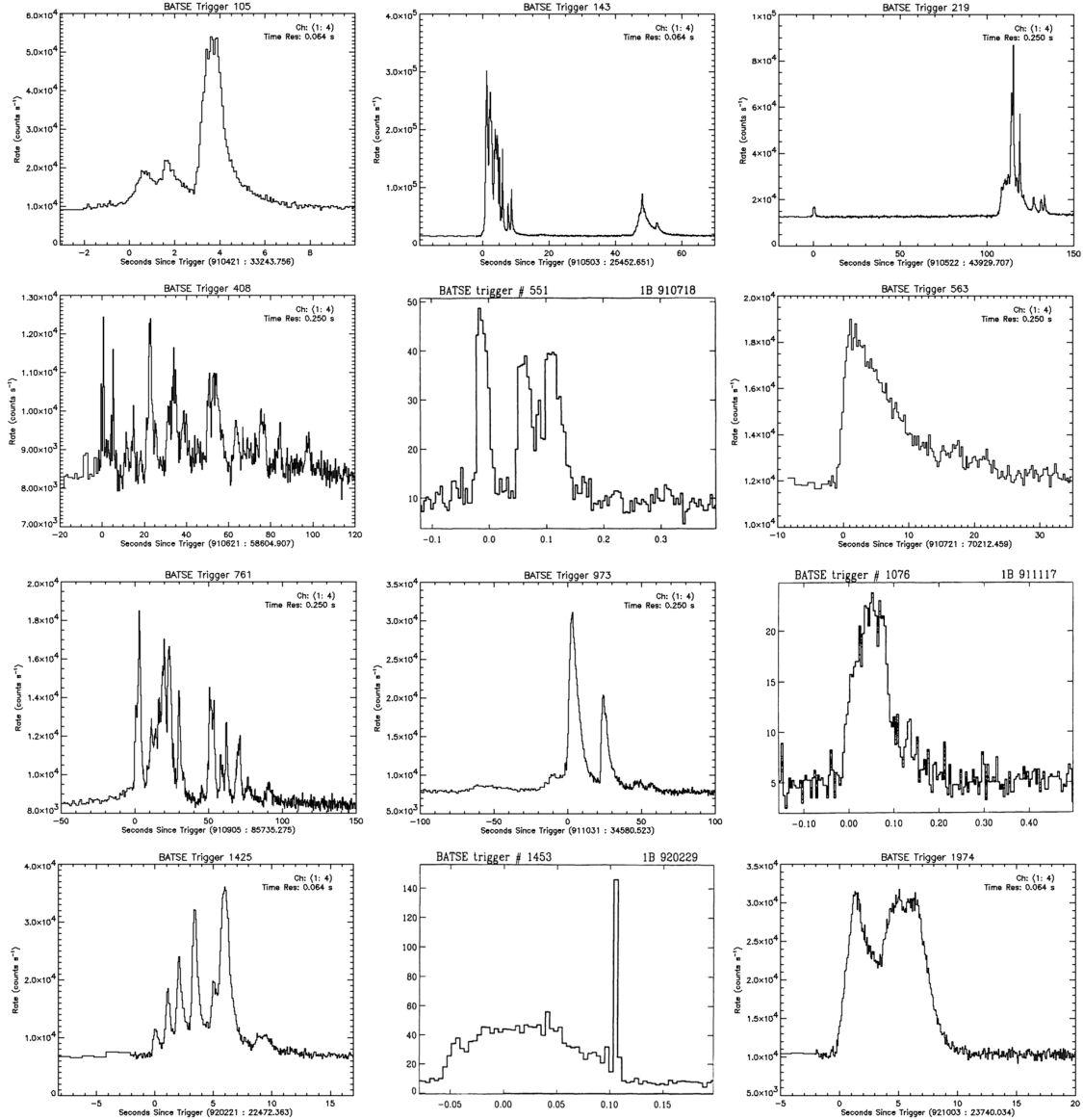

**Fig. 1.6.** A large variety of GRB light curves of the prompt emission as measured by the BATSE instrument (Credit: BATSE GRB Light Curves webpage[1], Fishman et al. 1994b).

## 1.3.4   Spectra

Contrary to light curves, the time-integrated GRB spectra do not show the same diversity. They are non-thermal and are usually well fitted by a so-called empirical Band function (Band et al. 1993). Let $dN/dE$ (sometimes denoted as $N_E(E)$) be the number of photons per energy bin and have the form of two power-laws with a smooth transition then:

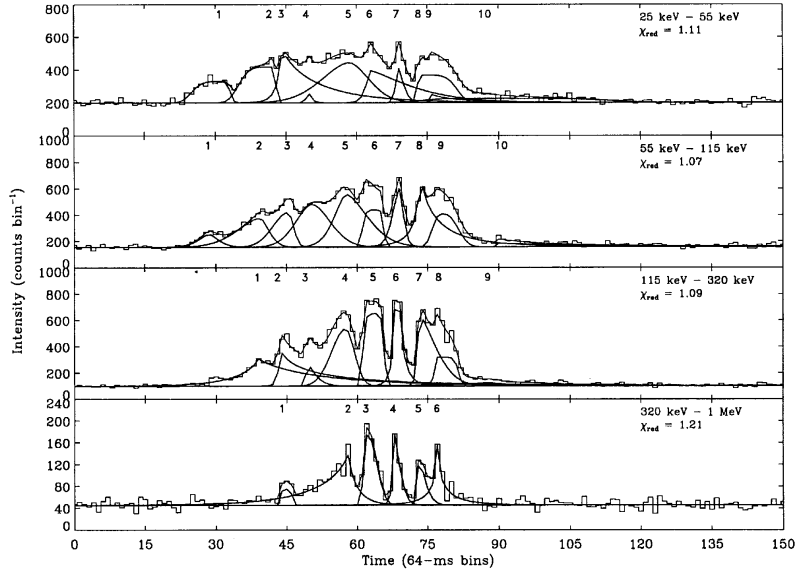

**Fig. 1.7.** The analysis of complex GRB light curves as the sum of several pulses (BATSE trig. 1683, Credit: Norris et al. 1996).

$$\frac{dN}{dE} = \begin{cases} A \left(\dfrac{E}{100\,\text{keV}}\right)^{\alpha} \exp\left(-\dfrac{E}{E_0}\right) & \text{if } E \leq E_{\text{b}}, \\[3mm] A \left[\dfrac{(\alpha - \beta)E_0}{100\,\text{keV}}\right]^{\alpha - \beta} \exp(\beta - \alpha) \left(\dfrac{E}{100\,\text{keV}}\right)^{\beta} & \text{if } E \geq E_{\text{b}}, \end{cases} \tag{1.1}$$

where the break energy $E_{\text{b}}$ is:

$$E_{\text{b}} = E_0(\alpha - \beta),$$

and $A$ is a normalization constant.

If the low-energy spectral index is $\alpha > -2.0$ and the high-energy spectral index $\beta < -2.0$ then the function $E^2\,dN/dE$ (sometimes denoted as $\nu F_{\nu}$) has the maximum at $E_{peak} = E_0(2 + \alpha)$. The most typical values of the parameters are $\alpha \approx -1.1$, $\beta \approx -2.4$ and $E_{peak} \approx 150\,\text{keV}$ (see Fig. 1.9). Sometimes the GRB spectrum is well fitted by a simpler model; the power-law with an exponential cutoff. As far as a measurement is provided only in a limited range, the spectrum can be well fitted by one power-law (simple power-law). For details see, for example, Wigger et al. (2008).

Figure 1.8 shows the time integrated spectrum of GRB 990123 with the best fitted Band function. Needed to say that this burst had an unusually hard spectrum. The GRB spectrum is, in general, strongly evolving during the burst. A global hard-to-soft evolution is observed in many GRBs (e.g., BATSE trig. 543 in Fig. 1.10).

Besides the continuum, cyclotron and annihilation lines were reported for many bursts. For example, the Konus instruments provided a population of GRBs presenting spectral line features around 400 keV in emission and 30 − 70 keV in absorption (Mazets et al. 1981b). Also other instruments observed such features, e.g.,

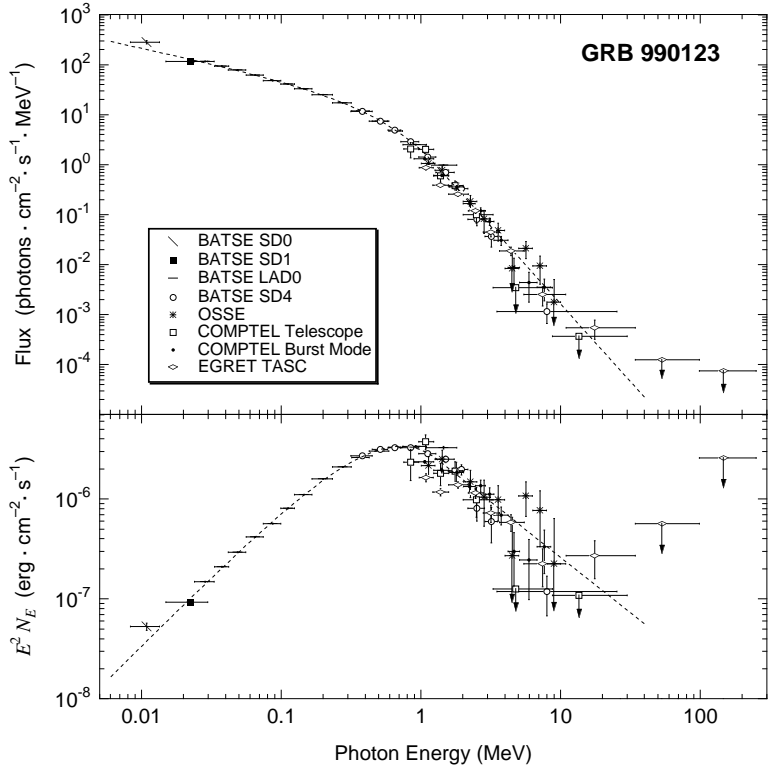

**Fig. 1.8.** The time integrated spectrum of GRB 990123 observed with the four instruments of CGRB. The dashed line is the fitted Band function with the model parameters: $\alpha = -0.6$, $\beta = -3.1$ and $E_{peak} = 720\,\text{keV}$ (Credit: Briggs et al. 1999).

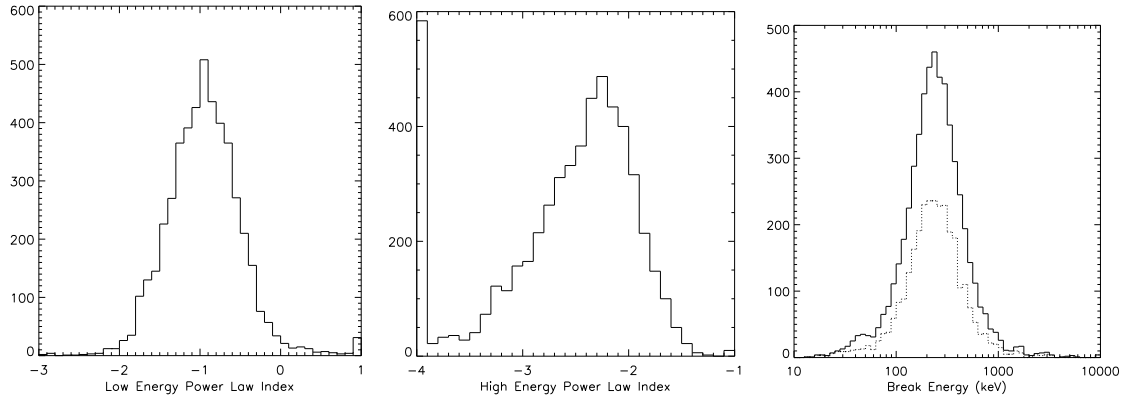

**Fig. 1.9.** A distribution of the low- and high-energy spectral indices and the break energy of the Band function from the BATSE data (Credit: Preece et al. 2000).

HEAO−1 (Hueter 1984), HXRBS on-board SMM (Dennis et al. 1982), ISEE−3 (Teegarden & Cline 1981), SIGNE experiments on the Venera 11 and 12 probes (Barat et al. 1984) or a Japanese satellite Ginga (Murakami et al. 1988; Fenimore et al. 1988). The interpretation of the absorption lines around 20−70 keV as the cyclotron lines of the electrons moving in a large magnetic field was a strong argument supporting highly magnetized galactic neutron stars as the progenitors of GRBs. Nev-ertheless the reality of these features raised a lot of debates. A line seen by one group was not always confirmed by another observing the same burst with a different instrument (Vedrenne & Atteia 2009, Chapter 1.3.2). Moreover, the spectral features reported by Konus and Ginga were not confirmed by BATSE by many authors (Palmer et al. 1994; Band et al. 1994, 1995, 1996), but on the other hand, Briggs et al. (1998) showed that there might be spectral lines also in the BATSE data.

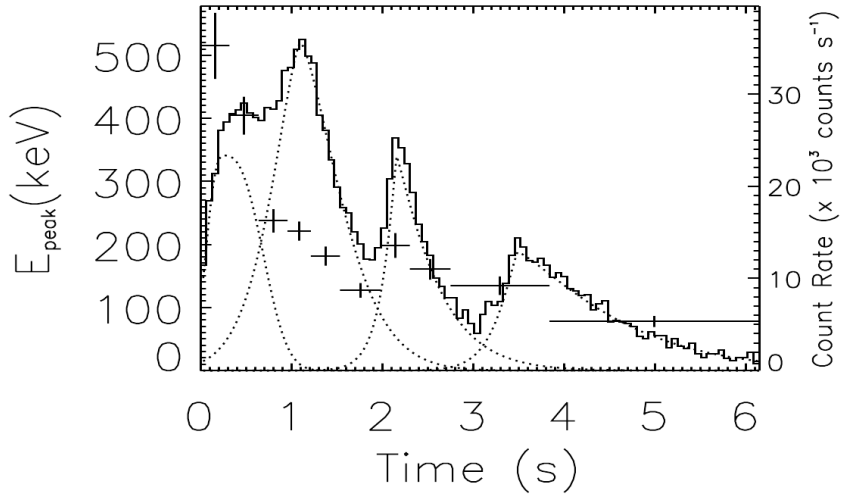

**Fig. 1.10.** The spectral evolution of BATSE trigger 543. A global hard-to-soft evolution as well as a hardness-intensity correlation in the individual pulses is seen (Credit: Crider et al. 1999).

### 1.3.5 Sky distribution

An important BATSE result was that the sky distribution of GRBs was almost exactly isotropic (Meegan et al. 1992), see Fig. 1.11. This was a strong evidence supporting the extragalactic origin of GRBs, because they were not clustered in the plane of Milky Way Galaxy or in the direction of the structures in the near universe.

However, Baláazs et al. (1998, 1999) found that BATSE short bursts ($T_{90} < 2$ s) are distributed anisotropically with a higher than 99 % probability, but the long bursts can still be distributed isotropically. Magliocchetti et al. (2003) also found within the short BATSE bursts an $\approx 2\sigma$ deviation from isotropy on angular scales $\theta \approx 2 - 4°$. Such an anisotropy is not observed in the distribution of long GRBs. Mészáros et al. (2000a) reported the anisotropy of BATSE intermediate GRBs at the $\geq 97$ % significance level. The following work Mészáros et al. (2000b) reported an intrinsic non-randomness in the angular distribution of the intermediate GRBs at a 96.4 % confidence level. The same work reported that the dim intermediate bursts have a non-randomness angular distribution even on a higher confidence level, i.e., 99.3 %. Litvin et al. (2001) confirmed the anisotropic sky distribution of short and intermediate GRBs. Surprisingly, it was also reported that even the long BATSE bursts may be distributed anisotropically at the $\geq 99\%$ significance level (Mészáros & Štoček 2003).

---

[1]http://f64.nsstc.nasa.gov/batse/grb/skymap/

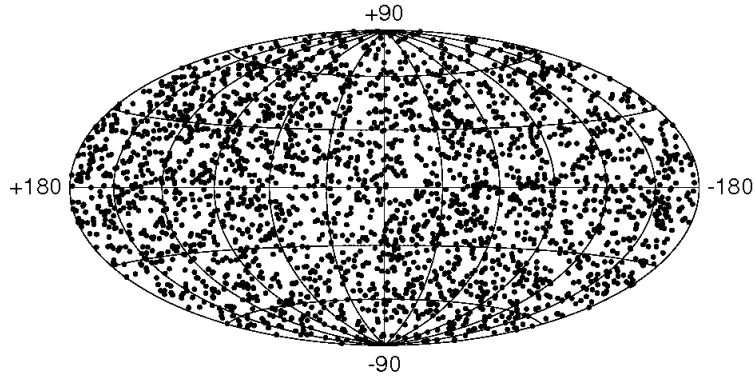

**Fig. 1.11.** This map shows the locations of 2704 GRBs recorded with BATSE on-board CGRO during the nine-year mission. The projection is in Galactic coordinates. The GRBs are not concentrated in the plane of the Milky Way Galaxy and the distribution is consistent with isotropy (Credit: BATSE Gamma-Ray Burst Skymaps webpage[1]).

Probably the most complex work investigating the eventual non-randomness in the sky distribution of BATSE GRBs is the article Vavrek et al. (2008), which provided 13 tests based on the Voronoi tesselation, minimal spanning tree, and on the multifractal spectrum technics. They concluded that the short and intermediate groups deviate significantly from the full randomness in the distribution of the squared Euclidean distances (defined and explained in the above-mentioned paper) at the $99.90 - 99.98\%$ and $98.51\%$ significance levels, respectively. They also found that the group of long GRBs does not deviate significantly from a random distribution.

### 1.3.6 Intensity distribution

A homogeneity distribution of the radial distances of GRB sources in space can be tested by comparing the measured peak fluxes with the so-called three-half law. If the sources are distributed homogenously in the Euclidean space, their number in the sphere with a certain radius and with the center at the observer's position is proportional to the radius cubed. The peak flux $P$, however, decreases with the distance squared. By the combination of these two dependences, it follows that the cumulative number of the bursts $N(> P)$ is proportional to $P^{-3/2}$. In this consideration it is assumed that the space is Euclidean and the intrinsic distribution of intensities $P$ is independent from the distance. From Fig. 1.12 it is seen that for the combined BATSE and PVO data the $N(> P)$ distribution follows the $-3/2$ law for most peak fluxes, however there is a lack of the weakest and thus the most distant bursts.

A direct statistical test for inhomogenity is the $V/V_{max}$ test (Schmidt 1968; Schmidt et al. 1988). For each burst the value $V/V_{max} = (C_p/C_{min})^{-3/2}$ is computed. $V$ is the spherical volume of the radius corresponding to the distance of a certain burst, $V_{max}$ is the maximal volume in which the given burst would be still detectable, $C_p$ is the peak count rate and $C_{min}$ is a minimal or limiting count rate that triggers detection. If the burst sources are distributed homogenously in the Euclidean space, this ratio will be uniformly distributed between 0 and 1, with an average value $\langle V/V_{max} \rangle = 1/2$.

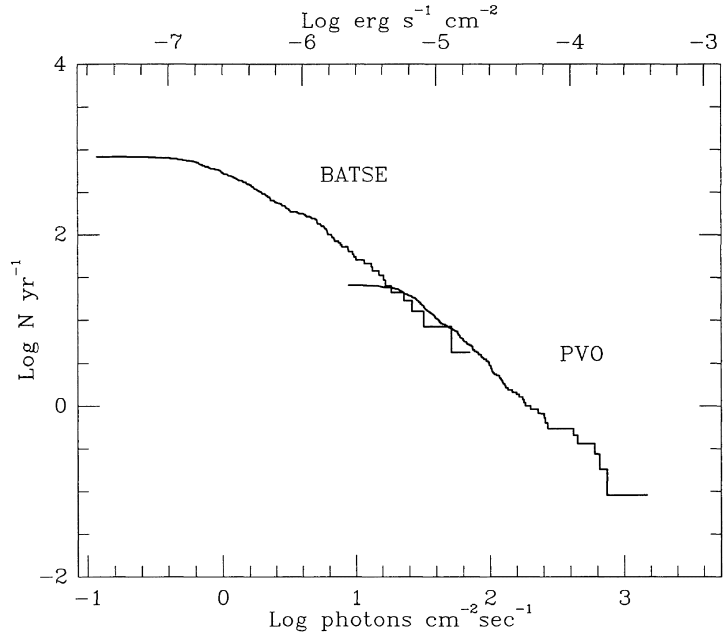

**Fig. 1.12.** The $\log N - \log P$ distribution from combined BATSE and PVO data. The distributions nicely match in the overlap region (Credit: Fenimore et al. 1993).

For the GRBs observed by BATSE it was found that $\langle V/V_{max} \rangle = 0.39 \pm 0.02$ for short bursts (Guetta & Piran 2005), and $\langle V/V_{max} \rangle = 0.29 \pm 0.01$ for long bursts (Guetta et al. 2005). Data from other missions give similar results, i.e., the average $V/V_{max}$ of all GRBs (short and long) are smaller than $1/2$. For example, the measurements from SMM gives $\langle V/V_{max} \rangle = 0.400 \pm 0.025$ (Higdon et al. 1992), from Ginga $\langle V/V_{max} \rangle = 0.35 \pm 0.035$ (Ogasaka et al. 1991), and from Granat-PHEBUS $\langle V/V_{max} \rangle = 0.376 \pm 0.017$ (Terekhov et al. 1994). This indicates that GRBs are in space distributed inhomogenously and/or are in cosmological distances (Mészáros & Mészáros 1995, 1996; Horváth et al. 1996; Reichart & Mészáros 1997).

## 1.4   Afterglow and modern era

### 1.4.1   BeppoSAX breakthrough

The breakthrough step in our understanding of GRBs was achieved thanks to the Italian-Dutch satellite BeppoSAX[1] (Boella et al. 1997). This satellite operated from April 1996 to April 2002 and collected 1082 GRB observations (Frontera et al. 2009). It carried on-board the Gamma-Ray Burst Monitor (GRBM) operating in the $40 - 700 \, \mathrm{keV}$ band, two Wide Field Cameras (WFC) in the band $2 - 28 \, \mathrm{keV}$ and four Narrow Field Instruments (NFI) in the band $0.1 - 10 \, \mathrm{keV}$ allowing observation of GRBs and its X-ray counterparts on the same platform. A typical BATSE error-box was a few degrees which made it very difficult to search for a GRB counterpart at longer wavelengths. The BeppoSAX satellite had the advantage that by the follow-up observations with WFC and NFI in the X-ray range it was able to reduce the localization down to a few arcminutes. Another advantage was the BeppoSAX's observational strategy. The satellite once per orbit transmitted data to the ground, thus in case of a GRB detection, the precise coordinates could have been distributed to ground-based observatories within a few hours. Needed to note that the real-time distribution of GRB locations detected by various spacecraft is provided by the GCN notices, which allows fast follow up with ground-based observatories and with robotic telescopes such as, e.g., BART[1], BOOTES[2].

---

[1]http://www.asdc.asi.it/bepposax/

The breakthrough moment came on February 28, 1997 when BeppoSAX detected the X-ray afterglow from GRB 970228 (Costa et al. 1997). The precise localization given by BeppoSAX led to the discovery of the very first optical afterglow (van Paradijs et al. 1997) (Fig. 1.13). The first radio afterglow of another burst (GRB 970508) was detected later (Frail et al. 1997).

### 1.4.2 Afterglow light curve

The intensity of the X-ray afterglow of the burst GRB 970228 was very fast decreasing. The fading followed a power-law with $F(t) \sim t^{-1.3 \pm 0.1}$ (Costa et al. 1997), as shown in Fig. 1.14. One week after the burst's discovery the ASCA[3] satellite observed the source (Yoshida et al. 1997) and reported that the afterglow was continuing to decay with the slope -1.4.

Figure 1.15 presents the light curves of the burst GRB 970228 in optical and near infrared bands and shows a power-law decline. After a few days the light curve presented a bump, which was later interpreted as the contribution from an underlying supernova (Galama et al. 2000). The connection between gamma-ray bursts and supernovae are discussed in more detail in the Section 1.4.7. It appeared in the BeppoSAX era that in comparison with the X-ray afterglows, which were present in most GRBs (90 %), the optical afterglows were less frequent (50 %) (Vedrenne & Atteia (2009), Chapter 3.3). GRBs without an optical afterglow are called "dark" GRBs. The percentage of "dark" bursts depends on the detection limit of the instruments used for the observation. Recent ground based results demonstrate, that the frequency of the bursts with optical transients could be much higher than 50 %, see, e.g., Greiner et al. (2011). In most cases the afterglow light curve cannot be described as one power-law but as a sequence of power-laws with different slopes and separated by one or more breaks (see Fig. 1.16).

At this point we should also mention the first observation of the optical emission of a GRB during the prompt phase. GRB 990123 was detected and localized quickly by BATSE with a precision of 10°. Only 22 seconds after the BATSE trigger, the ROTSE[4] robotic telescope pointed its 16° × 16° field of view instrument towards the

---

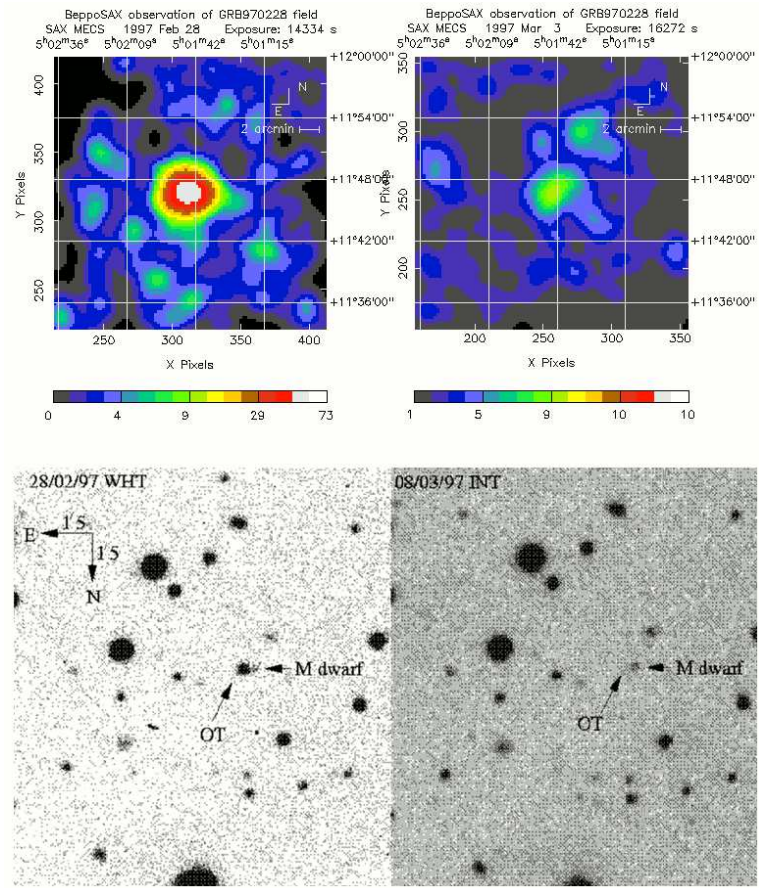

**Fig. 1.13.** A discovery of the X-ray and the optical afterglow/transient (OT) of the GRB 970228. *Top left:* The X-ray image of the afterglow 8 hours after the burst as seen by BeppoSAX. *Top right:* The same region 3 days afterwards. The brightness faded by a factor of ∼ 20. *Down left:* The V-band image of the GRB region obtained by the 4.2 m William Herschel telescope (La Palma) 20.8 hours after the burst's beginning. *Down right:* The same area of the sky a few days later taken by the 2.5 m Isaac Newton telescope (La Palma). The brightness dropped under the telescope's limit. An M dwarf star most likely at a distance of ∼ 3 kpc is also indicated (Credit: Costa et al. 1997, van Paradijs et al. 1997).

error-box an started with observation (Akerlof et al. 1999). This optical emission was extremely bright, reaching 9 mag for a source at redshift $z = 1.6$ (see more about the GRB redshifts in Section 1.4.4). Another GRB early optical emission observation was obtained for GRB 041219A, when the first optical detection started only 8 s after the INTEGRAL (see Section 1.6.1) alert with the Raptor[1] telescope (Vestrand et al. 2005). A concurrent optical and gamma observation of the prompt phase of GRB 080319B was achieved by TORTORA and "Pi of the Sky"[2] telescopes (Racusin et al. 2008). This burst is also known as a naked-eye GRB because its peak optical brightness reached amazing 5.3 mag at redshift $z = 0.94$.

---

[1] http://www.raptor.lanl.gov/
[2] http://grb.fuw.edu.pl/index.html

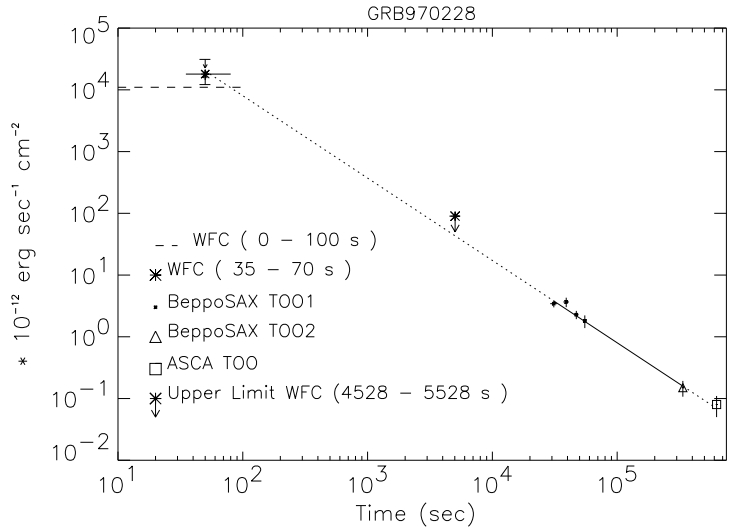

**Fig. 1.14.** The fading X-ray afterglow GRB 970228. Data from BeppoSAX TOO 1/2 in the range 2−10 keV are fitted by a power-law with a slop of -1.32 (solid line). The data extrapolation forward (dotted line) is in a good agreement with the ASCA satellite measurement. The extrapolation backward well matches the average flux detected with WFC during the last part of the GRB (Credit: Costa et al. 1997).

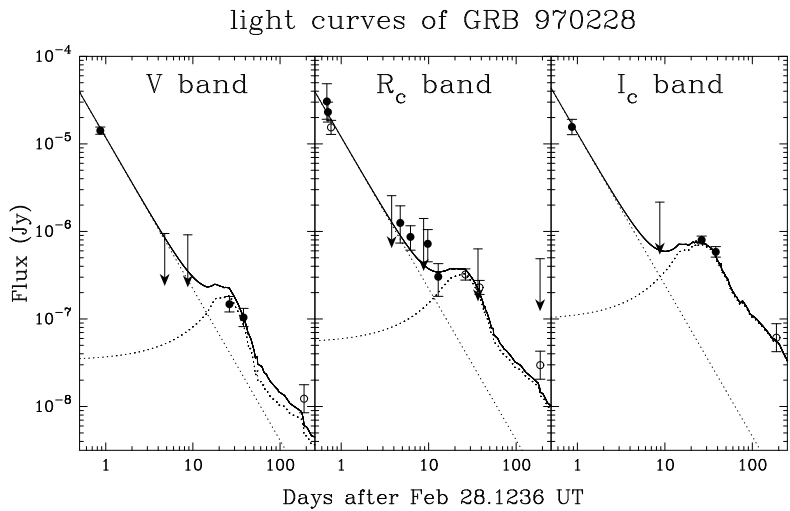

**Fig. 1.15.** The afterglow of GRB 970228 measured in V, $R_c$ and $I_c$ bands at late times. The light curves display a bump that was interpreted as the contribution from an underlying supernova. The afterglow decay is fitted with a superposition of a power-law ($F(t) \sim t^{-1.51}$) and the contribution of a supernova like SN 1998bw, redshifted at $z = 0.695$, which is the redshift of the burst (Credit: Galama et al. 2000).

### 1.4.3  Afterglow spectra

The afterglow of GRB 970508 was the first afterglow concurrently observed over the whole electromagnetic spectrum few hours after the burst. It was feasible to compose its multi-wavelength spectrum from X-rays to radio (Galama et al. 1998a). Figure 1.17 shows three spectral breaks connecting different power-law sections at three frequencies: $\nu_a, \nu_m, \nu_c$. This behaviour supports the fireball model discussed in Section 1.5.1. The whole spectrum together with its breaks supports the idea that GRB afterglows are driven by synchrotron emission of accelerated electrons in a relativistic shock.

### 1.4.4  GRB redshifts

The gamma-ray burst GRB 970508 holds, besides the first broadband afterglow observation from X-rays to radio waves, also another and probably more notable record. It was the first GRB whose distance was measured. The brightness of the optical counterpart remained high for more than days and allowed measurement of its spectrum with Keck Telescope (Metzger et al. 1997). The spectrum revealed absorption lines of Fe II, and Mg II at redshift $z = 0.835$ (Fig. 1.18). That was a firm evidence that at least this burst lay at a cosmological distance and it settled the scale of the energy released in GRBs as it had been predicted by Paczyński (1986). As soon as the afterglow faded under the detection limit a host galaxy was identified exactly at the same position where the afterglow lied. The spectrum of the galaxy exhibited emission lines of O II and Ne III at a redshift $z = 0.835$ (Bloom et al. 1998).

The redshift measurement of a short GRB came up several years later. GRB 050709 ($T_{90} \approx 70\,\text{ms}$) was at first noticed by the HETE-2 satellite, which determined its position with the accuracy of $0.5°$ (Butler et al. 2005). After that the Chandra X-ray Observatory focused on this area and specified the position to 81 arcsec. Then a new satellite Swift (see next section) detected its X-ray afterglow (Villasenor et al. 2005). Observations in the optical range from Hubble Space Telescope (HST) (Fox et al. 2005) (see Fig. 1.19) and from several ground-based observatories followed. The burst was localized at the edge of an irregular star-forming galaxy at redshift $z = 0.16$.

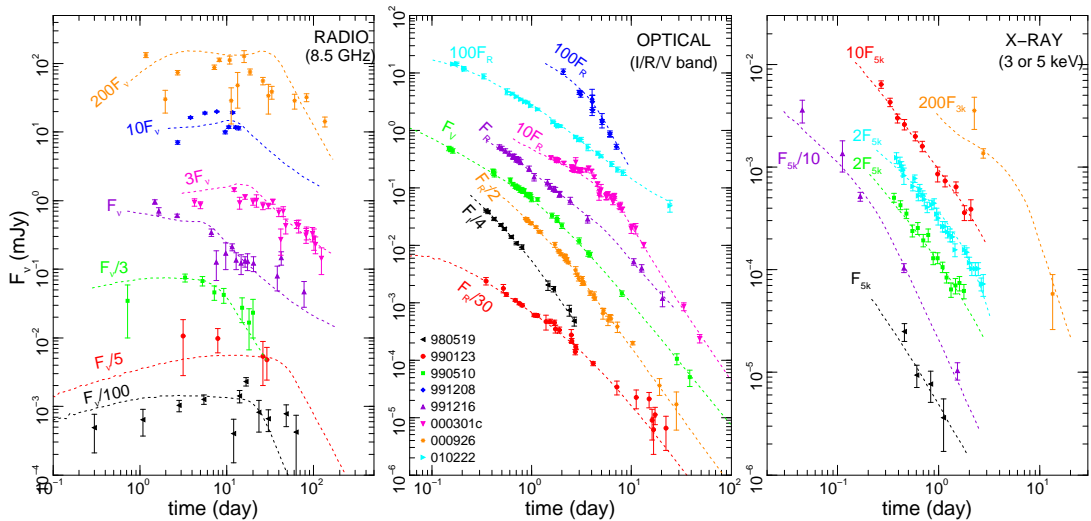

**Fig. 1.16.** Radio, optical, and X-ray afterglow light curves with fitted model curves for selected GRBs indicated in the middle panel. Optical data have been corrected for Galactic dust extinction. Fluxes have been multiplied by the indicated factors, for clarity (Credit: Panaitescu & Kumar 2001).

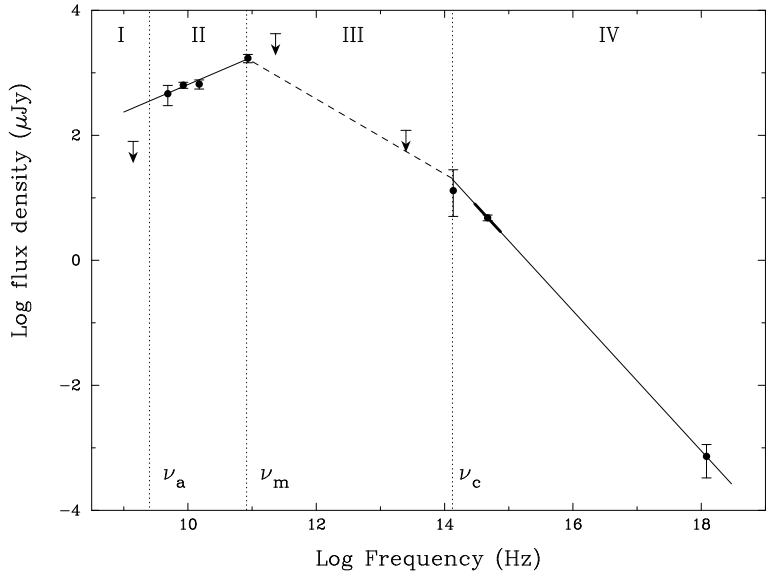

**Fig. 1.17.** Radio to X-ray energy spectrum of GRB 970508 as measured 12.1 days after the burst. Break frequencies are indicated (Credit: Galama et al. 1998).

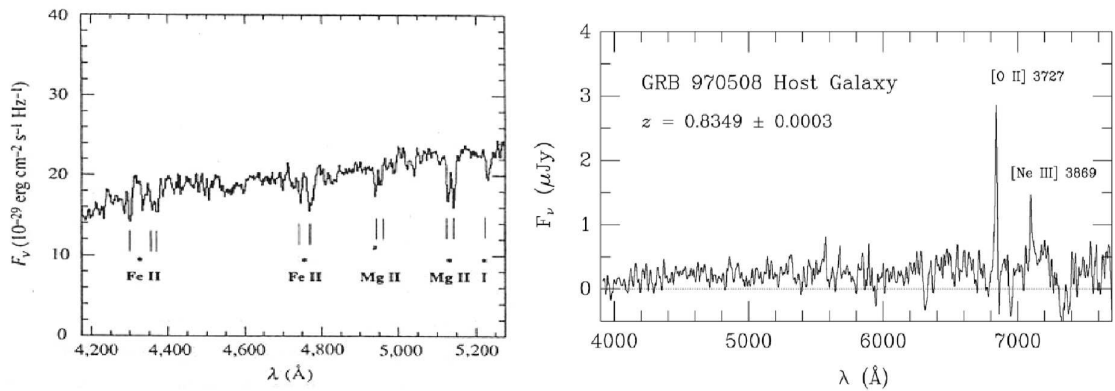

**Fig. 1.18.** *Left panel:* The spectrum of the optical afterglow of GRB 970508 showing Fe II and Mg II absorbtion lines at redshift $z = 0.835$. *Right panel:* The spectrum of the host galaxy showing O II and Ne III emission lines at the same redshift (Credit: Metzger et al. 1997, Bloom et al. 1998).

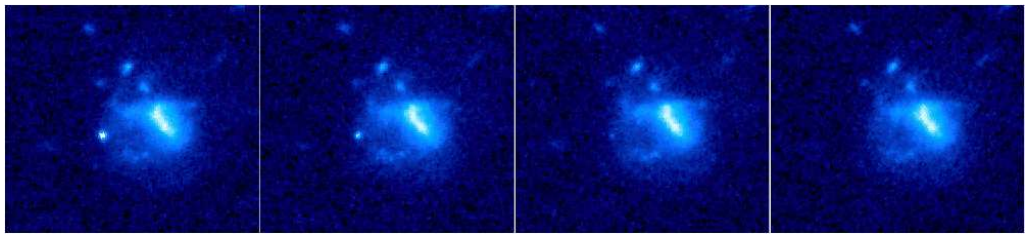

**Fig. 1.19.** The Hubble Space Telescope images showing the fading afterglow and the host galaxy of GRB 050709. The images were taken 5.6, 9.8, 18.6 and 34.7 days after the burst (Credit: Fox D., Penn State University).

Nowadays the measured range of all GRB redshifts goes from $z = 0.0085$ for GRB 980405 to $z = 8.2$ for GRB 090423. Figure 1.20 presents a distribution of $\sim 130$ redshifts measured by the Swift satellite.

The average redshift for short GRBs ($T_{90} < 2\,\mathrm{s}$) observed by Swift over first 5 years is $\bar{z} = 0.8$ and for long ones ($T_{90} > 2\,\mathrm{s}$) $\bar{z} = 2.2$, which indicates that the population of short bursts is closer than the population of long ones. The average redshift for all Swift GRBs is $\bar{z} = 2.1$. Bagoly et al. (2006) and Jakobsson et al. (2006) found that pre-Swift bursts had lower average redshift, i.e., $\bar{z} \approx 1.3$. The reason for this difference is the higher sensitivity of Swift compared to BeppoSAX and HETE-2 (Gehrels et al. 2009).

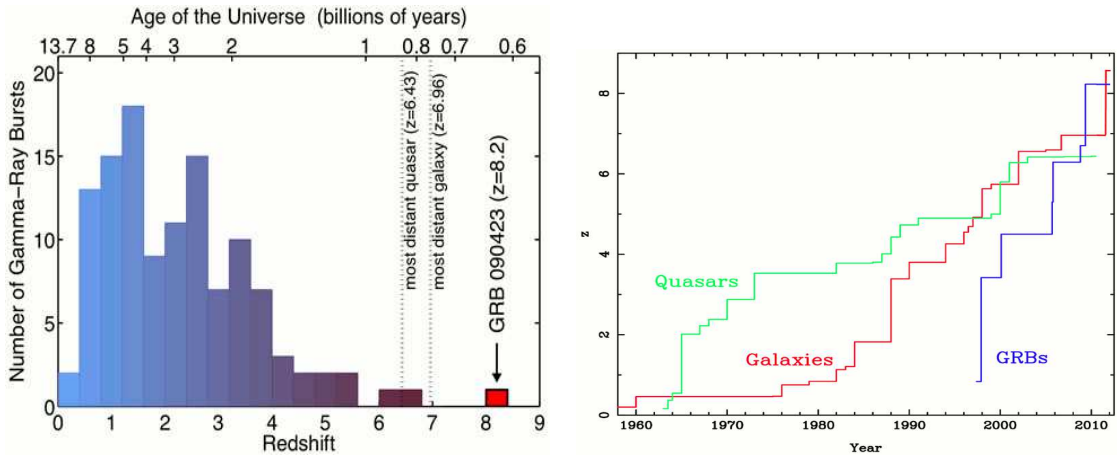

**Fig. 1.20.** *Left panel:* Distribution of redshifts and corresponding age of the Universe for GRBs detected by Swift over first four years (Credit: Berger E., Harvard/CfA). *Right panel:* The history of the most distant objects detected so far (Credit: Tanvir 2010).

### 1.4.5 The Swift satellite

Some great products of successful Swift mission[1] have already been mentioned, therefore it is worth introducing the spacecraft itself. This autonomous satellite was launched in November 2004, it is still active and it is specialized in observing and prompt in-flight localizing GRBs in gamma, X-ray, and optical wavebands (Gehrels et al. 2004). The satellite carries three observational instruments on-board:

1. Burst Alert Telescope (BAT), using a coded mask technique with a detection area $5240\,\mathrm{cm}^2$, with the field of view $1.4\,\mathrm{srad}$, the energy range $15 - 150\,\mathrm{keV}$ and with the spatial resolution $1 - 4'$. The number of GRBs observed annually is approximately 100.

2. X-Ray Telescope (XRT), having the energy range $0.2 - 10\,\mathrm{keV}$, $23'$ field of view and $18''$ spatial resolution.

---

[1] http://swift.gsfc.nasa.gov/docs/swift/swiftsc.html

3. Ultra-Violet/Optical Telescope (UVOT), which is co-aligned with the XRT, allowing broadband UV-visible photometry and low-resolution spectra of bright GRBs in the range $170 - 600$ nm.

If the BAT instrument detects a burst, Swift automatically slews to the BAT estimated position within from tens to hundreds of seconds to allow the prompt observation of the afterglow by the XRT and UVOT telescopes. The great Swift's benefit of the quick follow-up X-ray afterglow observations is shown in Fig. 1.21, which displays a complexity of a typical X-ray afterglow and its possible extrapolation to the prompt phase. The Swift satellite has detected more than 500 GRBs[1] to date (Sakamoto et al. 2008).

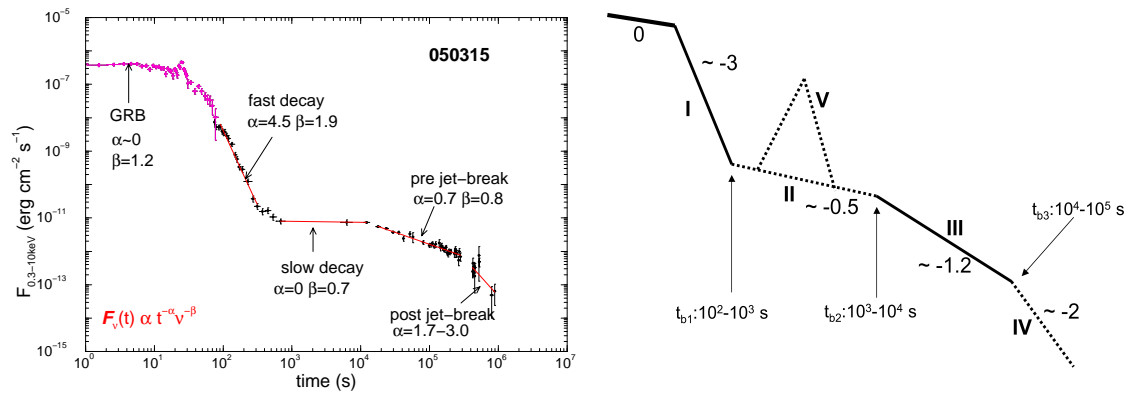

**Fig. 1.21.** *Left panel:* Phases of the afterglow of GRB 050315. This GRB provides a good illustration of the typical times and properties of each phase. Magenta points denote the $15 - 150$ keV Swift BAT measurements of the prompt emission. Data are extended to the $0.3 - 10$ keV Swift XRT measurements (black points). Red lines represent the fits of each phase. The characteristic values of $\alpha$ (the temporal decay index) and $\beta$ (the spectral index) of each phase are also indicated (Credit: Panaitescu 2006). *Right panel:* A cartoon of a typical X-ray afterglow based on the Swift XRT data. Phase 0 denotes the prompt emission, I denotes the prompt decay phase, II is the shallow phase with V a possible X-ray flare (a discovery of the Swift satellite). Phase III was already known from the BeppoSAX time with a possible break leading to phase IV (Credit: Zhang et al. 2006).

### 1.4.6  Host galaxies

Many surveys of GRB host galaxies have been written, see e.g., Fynbo et al. (2003, 2008); Chary et al. (2007); Le Floc'h et al. (2003, 2006) and references therein. The hosts have median apparent magnitude R $\approx 25$ mag, with detections reaching down to R $\approx 29$ mag. Figure 1.22 presents a sample of GRB host galaxies. A list of GRB hosts can be found at GHostS webpage[2].

It was suggested that the short-duration bursts occur rather in regions of low star formation, either in low star-forming elliptical galaxies as for GRB 050509B (Gehrels et al. 2005; Bloom et al. 2006) and GRB 050724 (Barthelmy et al. 2005; Berger et al. 2005) or in a region of a galaxy with low star formation (Fox et al. 2005; Hjorth et al. 2005; Villasenor et al. 2005). On the other hand, long-duration bursts seems to appear in high star-formation galaxies (Christensen et al. 2004) or in the high star-forming regions of a galaxy (Fruchter et al. 2006; Svensson et al. 2010). The hosts of short-duration bursts seems to have lower specific star formation rates by about an order of magnitude than the hosts of long-duration bursts (Berger 2009). However, the suggestion that hosts of short-duration GRBs are quiescent galaxies requires a larger sample to be confirmed (Savaglio et al. 2009).

Besides the star formation rate, it seems that host galaxies of short and long-duration bursts also differ in the average offset of the burst's position and the center of its host. A survey done by Fong et al. (2010) revealed that the projected physical offsets of 10 short GRBs range from about 1 to 50 kpc, with a median value of about 5 kpc. On the other side, the median projected offset in a survey of 20 long GRBs was found to be 1.3 kpc (Bloom et al. 2002). Similarly, no offsets of long GRBs are larger than about 7 kpc, whereas at least some short GRBs appear to have offsets in excess of 15 kpc. However, the distribution of host-normalized offsets of short GRBs is nearly identical to that of long GRBs. This is due to the systematically larger size of short GRB hosts (by about a factor of 2).

### 1.4.7 SN-GRB connection

During the last few years the understanding of type Ib/c supernovae (SNe) and their connection with long/duration GRBs has made a large progress. For reviews of a present knowledge of GRB-SN connection see Soderberg (2006); Woosley & Bloom (2006), with a comprehensive list of references therein.

The first GRB connected with a SN was GRB 980425, which was associated with SN 1998bw, a remarkably nearby and very bright type Ic supernova. From the very broad optical spectral features that were observed in SN 1998bw it was inferred that the explosion was unusually energetic, more than 10 times that of an ordinary supernova. GRB 980425 is the closest burst recorded to date with redshift $z = 0.0085$, which means a distance of about 37 Mpc (Galama et al. 1998b,c), and with a duration of about 30 s (Soffitta et al. 1998). The energy of the burst released in gamma-rays was only $8.5 \pm 0.1 \times 10^{47}$ erg, assuming isotropic emission (it is only a small fraction of the total released energy). This is more than three orders of magnitude lower than the majority of long-duration bursts (Frail et al. 2001; Bloom et al. 2003a).

The second GRB-SN association was GRB 030329 and SN 2003dh (Hjorth et al. 2003; Stanek et al. 2003). This GRB was very bright (with a fluence higher than 99.8 % of bursts detected by BATSE), at redshift $z = 0.1685$ (GCN 2020) and with the duration more than 25 s (GCN 1997). The optical afterglow was unusually bright with a magnitude of R = 12.6 1.5 hours after the burst and it was also very bright in radio and X-ray wavelengths. This burst was also the first one for which a spectroscopic signature of the underlying supernova in the optical afterglow was observed (Stanek et al. 2003), see Fig. 1.23. From the study of the optical curves of this su-pernova, it was suggested that the progenitor was a main sequence star with a mass of $\sim 25 - 40 \, M_{\odot}$ (Deng et al. 2005).

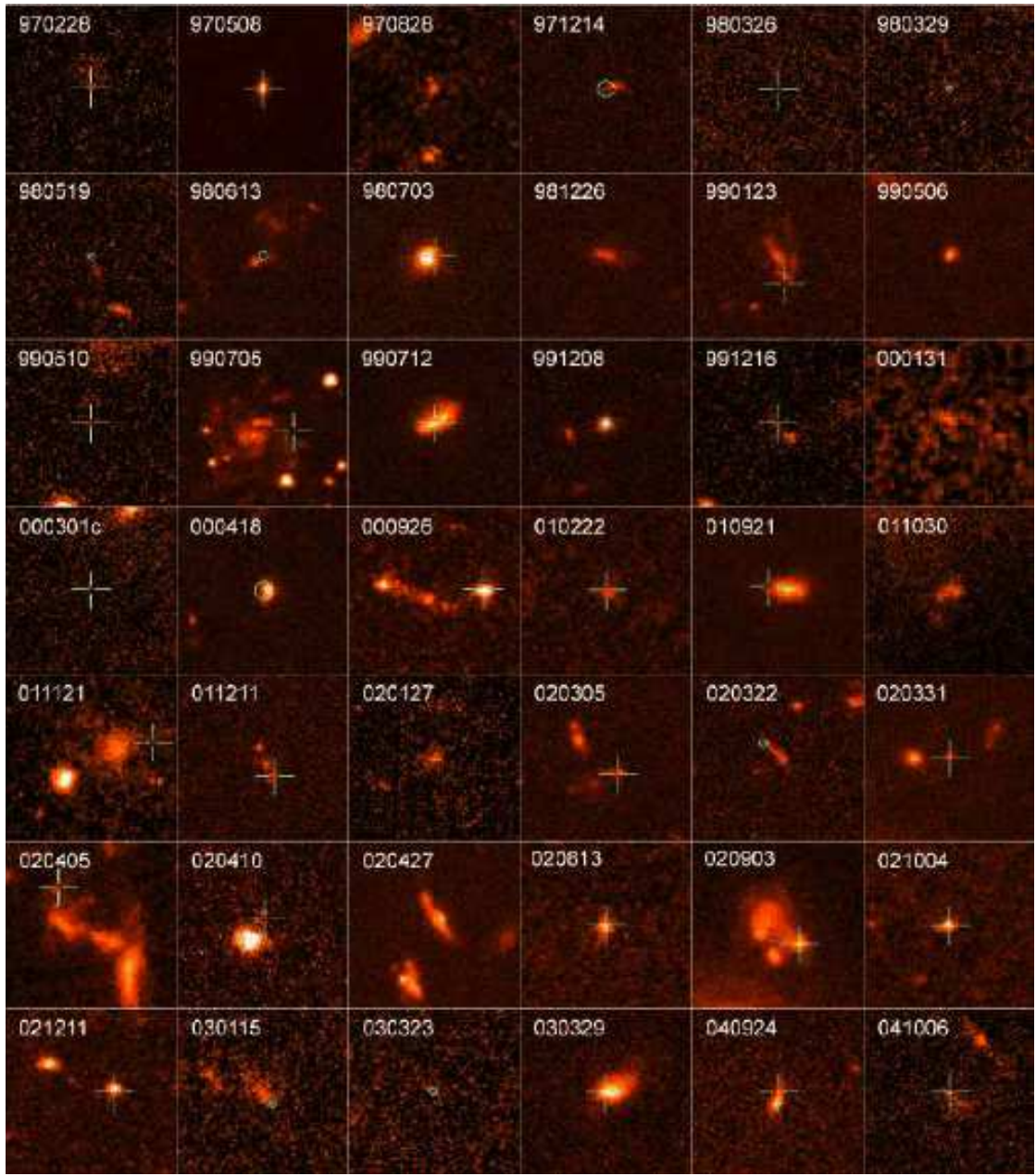

**Fig. 1.22.** A sample of GRB host galaxies imagined by HST. The positions of bursts are indicated by a green mark (Credit: Fruchter et al. 2006).

Besides these two cases of the association between long-duration busts and supernovae, there have been observed many of such connections, e.g., see Woosley & Bloom (2006). An example of the optical afterglow with a clear underlying supernova signature is shown in Fig. 1.24.

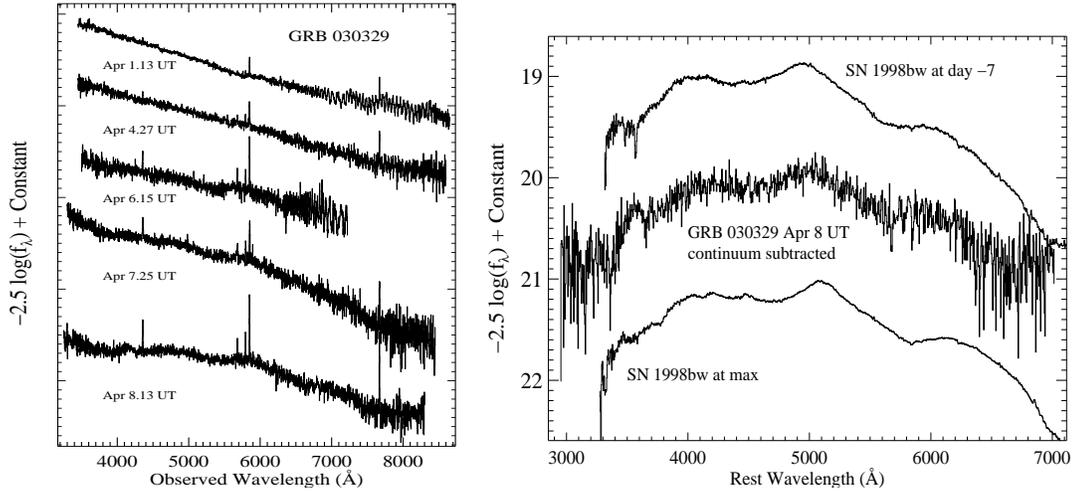

**Fig. 1.23.** *Left panel:* Evolution of the GRB 030329/SN 2003dh spectrum, from April 1.13 UT (2.64 days after the burst) to April 8.13 UT (9.64 days after the burst). The early spectra consist of a power-law continuum with narrow emission lines originating from HII regions in the host galaxy at a redshift of $z \approx 0.168$. Spectra taken at later times, after April 5, show a development of broad peaks in the spectra characteristic of a supernova. *Left panel:* The spectrum measured on April 8 with continuum subtracted. The residual spectrum shows broad bumps at approximately 5000 and 4200 Å (rest frame), which is similar to the spectrum of the peculiar Type Ic SN 1998bw a week before maximum light. (Credit: Stanek et al. 2003).

## 1.5 Models

### 1.5.1 The fireball model

In this section the most favored model, the fireball model or standard model, explaining GRBs and their afterglows is outlined. It is a theory describing the emission mechanisms of GRBs regardless the source progenitor (see next section). It assumes a release of an enormous amount of energy from the progenitor in a small region within a short time that powers a high-temperature relativistically expanding ejecta (fireball) beamed into a jet that interacts with a circum-burst medium. The first fireball shock model based on external and internal shocks in the circum-burst and ejected material was introduced by Rees & Mészáros (1992) and by Rees & Mészáros (1994), respectively.

According to this model, the variable prompt gamma emission is produced in internal shocks. They occur inside the fast-moving ejecta when the time-varying outflow from the central engine leads to the ejection of successive shells of material with different Lorentz factors and when the faster shells overtake and collide with the slower ones.

At later times, the external shocks appear when the relativistic flow is decelerated by the interaction with the ambient medium surrounding the source. This shock is responsible for the longer-wavelength afterglow radiation which decays with time (Katz 1994; Mészáros & Rees 1997; Paczyński & Rhoads 1993; Vietri 1997; Waxman 1997). The external shock generates a reverse shock crossing the ejecta (Daigne 2004; Piran 2004; Vedrenne & Atteia 2009). The main emission mechanism in both the internal and external shocks is supposed to be synchrotron emission from relativistic electrons twisting around the magnetic field amplified at the shocks (see Fig. 1.25). A schema of the fireball model and its phases is shown in Fig. 1.26.

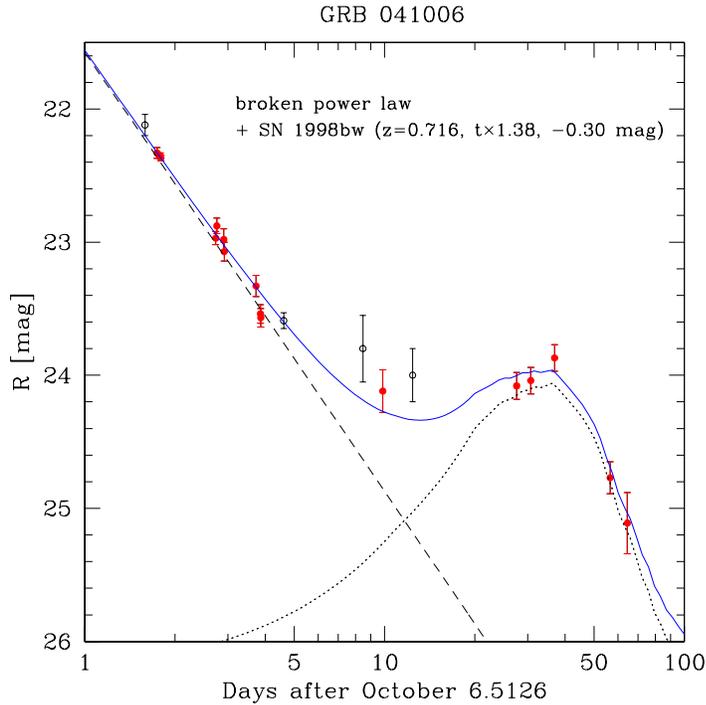

**Fig. 1.24.** The R band light curve of the afterglow of GRB 041006 (red and opened circles denote data from different authors). The power-law decay with a slope of -1.3 is shown by the dashed line. A supernova SN 1998bw light curve extended by a factor 1.38 is indicated by the dotted curve. The solid cure denotes the combination of both components. The supernova bump, powered by the radioactive decay $^{56}Ni \rightarrow ^{56}Co \rightarrow ^{56}Fe$ (by K-capture), is clearly evident (Credit: Stanek et al. 2005).

A beamed jet of the relativistic ejecta was for the first time pointed out by Rhoads (1997). Figure 1.27 illustrates the following effect: due to the relativistic beaming, the observer can only see photons coming from a small fraction of the emitting surface, this region being centered on the line of sight and having an opening angle $\sim 1/\Gamma$. Initially, except for a highly beamed ejecta, this opening angle is small compared to the geometrical opening angle of the ejecta $\Theta_0$. The deceleration due to the external medium makes the Lorentz factor $\Gamma$ decreases to $\Gamma_B$ and then the angle $1/\Gamma$ increases to $1/\Gamma_B$. The observer can see the whole emitting surface for the first time in the history of the jet propagation at the time when $1/\Gamma_B \simeq \Theta_0$. This should appear as a break in the light curve since a fraction of the photons emitted on the edge of the surface are now not sufficiently beamed to be detected. An additional physical effect has to be taken into account, the lateral expansion of the ejecta. The geometrical opening angle of the relativistic ejecta is not kept constant to $\Theta_0$ during the whole evolution. However, it can be shown that this lateral expansion becomes important more or less at the same period when $1/\Gamma_B \sim \Theta_0$ (Daigne 2004).

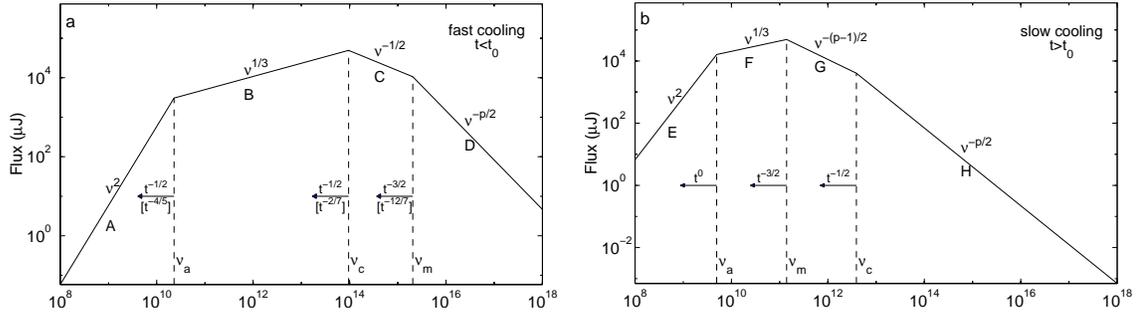

**Fig. 1.25.** Synchrotron spectrum emitted by electrons accelerated in a relativistic shock (x-axis denotes frequency in Hertz). The electrons have a power-law distribution of energies of index $p$. (a) *Fast cooling*, which is expected at early times. The spectrum consists of four segments. The frequencies, $\nu_m$, $\nu_c$, and $\nu_a$, decrease with time as indicated; the scalings above the arrows correspond to an adiabatic evolution, and the scalings in square brackets correspond to a fully radiative evolution. (b) *Slow cooling*, which is expected at late times. In this case the evolution is always adiabatic. In both cases self-absorption is important below $\nu_a$ (Credit: Sari et al. 1998).

Some observations indicate breaks in the light curves that can be interpreted as resulting from this effect. One important signature which confirms that a given break is due to this geometrical effect is that it is expected to be achromatic, i.e., observed at the same time at all frequencies. An example of the observed achromatic break in a GRB afterglow is shown in Fig. 1.28.

Nevertheless, in the recent years, many observations either not presenting an achromatic break or presenting brakes in optical and X-ray afterglows at different times came up. For example Melandri et al. (2008) showed that although the majority of the bursts in their sample (14 out of 24) were consistent with the standard model, for a significant fraction (10 bursts), the data could not be explained by the standard model, even if some modifications were made, i.e., energy injection or variation in the ambient matter (Fig. 1.29). A possible explanation beyond the standard model can be that the early X-ray afterglow is not due to forward shock emission, but is instead produced by late-time central engine activity (Ghisellini et al. 2007).

An alternative GRB model was developed by Dar & de Rújula (2004), a cannonball model. In the cannonball model a core-collapse supernova emits highly relativistic bullets (or blobs) of matter having about the mass of the Earth. The prompt gamma-ray emission is explained by inverse Compton scattering of light in the near environment of the supernova by the electrons in the plasma of the relativistic bullets.

### 1.5.2   Collapsar as a progenitor

The isotropic equivalent energies emitted by GRBs in gamma-rays reach $10^{50}$ to $10^{54}$ erg. As most GRBs are probably collimated with typical opening angles between

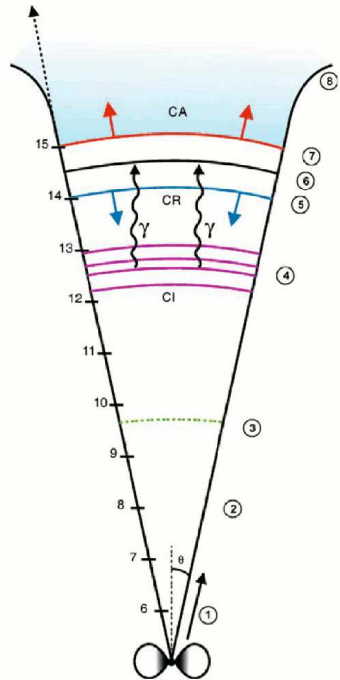

**Fig. 1.26.** A schema of the fireball model and its step-by-step phases. The left-hand scale is the logarithm of the distance (in meters) from the central GRB progenitor. The right-hand scale indicates the gradual phases: (1) Phase of acceleration: the thermal or magnetic energy is converted into kinetic energy of baryons within the ejecta that creates an ultrarelativistic jet with a certain beaming angle. (2) End of the acceleration phase: the Lorentz factor of the ejecta attains its maximum, typically few hundred. (3) The ejecta become transparent at the photospheric radius; if internal energy is available it can be radiated as quasithermal emission. (4) Internal shocks (CI) appear and progress within the relativistic ejecta. The matter heated by the internal shocks is the source of the prompt gamma-rays. (5), (6), (7) These phases characterize the various parts of the external shocks, which start with the deceleration of the ejecta as collide with the circumburst medium: (5) corresponds to the reverse shock (CR) which propagates backwards within the ejecta; (6) is the contact discontinuity; (7) is the forward shock (CA) that is responsible for the afterglow emission. (8) Corresponds to a region where the Lorentz factor highly decrease so that the expansion becomes non-relativistic (Vedrenne & Atteia 2009) (Credit: Atteia & Mochkovitch 2004).

$1°$ and $20°$, the true released energy in gamma rays is around $10^{50}$ to $10^{51}$ ergs. This amount of energy must be released during few seconds. Thus the central engine must be able to generate such high energies and accelerate typically $10^{-5} M_\odot$ to relativistic velocities. The short time variability of the prompt emission implies energy deposition in a very small volume, so it appears natural to invoke compact objects and probably black holes. Depending on the collimation of the ejecta, typical GRBs would have to occur once per $10^4 - 10^6$ years and galaxy, which is about $1 - 0.01\%$ of the rate of supernovae (Piran 2004). No repetition has been observed, which signals that during these events the progenitor is catastrophically destroyed. It was suggested that GRBs are associated with objects involving accretion of a massive ($\sim 0.1 M_\odot$) disk onto a compact object, most likely a newborn black hole.

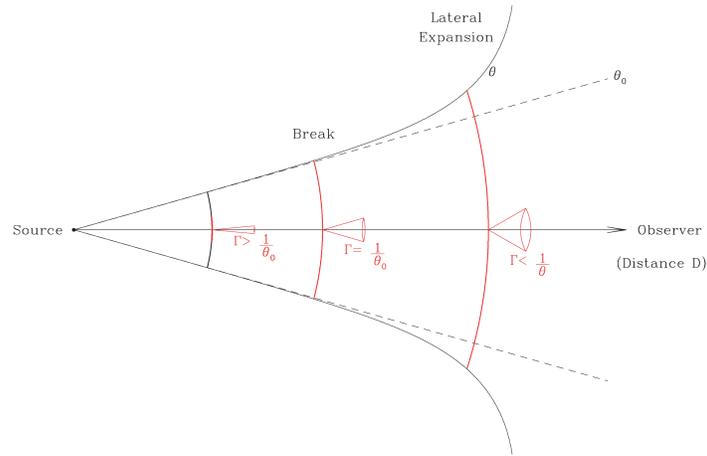

**Fig. 1.27.** A schema displaying how the geometrical and relativistic beaming of the ejecta form an achromatic break in the GRB afterglow. (Credit: Daigne 2004).

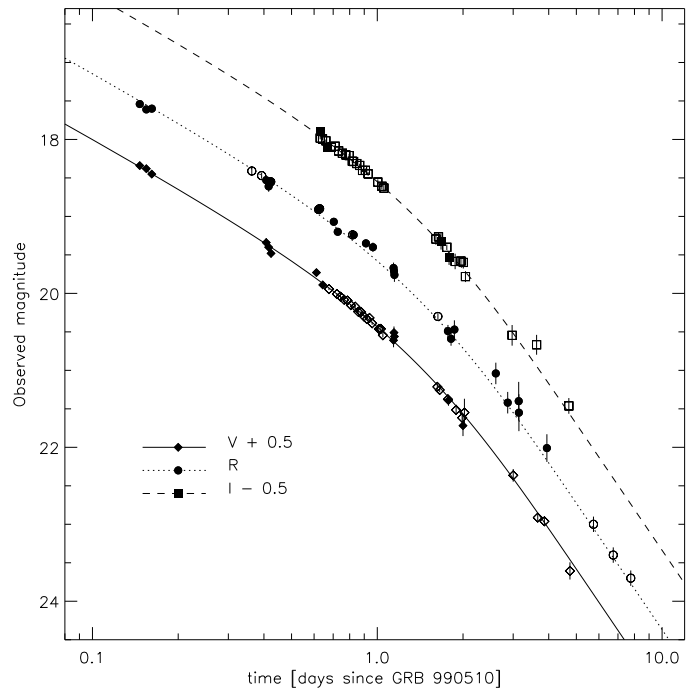

**Fig. 1.28.** Light curves of the optical afterglow of GRB 990510, measured at different wavelengths, clearly showing an achromatic break (Credit: Harrison et al. 1999).

Such conditions can result, either from the explosion of a massive star (collapsar model), or from a coalescence of binary compact stellar remnants (a model of merger). In both cases a spinning black hole is formed with a temporary accretion disk of debris. Hence the GRB can be powered by the gravitational energy of the in-falling matter. The spin energy of the black hole can be another source of energy through the Blandford-Znajek process (Blandford & Znajek 1977). Usov (1992) proposed an alternative to accretion, it involves pulsar-like activity of a rapidly rotating compact object with enormous magnetic field: a newborn 'magnetar'.

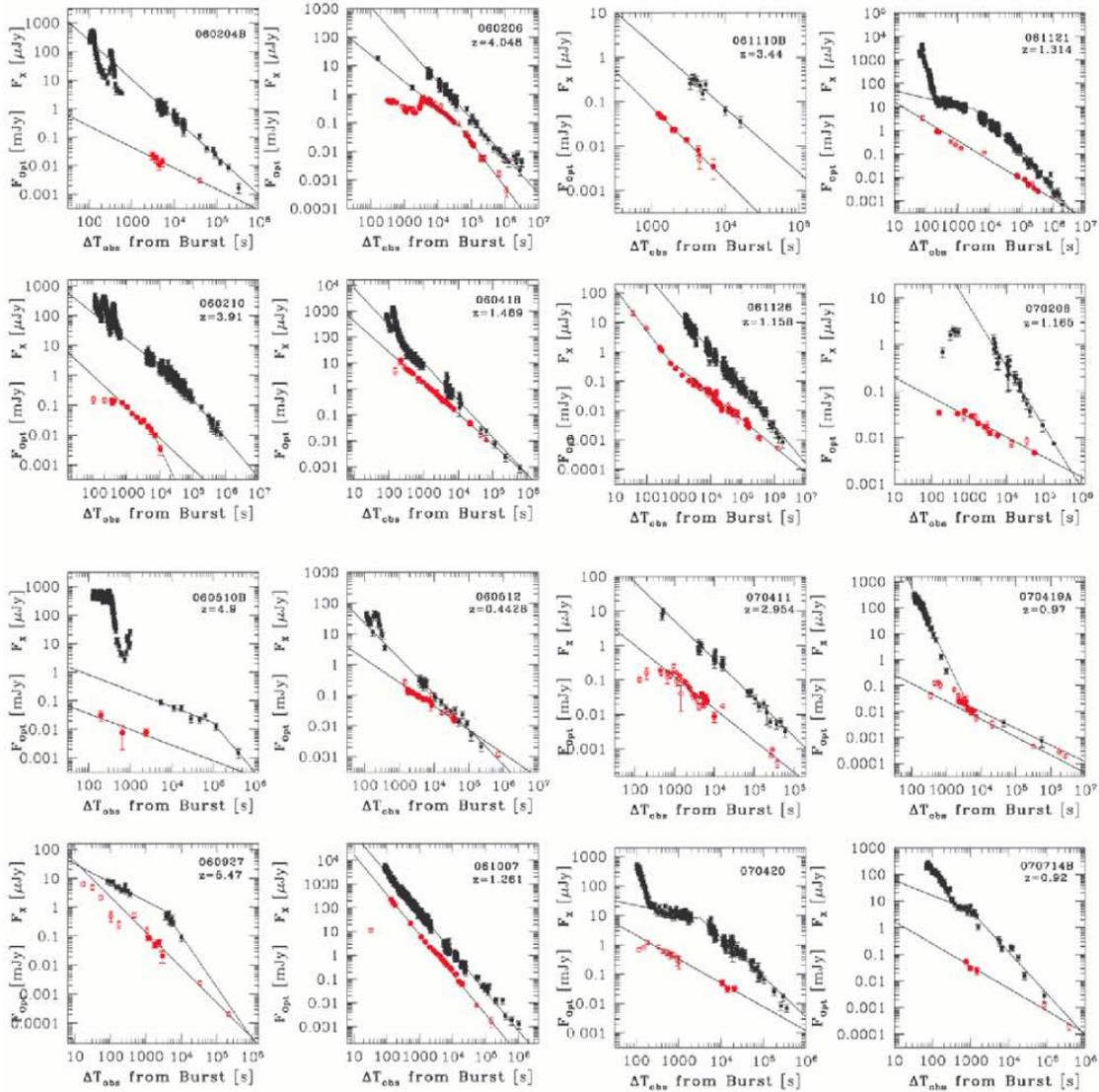

**Fig. 1.29.** A part of GRB afterglows from this sample of R-band optical (red circles) and X-ray (black squares) observations do not demonstrate achromatic breaks and thus challenge the standard model (Credit: Melandri et al. 2008).

The long-duration bursts ($T_{90} \gtrsim 2\,\mathrm{s}$), or Type II according to the definition by Zhang et al. (2009), are believed (mainly because of GRB-SN connection) to result from collapsars. They are rotating massive stars whose iron core eventually collapses directly to form a black hole (Woosley 1993). In the seconds/minutes following the collapse, the black hole accretes the residual matter and emits a relativistic jet that is detected as a GRB if directed towards the observer. The collapsars probably originate in Wolf-Rayet stars and are thus closely related to the hydrogen-deficient supernovae, i.e., type Ib/Ic (MacFadyen & Woosley 1999). Figure 1.30 illustrates a hydrodynamic simulation of the beamed ejecta emitted by a collapsar (Zhang et al. 2003). For another properties and details of collapsars see, e.g., Fryer et al. (1999); MacFadyen et al. (2001); Woosley & Heger (2006); Zhang et al. (2004).

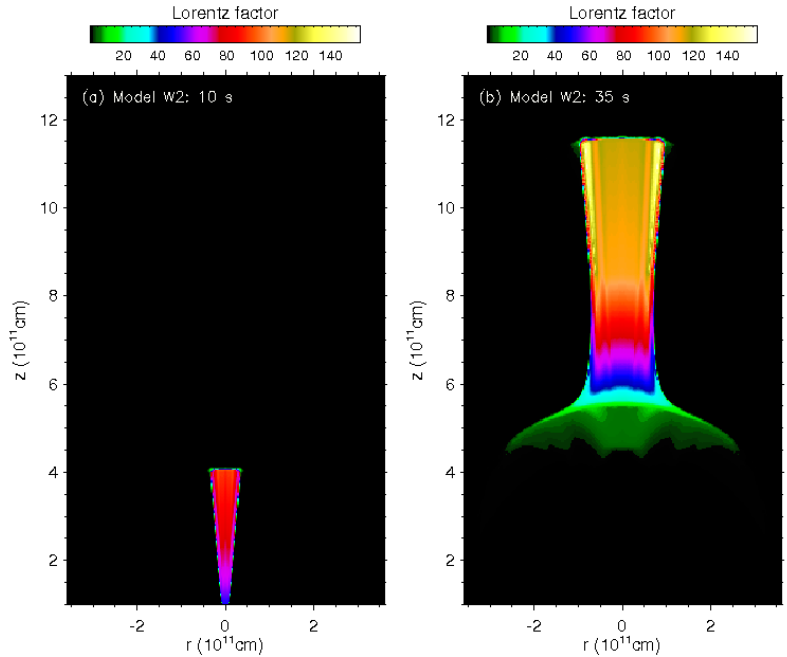

**Fig. 1.30.** A hydrodynamic simulation of the jet's Lorentz factor for the collapsar model W$_2$ of Zhang et al. (2003) at $t = 10$ s (left panel) and at $t = 35$ s (right panel) after the jets begin to propagate in the stellar wind. The initial opening angle is 3°. At $t = 35$ s the opening angle is 5°. As the Lorentz factor gradually decreases after 10 s, the jet has more lateral expansion (Credit: Zhang et al. 2003).

### 1.5.3 Merger as a progenitor

The short-duration bursts ($T_{90} \lesssim 2$ s), or Type I according to the definition by Zhang et al. (2009), might originate in a merger of compact objects. This was introduced by Paczyński (1986), and Eichler et al. (1989). The mergers can be of various types: neutron star + neutron star (NS-NS), black hole + neutron star, black hole + white dwarf, or black hole + helium star. The decrease of the separation between the two neutron stars due to gravitational radiation inevitably leads to the coalescence (Ruffert & Janka 2001; Taylor 1994).

Similarly to the collapsar model a spinning black hole is formed. The debris from the tidally disrupted compact object forms a temporary accretion disk, which sooner or later falls into the black hole releasing a fraction of its gravitational energy. Figure 1.31 illustrates a simulation of the NS-NS merger (Price & Rosswog 2006). Concerning a rate of NS-NS merger (Phinney 1991; Narayan et al. 1992; van den Heuvel & Lorimer 1996; O'Shaughnessy et al. 2008), it was estimated to be compatible with the GRB rate (Fenimore et al. 1993).

Nevertheless, a decisive evidence, e.g., detection of gravitational waves with a specific signal (see Section 1.7.3), of the GRB progenitor being a merger of compact objects, is still missing. To add on, a way how to explain short GRBs by the collapsar model was also proposed (Zhang et al. 2003).

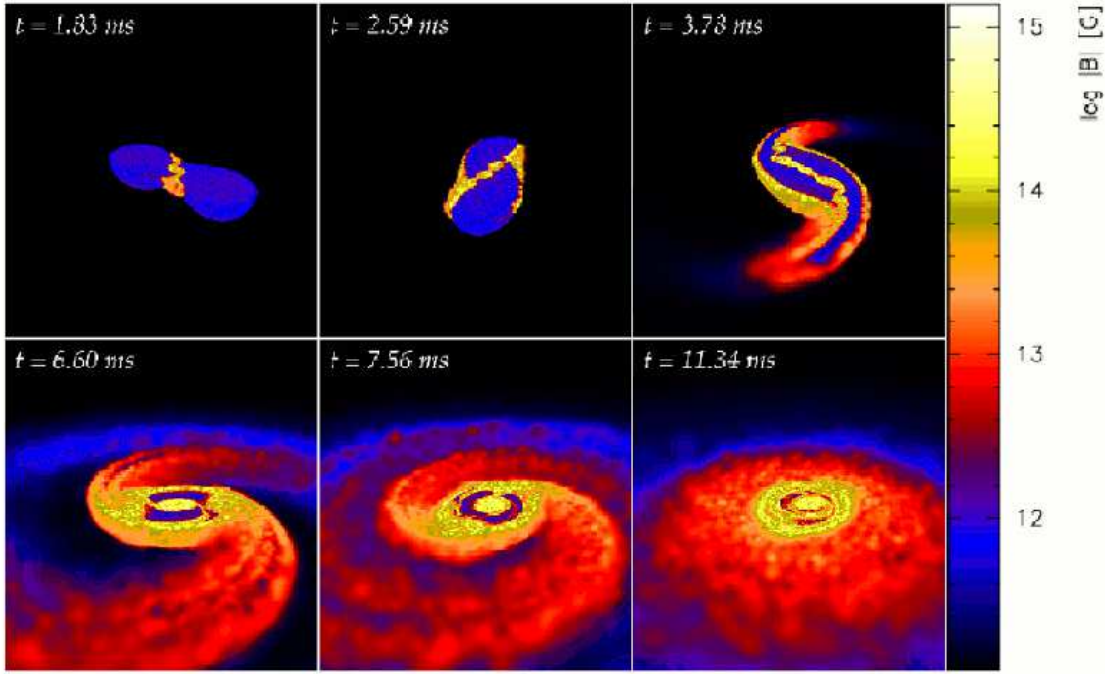

**Fig. 1.31.** A hydrodynamic simulation of a merger of two magnetized neutron stars. Colors indicate $\log(B)$ (in gauss), maximal magnetic field reaches $10^{15}$ G. The left to right dimension of the panels is $\sim 140$ km. The stars move gradually toward each other and then merge within about one orbital period ($\sim 2$ ms, the first two images). This new object sheds mass into spiral arms that are subsequently wrapped around the central object to form a hot torus (Credit: Price & Rosswog 2006).

## 1.6   Further missions

A brief information about other current satellites, besides Swift, focused on gamma-ray bursts is presented. The INTEGRAL mission was already mentioned in Section 1.4.2 and for example the latest mission Fermi must not be omitted.

### 1.6.1   INTEGRAL

The INTErnational Gamma-Ray Astrophysics Laboratory[1] (Winkler et al. 2003) was launched in October 2002 and provides a detailed high resolution sky mapping in the gamma range and also detects gamma-ray bursts. It carries an X-ray monitor JEM-X, an optical telescope OMC and two gamma instruments. One of them is SPI spectrometer which is made of germanium detectors and uses a coded mask technique. Its sensitivity ranges from 20 keV to 8 MeV, the field of view is 16° and the spatial accuracy 2°. The spectrometer is covered by the anti-coincidence shield ACS that also works as an efficient omnidirectional GRB detector[2] (Rau et al. 2005). The second gamma instrument aboard, IBIS, which also employs the coded mask technique, is sensitive from 15 keV to 10 MeV and has a field of view of 9°×9° with a fine spatial resolution of 12 arcmin.

---

[1] http://www.esa.int/esaMI/Integral/
[2] http://www.mpe.mpg.de/gamma/instruments/integral/spi/acs/grb/trigger/ACSTriggerTab.html

### 1.6.2 AGILE

AGILE[1] is an Italian Space Agency mission, launched in April 2007, and dedicated to the observation of the Universe in gamma-rays (Tavani et al. 2008). The spacecraft combines a gamma-ray imager (sensitive in the range $30\,MeV-50\,GeV$), a hard X-ray imager (sensitive in $18-60\,keV$) together with a Calorimeter ($300\,keV-100\,MeV$) and an anti-coincidence system. AGILE provides data for the study of various gamma-ray sources, e.g., active galactic nuclei, pulsars, supernova remnants, but also GRBs (Marisaldi et al. 2009).

### 1.6.3 Suzaku

Suzaku[2] (Astro-E2), the Japanese X-ray observatory, was launched in July 2005 (Mitsuda et al. 2007). There are four X-ray imaging CCD cameras (XIS) located in the focal plane of a dedicated X-ray telescope (XRT-I). One X-ray spectrometer (XRS, no longer operational) located behind the X-ray telescope (XRT-S). Next instrument is a non-imaging, collimated Hard X-ray Detector (HXD), which extends the bandpass of the observatory to higher energies with its $10-600\,keV$ bandpass. HXD is covered by anti-coincidence scintillators that are used as a Wideband All-sky Monitor (WAM) with $2\pi$ srad field of view (no positional capability), and with a bandpass of $50\,keV-5\,MeV$. This system detects $\approx 100$ GRBs[3] annually.

### 1.6.4 Fermi

The Fermi Gamma-Ray Space Telescope[4], formerly Glast (Gehrels & Michelson 1999), was launched in June 2008. It is the latest NASA gamma-ray burst mission to date. The payload comprises two science instruments, the Large Area Telescope (LAT) and the Gamma-Ray Burst Monitor (GBM). GBM consists of two types of detectors; NaI(Tl) with a sensitivity range $8\,keV-1\,MeV$ and BGO with a range of $150\,keV-40\,MeV$. The on-board localization accuracy is less than $15°$ and the final on-ground localization $\sim 3°$. GBM has a field of view $\sim 10$ srad (Meegan et al. 2009). The LAT instrument consists of anti-coincidence detector, particle tracker/converter and calorimeter (Atwood et al. 2009). The information from these three components is combined to estimate the energy and direction of gamma rays. LAT provides measurements in the energy range from $20\,MeV$ to $\sim 300\,GeV$. On-board localization accuracy is $< 15'$ and the field of view is $2.4$ srad. The Fermi satellite is expected to detect $\sim 250$ GRBs[5] annually.

## 1.7 Additional remarks

At the end of this introductory chapter, some additional remarks concerning the recent observations provided by the Fermi satellite, remarks about the polarization measurements, about other observational windows such as, e.g., gravitational waves, and about expectations of the detections of GRBs from the population III stars are presented.

---

[1]http://agile.rm.iasf.cnr.it/
[2]http://www.astro.isas.ac.jp/suzaku/index.html.en
[3]http://www.astro.isas.ac.jp/suzaku/research/HXD-WAM/WAM-GRB/
[4]http://fermi.gsfc.nasa.gov/
[5]http://fermi.gsfc.nasa.gov/ssc/observations/types/grbs/grb_table/

### 1.7.1   Polarization

A measurement of the linear polarization of the GRB prompt emission would provide a useful information of the emission mechanism and of the jet's geometry. The first detection of linear polarization ($80\,\% \pm 20\,\%$) of the prompt emission was claimed by Coburn & Boggs (2003), who analysed the extremely bright burst GRB 021206 measured by the RHESSI satellite (see Section 2). Later on, however, Rutledge & Fox (2004) and - independentely - Wigger et al. (2004), reanalysed these data and found several inconsistencies within the used methodology. Wigger et al. (2004) reached a higher precision with their analysis, but the result (a polarization level of $40\,\% \pm 60\,\%$) does not allow to make a decisive statement about the polarization level of GRB 021206.

Another case of a putative polarization detection was the burst GRB 040219A. McGlynn et al. (2007) measured the polarization over several energy ranges and time intervals using INTEGRAL-SPI and found a signal consistent with a polarization of about $60\,\%$, but again at not a high level of significance ($\sim 2\sigma$). Kalemci et al. (2007) did a similar analysis which yielded a polarization of $98\,\% \pm 33\,\%$. However, they concluded that, statistically, they could not claim a polarization detection from this burst and that also some instrumental systematic effects could play a role.

The polarization of the prompt GRB emission was also studied with BATSE using the GRB photons Compton-scattered off the Earth's atmosphere. Two bursts showed a clue of polarization, but many systematic effects involved in the analysis did not allow definite conclusions to be reached (Willis et al. 2005).

The polarization of the GRB prompt emission could become a main aim to the future already proposed space missions, e.g., POLAR (Produit et al. 2005), POET (Hill et al. 2008) or TSUBAME (Arimoto et al. 2008).

### 1.7.2   Spectrum hardening and delayed high-energy component

For the majority of GRBs time-integrated spectrum peaks around 150 keV and above this point the energy flux decreases following a power-law. However, in some cases there was found a hardening of the spectrum above several MeV. For example, González et al. (2003) reported a high-energy component in the compound BATSE-EGRET spectrum of GRB 941017 between 1 and 200 MeV and which was not the power-low extension of the emission detected below 1 MeV. The authors claimed that this high-energy component was inconsistent with the standard synchrotron shock model.

Another case of observed spectral hardening was the very bright burst GRB 021206 as reported by Wigger et al. (2008). Its prompt emission spectrum, as measured by RHESSI, can be described by a Band function only up to 4.5 MeV. Above 4.5 MeV, the spectrum hardens, so the Band function fails to fit the whole RHESSI energy range up to 17 MeV. On the other hand, it is shown that the cannonball model, which predicts such a hardening, well fits the whole measured spectrum.

A recent spectral analysis of GRB 090902B using GBM and LAT instruments on-board the Fermi satellite revealed a significant power-law component that is distinct from the usual Band model emission seen in the sub-MeV energy range. This power-law component appears to extrapolate from the GeV range to the lowest energies and is more intense than the Band component, both below $\sim 50\,\text{keV}$ and above $100\,\text{MeV}$ (Abdo et al. 2009). It is not clear what processes lead to such high-energy spectral component. From the recent observations provided by Fermi-LAT it also seems that the onset of $100\,\text{MeV}$ detection is delayed with respect to the detection of lower-energy photons.

These findings challenge the standard GRB model and suggest that some new modifications of the theory might be necessary.

### 1.7.3 Further observational windows

In this chapter the emission of very high-energy photons (TeV) and non-electro-magnetic signatures which accompany or might accompany GRBs are shortly discussed.

Many ground-based Cherenkov detectors have participated in the search for TeV photons from GRBs. Milagrito was one of them. It was a wide-field-of-view TeV water Cherenkov detector located at altitude $2650\,\text{m}$ near Los Alamos. Fifty-four GRBs were detected by BATSE during Milagrito activity ($1997-1998$), but only one, GRB 970417A, was associated with an excess of counts in the Milagrito detector (Atkins et al. 2000). The excess had a chance probability $2.8 \times 10^{-5}$ of being a background fluctuation, but this probability reduces to $1.5 \times 10^{-3}$ if one takes into account the fact that 54 GRBs were searched (Atkins et al. 2005; Abdo et al. 2007). Furthermore, this detection was not confirmed by the GRAND muon detector array, which can also detect gamma-rays ($0.01 - 10\,\text{TeV}$) through the secondary muons produced in the gamma-ray-hadronic cascade in the atmosphere with pions decaying to muons. However, GRAND may have detected sub-teraelectronvolt emission from GRB 971110 (Poirier et al. 2003), among the eight GRB candidates which were observed. The muon event was coincident in time and direction obtained by BATSE, but at a low level of significance ($2.7\sigma$). On the other hand, the search for the accompanying TeV emission from the BATSE GRBs within the period from October 1995 to March 1999 by the Tibet Air Shower Array (Amenomori et al. 2001) did not reveal any significant coincidence.

A search for high-energy neutrinos accompanying GRBs has been also accomplished. For example, AMANDA[1] (Antarctic Muon And Neutrino Detector Array), an array of photomultiplier tubes placed beneath the surface of the ice at the South Pole, searched for neutrino excesses that would be coincident with 400 GRBs occurred during its 7-year-long observation without positive detection (Achterberg et al. 2008).

There are also other neutrino detectors involved in the search for GRB neutrinos. For example, IceCube[2] which consists of $1\,\text{km}^3$ of Antarctic ice instrumented with 4800 photomultiplier tubes (Karle 2006), or Antares[1] a large Cherevkov detector in Mediterranean Sea (Aguilar et al. 2006). Nevertheless any significant neutrino point source being coincident with a GRB has not been detected so far, to the author's knowledge.

---

[1]http://amanda.uci.edu/
[2]http://icecube.wisc.edu/

In the internal-external shock model electrons are accelerated to relativistic speeds. If protons are also accelerated by the same shocks, they might be accelerated to Lorentz factors up to $< 10^{11}$ in the observer frame (Waxman 1995; Vietri 1995), i.e., to energies $E_p < 10^{20}$ eV. This is the domain of Ultra-High-Energy Cosmic Rays (UHECRs). Although the origin of UHECRs is still a mystery, their isotropic distribution suggests an extragalactic origin. Milgrom & Usov (1995) reported the possible association of two UHECR events with a GRB. They concluded that this coincidence was sufficiently remarkable to think of the production of these two events in the same explosion that created the GRB. The best arguments for a potential association of UHECRs with gamma-ray bursts have been summarized by Waxman (2006). The Pierre Auger Observatory[2] (Abraham et al. 2004; Šmída 2009) could confirm this origin. It was, however, announced by The Pierre Auger Collaboration et al. (2007) that among 27 energetic events collected between January 2004 and August 2007 and detected above $5.7 \times 10^{19}$ eV, 20 come from the direction of known active galactic nuclei (AGNs). They also reported that there is less than 1% chance that this correlation between AGNs and UHECRs is random. This result, if it is confirmed with better statistics, will have weighty consequences for the hypothesis of the same origin of UHECRs and GRBs.

Concerning gravitational waves (GW), they are expected to be associated with GRBs, especially with short-duration GRBs if those originate from the coalescence of two compact objects. Nonetheless, also collapsars of rapidly rotating stars may lead to a formation of a bar and thus produce similar GW signals. The detection of a GW signal in coincidence with a GRB would be a great discovery. GWs are emitted from the immediate vicinity of the central engine and they constitute the most direct probes of the progenitor, of its mass distribution and of its nature. For an overview of GW sources see Cutler & Thorne (2002). Major present facilities dedicated to measure GWs are, e.g., LIGO[3] (Laser Interferometer Gravitational Wave Observatory) in the USA (Abramovici et al. 1992), VIRGO[4] (Variability of Irradiance and Gravity Oscillations) in Italy (Acernese et al. 2004), GEO 600[5] in Germany (Strain et al. 2004), or TAMA[6] in Japan (Tatsumi et al. 2007). There is also a plan for the Advanced LIGO[7] (LIGO II) facility with more than a factor of 10 greater sensitivity than LIGO and a proposed project for a three-spacecraft laser interferometer in solar orbit LISA[8] (Laser Interferometer Space Antenna).

---

[1] http://antares.in2p3.fr/

[2] http://www.auger.org/

[3] http://www.ligo.caltech.edu/

[4] http://www.virgo.infn.it/

[5] http://www.geo600.org/

[6] http://tamago.mtk.nao.ac.jp/

[7] http://www.ligo.caltech.edu/advLIGO/

[8] http://lisa.nasa.gov/

Nakar et al. (2006) evaluated the capability of LIGO, and LIGO II of detecting GWs from a merger of neutron star (NS) and $10\,M_\odot$ black hole (BH). They found that such systems might be detectable by LIGO up to a distance $\sim 40\,\mathrm{Mpc}$. The expected GWs detection rate would be $\sim 3\,\mathrm{yr^{-1}}$ for an assumed rate of $10\,\mathrm{Gpc^{-3}\,yr^{-1}}$ of short GRBs. The expected rate of simultaneous short GRB + GW events is $0.3\,\mathrm{yr^{-1}}$. LIGO II would be able to detect the coalescence of BH$-$NS systems up to a distance of $650\,\mathrm{Mpc}$. If short GRBs originate in such mergers a lower limit for the detection of GW is $\sim 20\,\mathrm{yr^{-1}}$. The coincident event rate for LIGO II might be a few per year.

### 1.7.4 GRBs and Population III stars

The first stars, called Population III (Pop III) stars, are expected to be different from present stars because they are formed from completely metal-free gas. They are considered to consist mainly of very massive stars in the the range of hundreds of solar masses (Ohkubo et al. 2006; Yoshida et al. 2006). To date, however, Pop III stars have not been detected. The question of whether the progenitors of the highest redshift bursts can be due to Pop III stars is open.

A model of Poynting-dominated gamma-ray bursts from the collapse of very massive first generation stars at redshifts of order 20 is discussed by Mészáros & Rees (2010). The resulting relativistic jets would radiate in the hard X-ray range around $50\,\mathrm{keV}$ and above, followed after roughly a day by an external shock component peaking around a few keV. The fluences of this emission would be above the threshold for detectors such as Swift and Fermi and thus it might be possible to observe such flashes from the first stars.

## 2 The RHESSI Satellite

The Reuven Ramaty High-Energy Solar Spectroscopic Imager[1,2] (RHESSI) is a NASA Small Explorer satellite (Fig. 2.32) designed to study solar flares by use of imaging and spectroscopy of the continuous hard X-rays and gamma rays emitted by the upper layers of the Sun's surface.

A plan to develop RHESSI was accepted by NASA in October 1997 with its launch scheduled to July 2000 near maximal solar activity. After several technical problems (including a fatal failure during a vibration test in March 2000) and after several postponements, RHESSI was finally launched on the 5[th] of February 2002 using the Pegasus XL carrier rocket hanged under the L-1011 airplane. It was released at an altitude of about 12 000 m and launched on an almost circular orbit around the Earth with inclination 38° and altitude 600 km. An observing program started a week later. At the beginning of the mission the name of the satellite was HESSI. Later on it was renamed to RHESSI in commemoration of the eminent astrophysicist of high energies Dr. Reuven Ramaty, who had been working in Goddard Space Flight Center and participated on the development of this satellite. The development of the satellite, maintenance and processing of the observational data was also managed by other participating institutes, e.g., SSL Berkeley, ETH Zurich, PSI Villigen.

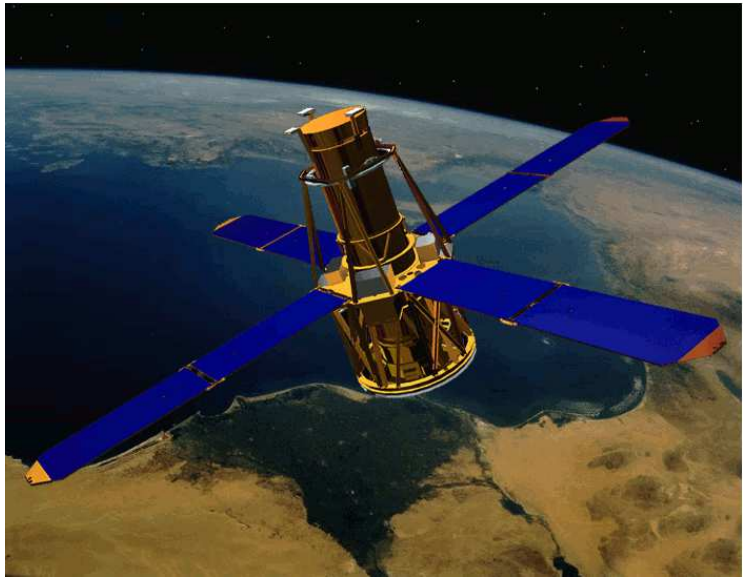

**Fig. 2.32.** A cartoon of the RHESSI satellite (Credit: RHESSI webpage[2]).

The gamma imaging system that constantly points to the direction of the Sun consists of an imaging tube with nine front and nine rear plane-parallel imagining grids. Behind the rear grids there are attenuators followed by the spectrometer. The spectrometer is composed of nine germanium detectors (each 7.1 cm in diameter and 8.5 cm in height), electronically separated to front and rear segments, and cooled by a Stirling cryostat to a temperature of $\approx 75$ K (Fig. 2.33). This arrangement allows imaging spectroscopy in a high-energy X-ray band with precise resolution, and allows line gamma spectroscopy with high resolution above $100\,\text{keV}$. The spatial resolution goes down to $2.3\,\text{arcsec}$ with a field of view of about $1°$. The spectral resolution ability is $1-10\,\text{keV}$ FWHM (full width at half maximum) in the energy range from $3\,\text{keV}$ to $17\,\text{MeV}$. Photon hits are stored event-by-event in an onboard memory with a time sampling of $\Delta t = 1\,\mu s$. (Tab 2.34).

---

[1]http://hesperia.gsfc.nasa.gov/hessi/
[2]http://hessi.web.psi.ch/

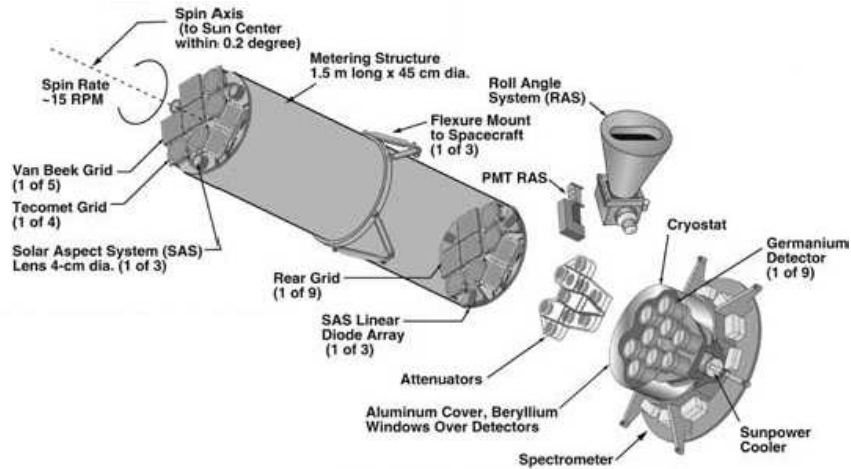

**Fig. 2.33.** A schema of the RHESSI satellite with the most important components indicated (Credit: Lin et al. 2002).

The detectors of the spectrometer are only lightly shielded ($4\,\text{mm}$ of aluminium), ensuring that RHESSI is also useful for detecting non-solar photons from any direction (Hajdas et al. 2004; Smith et al. 2000, 2003). With a field of view of about half of the sky and an effective area of about $150\,\text{cm}^2$, RHESSI observes about one or two GRBs per week[1].

RHESSI characteristics.

| Ia. Instrument characteristics: | | Ib. Spacecraft characteristics: | |
|---|---|---|---|
| Energy range | 3 keV to 17 MeV | Mass | Total 291.1 kg, instrument 130.8 kg |
| Energy resolution (FWHM) | $\lesssim 1$ keV at 3 keV, increasing to $\sim 5$ keV at 5 MeV | Power | Total 220.4 W, instrument 142.3 W |
| Angular resolution | 2.3 arc sec to 100 keV, 7 arc sec to 400 keV, | Size | 1.18 m diameter, 2.06 m height, |
| | 36 arc sec to 15 MeV | | 5.74 m tip-to-tip with solar panels deployed |
| Temporal resolution | 2 s for detailed image, tens of ms for basic image | Telemetry | 4 Mbps, downlink, 2 kbps command uplink |
| Field of view | full sky ($\sim 1°$) | On-board storage | 4 Gbyte solid state memory |
| Effective area (photopeak) | $\sim 10^{-3}$ cm$^2$ at 3 keV, $\sim 32$ cm$^2$ at 10 keV | Attitude | 15 rpm spin rate, pointing to 0.2° of Sun center |
| | (with attenuators out), $\sim 60$ cm$^2$ at 100 keV, | | |
| | $\sim 15$ cm$^2$ at 5 MeV | Ic. Mission characteristics: | |
| Detectors | 9 germanium detectors (7.1-cm dia. $\times$ 8.5 cm), | Launch date | 5 February 2002 |
| | cooled to $< 75$ K with Stirling-cycle | Launch vehicle | Pegasus XL |
| | mechanical cooler | Orbit | 38° inclination, 600 km altitude apogee, |
| Imager | 9 pairs of grids, with pitches from 34 microns | | 586 km perigee |
| | to 2.75 mm, and 1.55-m grid separation | Nominal mission lifetime | 2 years, 3rd year highly desirable |
| Aspect system | Solar Aspect System: Sun center to $< 1$ arc sec | | |
| | Roll Angle System: roll to $\sim 1$ arc min | | |
| Number of flares expected | $\sim 1000$ imaged to $> 100$ keV | | |
| | $\sim$ tens with spectroscopy to $\sim 10$ MeV | | |

**Fig. 2.34.** Technical specifications of the RHESSI satellite (Credit: Lin et al. 2002).

---

[1] http://grb.web.psi.ch/

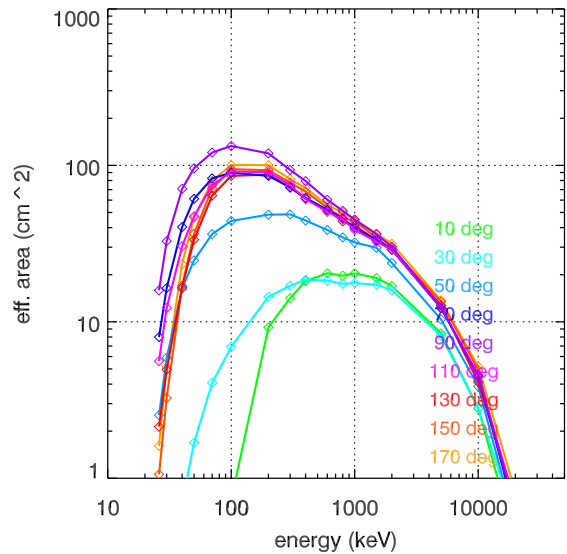

**Fig. 2.35.** Summed effective area of rear segments for observed energy = incoming energy (i.e., photopeak) for different off-axis angles (zero angle corresponds to the front direction, i.e., towards the sun) (Credit: Wigger C.)

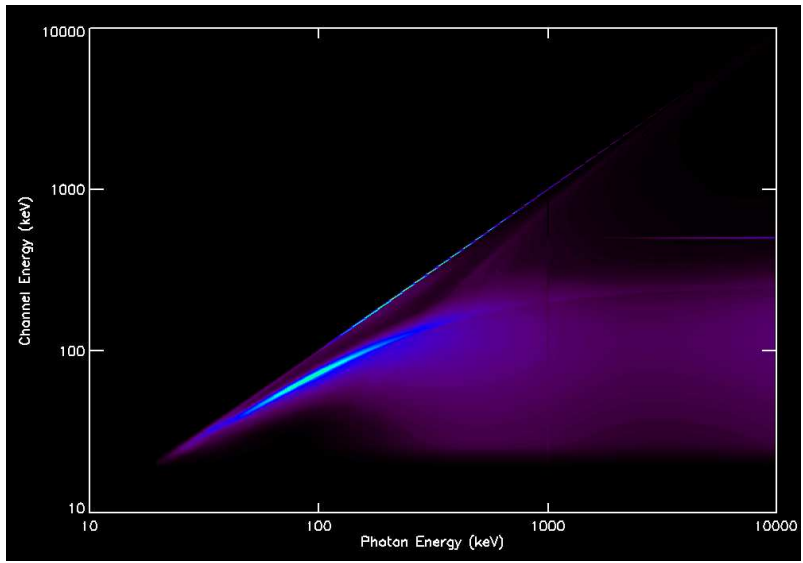

**Fig. 2.36.** The detector response matrix for the summed all rear segments except number 2 and for the zero off-axis angle. The input photon energy is on the x-axis. The channel energy (observed count energy) is on the Y-axis. The brighter color the higher response (count rate).

The detector effective area depends on the photons' incoming angle, see Fig. 2.35. The detector response is not the same for photons of different energies, see the detector response matrix for rear segments displayed in Fig. 2.36. Fig. 2.37 presents an example of the RHESSI response function for the input photon energy of 300 keV and 2 MeV.

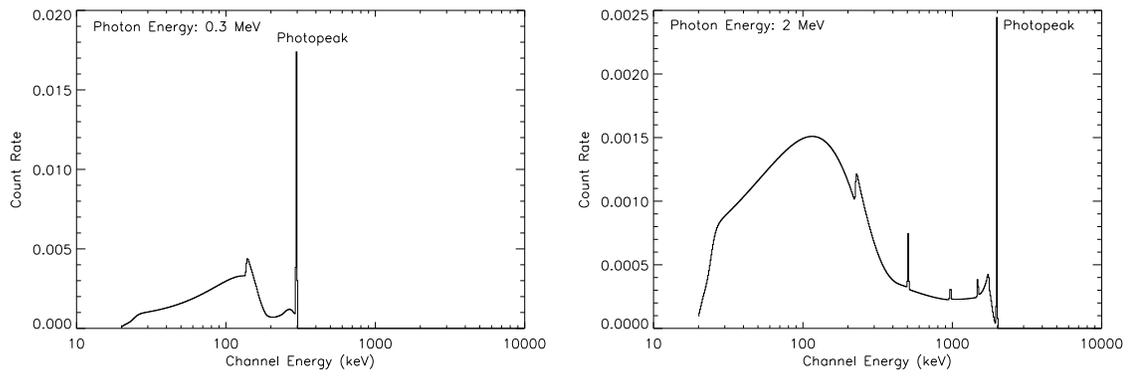

**Fig. 2.37.** An example of the RHESSI response function for the input photon energy of 300 keV (left panel) and 2 MeV (right panel).

# 3 GRB Groups and Motivation

In this section an insight into the issue of GRB groups/classes is presented. Differences between the short and long-duration GRBs (SGRBs, LGRBs), either observational or concerning their progenitors, have already been mentioned. However, there are also further types of x-ray or gamma-ray flashes, either originating in distinct astrophysical objects or having probably the same or similar progenitors like SGRBs or LGRBs, but in some aspects differing from these two GRB groups. Several authors dealt with a quest for GRB groups, their number, features and even with a possibility that in duration intermediate GRBs form another distinct group. This issue of intermediate-duration GRBs being a significantly distinct group became a main motivation for this thesis and it is investigated in next Sections 4 and 5 by employing data from the RHESSI satellite.

## 3.1 Soft gamma repeaters

Soft gamma repeaters (SGR) first recorded by (Mazets et al. 1979) are observed as X-ray to low-energy gamma-ray periodical flashes of width $\sim 0.1\,\mathrm{s}$ (Kouveliotou et al. 1993b), coming from a direction of the Milky Way Galaxy, e.g., SGR 1806-20, or from the Large Magellanic Cloud, and having the spectral shape of optically thin thermal bremsstrahlung ($T \approx 30 - 40\,\mathrm{keV}$). These flashes are known to originate in different astrophysical objects than classical GRBs. Concretely its source is a magnetar (Kouveliotou et al. 1994; Kulkarni et al. 1994; Duncan & Thompson 1992), a neutron star, born at a supernova explosion, with very fast rotation period ($< 20\,\mathrm{ms}$) and with huge magnetic field (up to $10^{15}\,\mathrm{G}$) and the X-ray and gamma-ray radiation is released via magnetic reconnection.

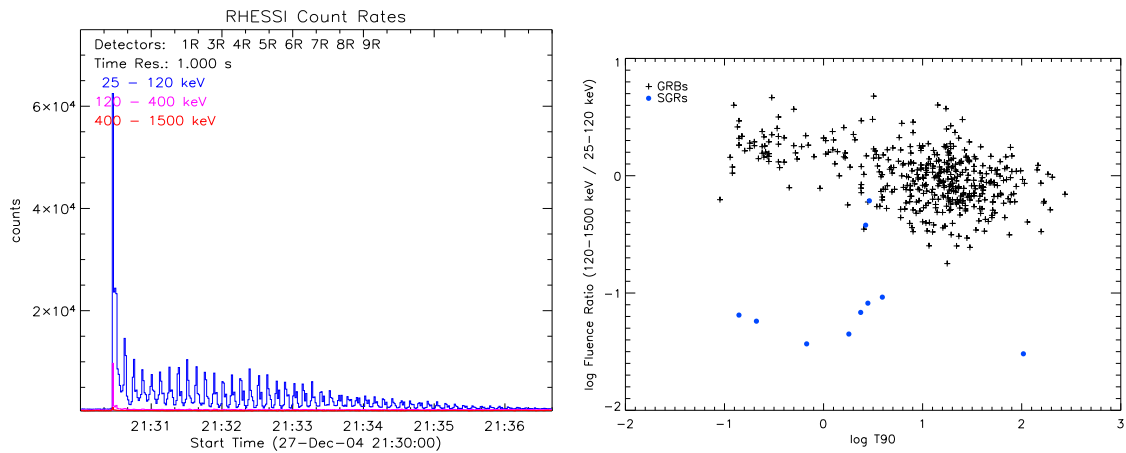

**Fig. 3.38.** *Left panel:* Profile of the 27 December 2004 giant flare of SGR 1806-20. *Right panel:* A comparison of hardness ratios of GRBs and of flashes from SGRs mainly from SGR 1806-20 or SGR 1900+14. All data were obtained with the RHESSI satellite.

Even if SGRs are different objects than classical GRBs, they are mentioned in this chapter because it is necessary to account for these flashes in any database of gamma-ray transients and beware of admixing SGRs to a GRB sample intended to be further statistically analysed.

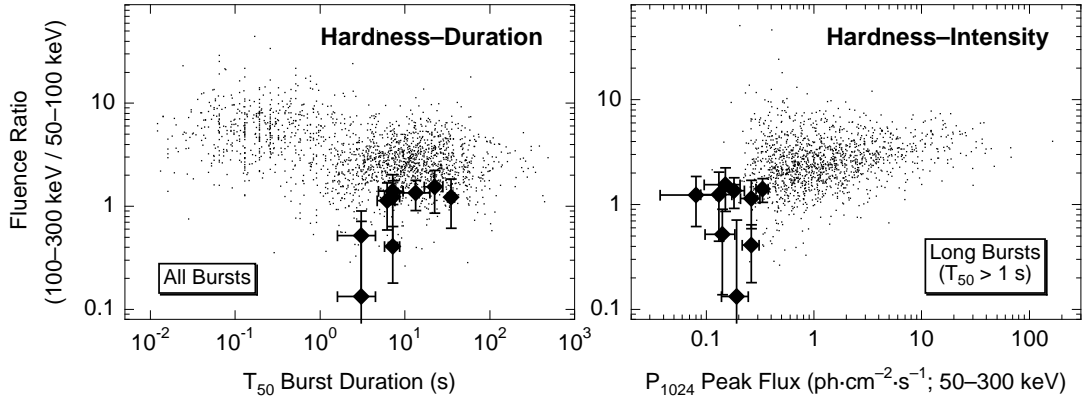

**Fig. 3.39.** Gamma-ray parameters of nine XRFs (diamonds) compared to GRBs from the BATSE catalog (dots). The plot at right shows only long-duration GRBs ($T_{50} > 1$ s). XRFs are consistent with the extrapolation of the GRB data towards faint and soft events (Credit: Kippen et al. 2001).

## 3.2 X-ray flashes

X-ray flashes (XRF), sometimes called fast X-ray transients, belongs to a class of soft GRBs. They are not recurrent flashes of X-rays, sometimes accompanied by weak gamma-ray emission with no properties common to any known catalogued X-ray sources. XRFs were first detected by the Ginga and later by the BeppoSAX satellites (Strohmayer et al. 1998; Heise et al. 2001). The unique property of these bursts is their large X-ray flux (up to $\sim 30$ keV) comparable to or even dominant over their gamma-ray emission.

Kippen et al. (2001, 2003, 2004) discovered and investigated nine BeppoSAX XRFs seen also in the BATSE continuous data and found that the durations and spectra of BeppoSAX XRFs were consistent with the extrapolation of the BATSE GRB data towards faint and soft events, suggesting that XRFs constitute a natural extension of the known GRB population (see Fig. 3.39). Barraud et al. (2003) and Sakamoto et al. (2005) studied the properties of the XRF prompt emission and also showed that they form an extension of the GRB population towards lower energy events. Sakamoto et al. (2005) found that the HETE sample contains about 1/3 XRFs, 1/3 GRBs, and 1/3 intermediate events called X-ray-rich GRBs.

Important efforts were made to study the other properties of XRFs, like their afterglows or host galaxies (Frail et al. 1999; Bloom et al. 2003b). Despite some differences, the basic properties of XRFs were found to be similar to those of classical GRBs. Also the association of SNa and XRFs was discovered, like the LGRBs-SNa association, e.g., in the case of GRB (XRF) 060218 and supernova SN 2006aj, implying a common origin (Pian et al. 2006; Sollerman et al. 2006).

However, the question why are XRFs much softer than ordinary GRBs still remains. Different models were suggested (Heise et al. 2001; Heise 2003), but two explanations are favored at the present time; XRFs are ordinary GRBs but the jet is viewed slightly off-axis, or XRFs are intrinsically soft. The off-axis model was studied, e.g., by Zhang & Mészáros (2002); Yamazaki et al. (2002, 2003) and it was found that XRFs could be explained by off-axis GRBs at low redshifts ($z < 0.2$). Dado et al. (2004) analyzed XRFs in the frame of the cannonball model and they concluded that XRFs are intrinsically identical to GRBs, but viewed off-axis. On the other hand, Gendre et al. (2007) studied XRFs observed by Swift and they came to the conclusion that XRFs are intrinsically soft because of different thermal energy per particle in the outflow.

## 3.3  Short GRBs with extended emission

Another group of GRBs that might constitute a separate class are recently recognized and discussed bursts called short gamma-ray bursts with extended emission (SGR-BEE). Their basic feature is that a short ($\sim 0.1 - 5$ s) hard spike of gamma emission is followed, sometimes after a certain lull ($\sim 3 - 10$ s), by an X-ray or soft gamma-ray much longer hump ($\sim 30 - 100$ s) as seen in Fig. 3.40 (Norris & Bonnell 2006; Minaev et al. 2010; Perley et al. 2009). Figure 3.41 shows their hardness/duration properties.

SGBEEs show no evidence for an associated supernova and, in conflict with compact object merger models, appear to originate exclusively inside the disk of their host galaxies. Troja et al. (2008) found that they are characterized by a median offset of $2.5 \pm 3.4$ kpc whereas SGRBs display a more heterogeneous displacement around the host of a median offset of $22 \pm 26$ kpc. It was suggested that they may be produced by the accretion-induced collapse of a white dwarf (WD) or the merger and collapse of a WD-WD binary, perhaps thus sharing a close relation to the progenitors of Type Ia SNe (Metzger et al. 2008a,b).

## 3.4  Very short GRBs

Cline et al. (1999) studied very short BATSE GRBs ($T_{90} < 100$ ms) and came to the conclusion that they are very different from the rest of GRBs and suggested a possible model of primordial black hole evaporation as their source. In the later work (Cline et al. 2005) it was introduced that very short GRBs indicate a significant angular asymmetry and a uniform $V/V_{max}$ distribution. The $V/V_{max}$ distribution for short and long GRBs, however, had been found nonuniform (Section 1.3.6.) The origin of very short GRBs and their temporal properties are also discussed by Czerny et al. (2011).

## 3.5  Quest for GRB groups and open questions

It is a deep history when the first classifications of GRBs appeared. Mazets et al. (1981); Norris et al. (1984); Kouveliotou et al. (1993a); Aptekar et al. (1998) suggested a separation of bursts into short/long according to their duration (at $\sim 2$ s). Many observational evidences demonstrating different properties of short/long bursts were found so far, as mentioned in the previous overview chapter (Section 1). At present it is accepted that they are physically different phenomena (Norris et al. 2001; Baláis et al. 2003; Fox et al. 2005; Kann et al. 2008). Nevertheless, since 1998 there are statistical indications for an "intermediate" GRB group. This separation into three groups was confirmed by statistical studies of different databases: BATSE (Horváth 1998; Mukherjee et al. 1998; Horváth et al. 2006); BeppoSAX (Horváth 2009); Swift (Horváth et al. 2008; Huja et al. 2009).

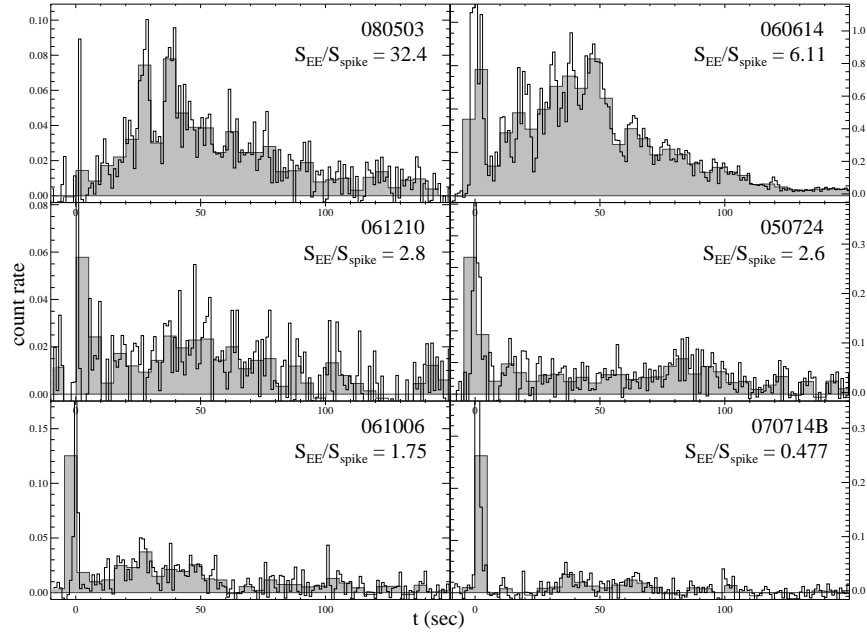

**Fig. 3.40.** BAT 25–100 keV light curves of several different Swift short bursts with high signal-to-noise ratio extended emission. The 1 s binned curve is plotted as a black line and a 5 s binning is plotted in solid gray (Credit: Perley et al. 2009).

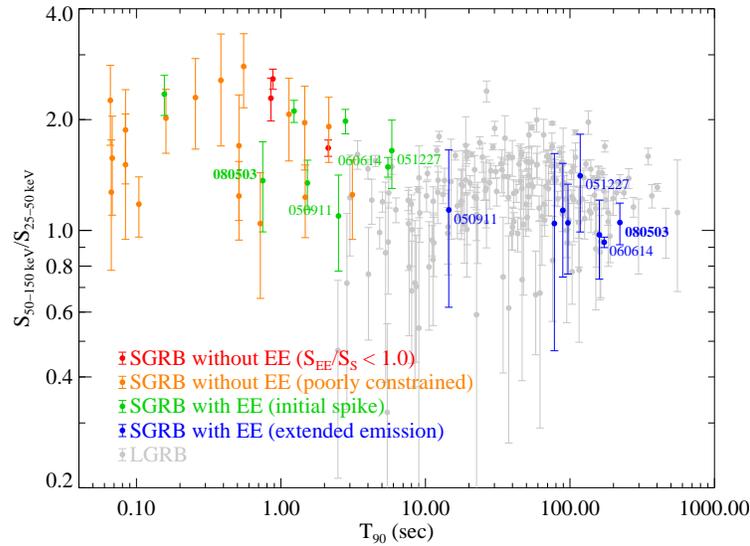

**Fig. 3.41.** Duration-hardness plot for bursts detected by the Swift BAT instrument. Long bursts are shown in gray. Short bursts ($T_{90} < 2$ s) are colored based on the presence or absence of extended emission (EE): bursts without EE are shown in red, faint bursts for which the presence of EE is poorly constrained are orange, and short bursts with observed EE are plotted for the initial spike (green) and for the EE itself (blue) (Credit: Perley et al. 2009).

In essence, there are different types of statistical methods used in these studies. The classical $\chi^2$ test (see Appendix A) applied on the GRB durations confirmed three groups in the BATSE (Horváth 1998), see Fig. 3.42, and Swift (Huja et al. 2009) databases. This method is not highly sensitive. More effective is the maximum likelihood (ML) method applied on durations (Appendix A). This method confirmed the third group in three different databases; BATSE (Horváth 2002); Swift (Horváth et al. 2008), see Fig. 3.42; BeppoSAX (Horváth 2009). The third method is again ML test, but applied on the joint probability of durations and hardness ratios (Appendix A). This method has proven three groups in two databases; BATSE (Horváth et al. 2004, 2006), see Fig. 3.43, and Swift (Horváth et al. 2008), see Fig. 3.43. Other statistical procedures, such as multi-variate analyses and other sophisticated methods (Mukherjee et al. 1998; Balastegui et al. 2001; Chattopadhyay et al. 2007) have also been applied and confirmed the existence of three groups in the BATSE sample, from the statistical point of view. However, there are also works (Hakkila et al. 2000, 2004; Rajaniemi & Mähönen 2002) concluding that the third group (intermediate fluence, intermediate duration, soft) identified by Mukherjee et al. (1998) in the BATSE data might be a systematic or instrumental bias and that it does not have to be a distinct source population. For a survey, the results of the above-mentioned statistical methods are collected in Tab. 3.1.

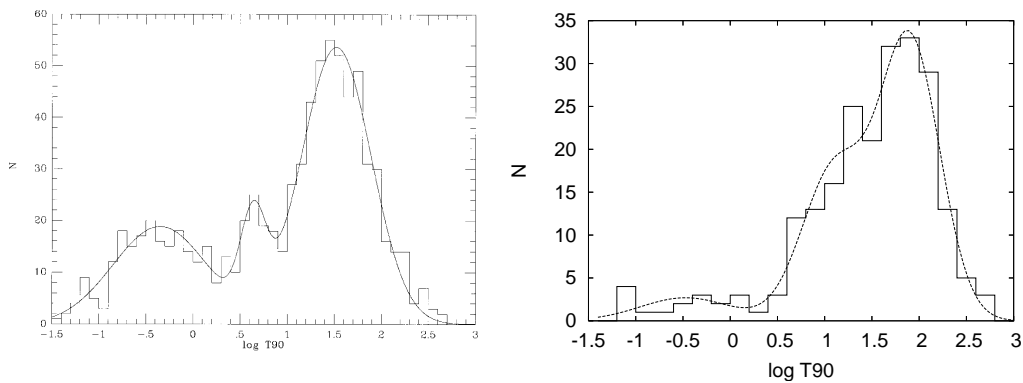

**Fig. 3.42.** A distribution of durations $T_{90}$ of the BATSE (*left panel*) and the Swift (*right panel*) GRBs with the best-fit model of the sum of three log-normal functions. In both cases an intermediate group is significant, see Tab. 3.1 (Credit: Horváth 1998, Horváth et al. 2008.)

**Tab. 3.1.** Probabilities (in %) of the occurrence of an intermediate GRB group being accidental as gained from different methods and data samples. Lower percentage means higher significance of the intermediate group. F-test compares the best $\chi^2$ fits (two and three Gaussian curves) of the $\log T_{90}$ duration histogram. ML is the Maximum Likelihood ratio test applied either on the $\log T_{90}$ durations or on the durations−hardness pairs. "?" means that authors did not explicitly mention the significance level. "—" means that a test has not been done on such data sample so far. The cited references are as follows: [1] Mukherjee et al. (1998), [2] Horváth (1998), [3] Balastegui et al. (2001), [4] Horváth (2002), [5] Horváth et al. (2006), [6] Chattopadhyay et al. (2007), [7] Horváth et al. (2008), [8] Řípa et al. (2009), [9] Huja et al. (2009), [10] Horváth (2009), [11] Horváth et al. (2010). Section 4 will deal with the work [8].

| Method | BATSE | Swift | RHESSI | BeppoSAX |
|---|---|---|---|---|
| F-test ($T_{90}$) | $< 10^{-2}$ [2] | $\simeq 3$ [9] | 6.9 [8] | — |
| ML ($T_{90}$) | 0.5 [4] | 0.46 [7] | 0.036 [8] | 3.7 [10] |
| ML ($T_{90}$ vs. $H$) | $< 10^{-8}$ [5] | $10^{-6}$ [11] | 0.13 [8] | — |
| Other methods | $\ll 0.01$[1] | — | — | — |
| | ? [3,6] | | | |

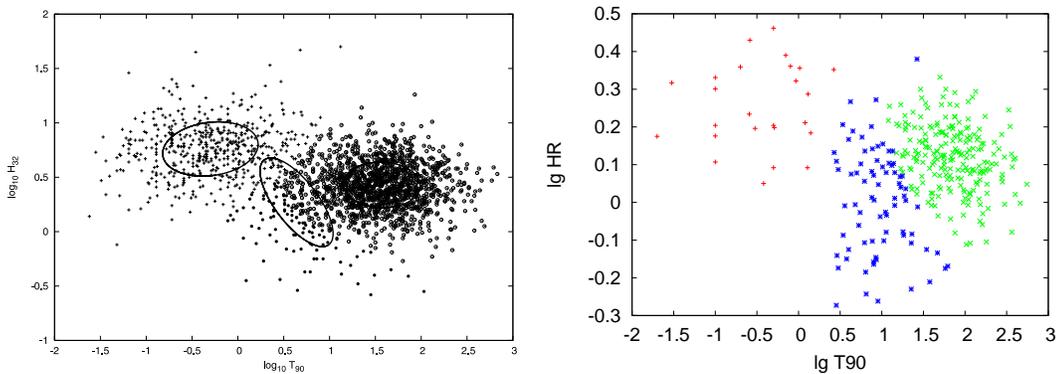

**Fig. 3.43.** A distribution of duration−hardness pairs of the BATSE (*left panel*) and the Swift (*right panel*) GRBs with an indication to which group each burst belongs by maximal probability. In both cases an intermediate group was found to be statistically significant, see Tab. 3.1 (Credit: Horváth 2006, Horváth et al. 2010.)

All works mentioned in Tab. 3.1 dealing with the intermediate GRB group were focused on duration and spectral hardness and did not tell anything about other aspects like, e.g., their spectral lags, host galaxies, afterglows, redshifts, etc. Such a complex survey, on the Swift data, was given by de Ugarte Postigo et al. (2010). They concluded that intermediate bursts clearly differ in luminosity of the X-ray afterglows with respect to the short and long types. The measured luminosity for Swift intermediate bursts is on average one order of magnitude fainter than for long bursts and one order of magnitude brighter than for short bursts, but less significant in the optical afterglow. They also found that spectral lags of intermediate burst are similar to those of long bursts, but significantly different from those of short bursts. They also concluded that intermediate and long bursts in their sample appear to have the same type of progenitors.

Veres et al. (2010) studied Swift GRBs and found a significant intermediate group. They concluded that these Swift intermediate bursts are closely related to XRFs.

It is clear that any new result in the classification scheme of GRBs is highly desired. Therefore in what follows a new GRB database, particularly from the RHESSI satellite, is studied.

# 4  GRB Groups by the RHESSI Satellite

## 4.1  Statistics

This chapter presents results of the statistical tests performed on the RHESSI GRB durations and spectral hardnesses in order to determine whether an intermediate GRB group revealed in BATSE, Swift, and BeppoSAX data also appears in the RHESSI GRB sample and it draws from the article Řípa et al. (2009).

### 4.1.1  Data sample

The RHESSI GRB Catalog[1] and the Cosmic Burst List [2] were used to detect 487 GRBs in the RHESSI data between February 14, 2002 and April 25, 2008. The strategy by which GRBs were identified was the following: There is no automatic GRB search routine. Only if a message from any other instrument of IPN occurs, the RHESSI data are searched for a GRB signal. Therefore, in our data set there are only GRBs, which are also confirmed by other instruments. Any flashes from SGRBs or responses due to, e.g., high-energy particles are excluded from our sample. The biggest overlap is with Konus−Wind. About 85 % of all RHESSI GRBs are also observed by that instrument[3].

For a deeper analysis, we chose a subset of 427 GRBs with data of a signal-to-noise ratio higher than 6. The signal-to-noise ratio was defined as $S/\sqrt{2B+S}$, where $S$ is a GRB signal over the background level $B$ (both are counts in a $T_{90}$ time interval over the range $25\,\mathrm{keV}-1.5\,\mathrm{MeV}$).

The SolarSoftWare[4] program developed in IDL[5] programming application as well as routines written by the RHESSI team at Paul Scherrer Institute[6] were used to derive background-subtracted count light curves (with a time resolution higher than 10 % of each burst's duration for the vast majority of our entire data set) and GRB count fluences from the rear segments (except malfunctioning segment R2) of the RHESSI spectrometer's detectors in the energy band $25\,\mathrm{keV}-1.5\,\mathrm{MeV}$ (see Fig. 4.44). This data set with the time resolutions of derived count light curves are listed in Appendix B in Tab. $9.15-9.26$.

Count light curves, count fluxes and fluences were derived instead of those in energy units, because only for a small portion of RHESSI GRBs the localization is known ($\approx 18\,\%$), e.g., from Swift or IPN. The localization is needed to derive the exact fluxes in energy units, because the instrument's response depends on the incoming direction of the photons (see Section 2). RHESSI itself is not able to localize GRBs. It is quite common in the GRB field that authors use count fluxes if the energy fluxes can not be calculated. In what follows, this approach is also taken.

Needed to note that the classification methods, used in this chapter, are based on the observer frame properties of bursts. A detailed rest-frame classification would allow to obtain intrinsic properties and would be more physically meaningful, however such a classification requires a large number of bursts with measured redshift and, for the RHESSI GRB database, there is only a very limited number of such bursts (19 from all 427 bursts used in this analysis, i.e., $\approx 4.5\,\%$).

---

[1]http://grb.web.psi.ch/
[2]http://www.ssl.berkeley.edu/ipn3/masterli.html
[3]see http://grb.web.psi.ch/publications/talk_venice.pdf
[4]http://sohowww.nascom.nasa.gov/solarsoft/
[5]http://www.ittvis.com/ProductServices/IDL.aspx
[6]http://www.psi.ch/

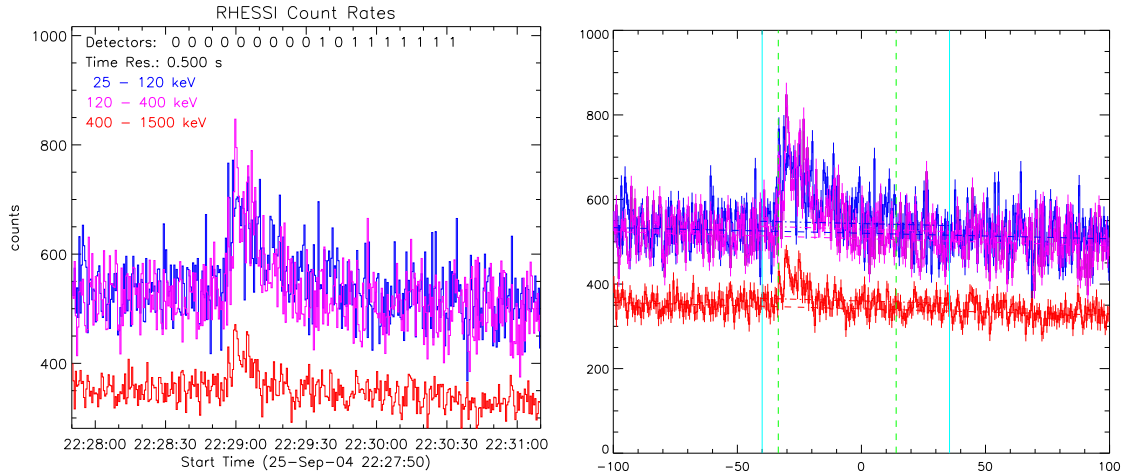

**Fig. 4.44.** *Left panel:* An example of GRB count light curve in three energy bands as obtained from RHESSI. *Right panel:* A demonstration of a background fit (horizontal dashed line). The background is fitted from the data before and after the GRB that are denoted by the vertical solid lines. The dashed green lines denotes the start and end times of the $T_{90}$ interval.

### 4.1.2  Duration distribution

First, the one-dimensional duration distribution is studied. The $T_{90}$ uncertainty consists of two components. We assume that one is given by the count fluence uncertainty during $T_{90}$ ($\delta t_s = T_{90}\sqrt{2B+S}/S$), which is given by Poissonian noise, and that the second one is the time resolution of derived light curves ($\delta t_{res}$). The total $T_{90}$ uncertainty $\delta t$ was calculated to be $\delta t = \sqrt{\delta t_s^2 + \delta t_{res}^2}$.

The histogram of the times $T_{90}$ is a distribution with two maxima at approximately $0.2\,\mathrm{s}$ and $20\,\mathrm{s}$ (Fig. 4.45). It consists of 19 equally wide bins on a logarithmic scale (of base 10) starting at $0.09\,\mathrm{s}$ and ending at $273.4\,\mathrm{s}$. We follow the method completed by Horváth (1998) and fitted one, two, and three log-normal functions (Fig. 4.45) and used the $\chi^2$ test to evaluate these fits (for details see Appendix A). The minimal number of GRBs per bin is 5, only in the last bin it is 4, and hence the use of the $\chi^2$ test is possible. The best fitted parameters, the values of $\chi^2$, the degrees of freedom, and the goodness-of-fits are listed in Tab. 4.2.

Among the statistical properties of the observed quantities of gamma-ray bursts the log-normal distributions are often seen. McBreen et al. (1994); Kouveliotou et al. (1996) and Aptekar et al. (1998) pointed out that the duration of short and long GRBs are consistent with log-normal distributions. McBreen et al. (1994) also pointed out that the time interval between the GRB pulses are consistent with a log-normal distribution. It was also found that the durations of the GRB pulses (Nakar & Piran 2002) and the break energies in the GRB spectra follow log-normal distributions (Preece et al. 2000).

Many examples of log-normal distributions occur in nature, e.g., the propagation of a laser beam in a turbulent atmosphere (Majumdar & Gamo 1982). In general a log-normal distribution, under certain conditions, results if the variable is made from the product of a large number of independent, identically-distributed variables, in the same way as a normal distribution results if the variable is the sum of a large number of independent, identically-distributed variables, e.i., by the the central limit theorem (Aitchison & Brown 1957; Crow & Shimizu 1988; Topinka 2009). The possible origin of the observed log-normal distributions in GRBs from the viewpoint of the central limit theorem is discussed by Ioka & Nakamura (2002).

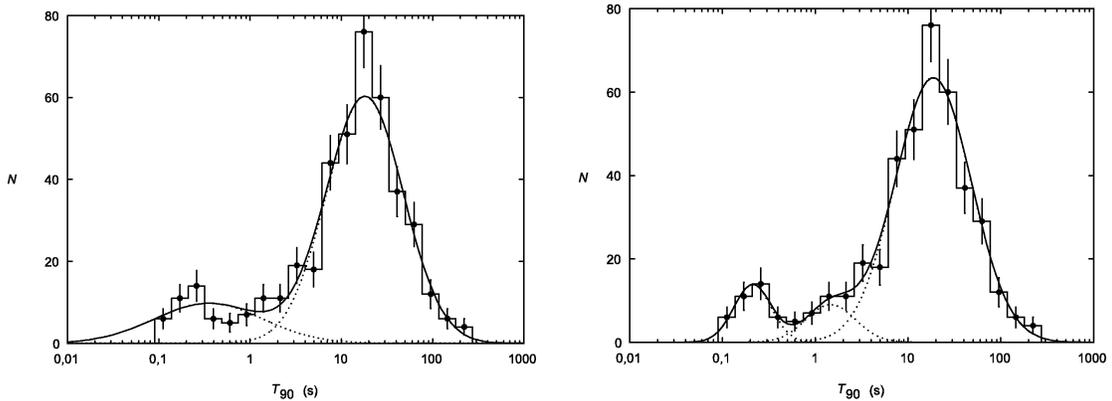

**Fig. 4.45.** *Left panel:* A duration distribution of the 427 RHESSI bursts with the best $\chi^2$ fit of two log-normal functions. The number of bins is 19, d.o.f. $= 14$ and $\chi^2 \simeq 19.1$ which implies the goodness-of-fit $\simeq 12\,\%$. The error bars are equal to $\pm\sqrt{N_i}$, where $N_i$ is the number of events in i-th bin. *Right panel:* The same duration distribution with the best $\chi^2$ fit of three log-normal functions. d.o.f. $= 11$ and $\chi^2 \simeq 10.3$ which implies the goodness-of-fit $\simeq 42\,\%$.

In the case of the fit with one log-normal function, $\chi^2 \simeq 157$ for 17 degrees of freedom (d.o.f.) was obtained. Therefore, this hypothesis is rejected on a smaller than 0.01 % significance level.

The assumption of two groups being represented by two log-normal fits is acceptable, the fit with three log-normal functions even more. The important question is whether the improvement in $\chi^2$ is statistically significant. To answer this question, we used the F-test, as described by Band et al. (1997), in their Appendix A (see also Appendix A in this work). The F-test gives a probability of 6.9 % of the improvement in $\chi^2$ being accidental. This value is remarkably low, but not low enough to reject the hypothesis that two log-normal functions are still enough to describe the observed duration distribution.

**Tab. 4.2.** Parameters of the best $\chi^2$ fits of 2 and 3 log-normal functions of the RHESSI GRB $T_{90}$ distribution. $\mu$ are the means, $\sigma$ are the standard deviations and $w$ are the weights of the distribution. Given uncertainties are $(\text{Max}(p_i)\text{-Min}(p_i))/2$, where $p_i$ are the parameters obtained by fitting of ten different histograms of durations randomly changed by their uncertainties as described in the text.

| Parameter | 2 log-normal | 3 log-normal |
|---|---|---|
| $\mu_{\text{short}}$ | -0.46 ± 0.22 | -0.68 ± 0.15 |
| $\sigma_{\text{short}}$ | 0.60 ± 0.09 | 0.20 ± 0.05 |
| $w_{\text{short}}$ (%) | 18.7 ± 2.7 | 8.9 ± 1.6 |
| $\mu_{\text{long}}$ | 1.26 ± 0.04 | 1.27 ± 0.04 |
| $\sigma_{\text{long}}$ | 0.42 ± 0.02 | 0.41 ± 0.02 |
| $w_{\text{long}}$ (%) | 81.3 ± 2.7 | 83.4 ± 1.5 |
| $\mu_{\text{middle}}$ | | 0.17 ± 0.09 |
| $\sigma_{\text{middle}}$ | | 0.27 ± 0.09 |
| $w_{\text{middle}}$ (%) | | 7.7 ± 1.5 |
| d.o.f. | 14 | 11 |
| $\chi^2$ | 19.13 | 10.30 |
| goodness (%) | 11.9 | 41.5 |
| $F_0$ | 3.14 | |
| $P(F > F_0)$ (%) | 6.9 | |

To determine how the $T_{90}$ uncertainties affected the obtained result, we randomly selected one half of the measured GRB durations and shifted them by the full amount of their uncertainties to lower values and the second half to higher values. Then a new histogram with the same number of bins and the same boundaries was compiled. The new best-fit model parameters, $\chi^2$, and F-test were recalculated. This was repeated ten times. Obtained results are listed in Tab. 4.3. This procedure gives an information how the fitted parameters vary (see Tab. 4.2). In seven cases out of ten the F-test gave the probability higher than 5 %, i.e., the improvement in $\chi^2$ being insignificant while introducing a third group.

As the number of GRBs is low for many bins, it is worthwhile using a maximum likelihood (ML) method (see Horváth (2002) or Appendix A). The the best-fit parameters obtained by this method are listed in Tab. 4.4. Since the difference of the logarithms of the likelihoods $\Delta \ln L = 9.2$ should be half of the $\chi^2$ distribution for 3 degrees of freedom, one can infer that the introduction of a third group is statistically significant on the 0.036 % level (of being accidental).

To derive an image of how the $T_{90}$ uncertainties affected the result, it was proceeded similarly to $\chi^2$ fitting and ten different data sets randomly changed in durations by the full amount of their uncertainties were generated. The results are presented in Tab. 4.5. From this table, it can be seen that all ten simulations give probabilities, of introducing a third group being accidental, much lower than $5\,\%$. Thus, the hypothesis of introducing a third group is highly acceptable.

**Tab. 4.3.** The minimal $\chi^2$, corresponding goodness-of-fits and F-tests for fitted two and three log-normal functions on the RHESSI GRB $T_{90}$ distribution for ten different changes of durations by their uncertainties.

| 2 log-normal | | 3 log-normal | | F-test |
|---|---|---|---|---|
| $\chi^2$ | goodness (%) | $\chi^2$ | goodness (%) | $P(F > F_0)$ (%) |
| 23.97 | 4.6 | 14.19 | 22.3 | 11.1 |
| 18.03 | 20.6 | 10.65 | 47.3 | 11.0 |
| 15.99 | 31.4 | 5.13 | 92.5 | 0.5 |
| 13.52 | 48.6 | 7.50 | 75.8 | 8.0 |
| 17.87 | 21.3 | 7.55 | 75.3 | 2.0 |
| 16.36 | 29.2 | 9.66 | 56.1 | 11.0 |
| 11.89 | 61.5 | 6.95 | 80.3 | 10.4 |
| 21.86 | 8.2 | 13.77 | 24.6 | 15.1 |
| 20.07 | 12.8 | 9.49 | 57.7 | 3.5 |
| 20.40 | 11.8 | 12.73 | 31.1 | 14.4 |

### 4.1.3  Duration-hardness pairs

A two-dimensional scatter plot is shown in Fig. 4.46. One axis is the duration $T_{90}$, used in the previous section, another axis is the hardness ratio, i.e., the ratio of GRB counts over $T_{90}$ in the energy bands $(25 - 120)\,\mathrm{keV}$ / counts in the band $(120 - 1500)\,\mathrm{keV}$.

Using the maximum likelihood method, two and three bivariate log-normal functions were fitted to search for clusters (see Horváth et al. (2004, 2006) or Appendix A). In the left panel of Fig. 4.46, the best-fit solution of two bivariate log-normal functions is presented (11 independent parameters, since the two weights must add up to $100\,\%$). The parameters are listed in Tab. 4.6. One result is that the short GRBs are on average harder than long GRBs. After a closer look at the GRB distribution within the short class, one can see that the points within the $80\,\%$ CL ellipse are not evenly distributed. They cluster towards the shortest durations.

The fit with the sum of three log-normal functions (17 independent parameters) is presented in the right panel of Fig. 4.46. The best-fit model parameters are listed in Tab. 4.6. The former short group is clearly separated into two parts. As far as one can tell by sight, the data points scatter evenly within (and around) the 80% CL ellipses. Since the difference in the logarithms of the likelihoods $\Delta \ln L = 10.9$ should equal one half of the $\chi^2$ distribution for 6 degrees of freedom (Horváth et al. (2006)), it follows that the introduction of a third group is statistically significant at the 0.13% level (of being accidental).

**Tab. 4.4.** Parameters of the best fit with two and three log-normal functions done by the maximum likelihood method. $\mu$ are the means, $\sigma$ are the standard deviations, $w$ are the weights of the distribution and $L_2, L_3$ are the likelihoods. Given uncertainties are $(\text{Max}(p_i)\text{-Min}(p_i))/2$, where $p_i$ are the parameters obtained by fitting of data sets, in which the durations were randomly changed by their uncertainties as described in the text.

| Parameter | 2 log-normal | 3 log-normal |
|---|---|---|
| $\mu_{\text{short}}$ | $-0.60 \pm 0.11$ | $-0.64 \pm 0.03$ |
| $\sigma_{\text{short}}$ | $0.25 \pm 0.09$ | $0.20 \pm 0.04$ |
| $w_{\text{short}}$ (%) | $10.2 \pm 2.0$ | $9.4 \pm 0.7$ |
| $\mu_{\text{long}}$ | $1.20 \pm 0.03$ | $1.26 \pm 0.01$ |
| $\sigma_{\text{long}}$ | $0.47 \pm 0.02$ | $0.41 \pm 0.01$ |
| $w_{\text{long}}$ (%) | $89.8 \pm 2.0$ | $84.4 \pm 1.5$ |
| $\mu_{\text{middle}}$ | | $0.17 \pm 0.05$ |
| $\sigma_{\text{middle}}$ | | $0.22 \pm 0.10$ |
| $w_{\text{middle}}$ (%) | | $6.2 \pm 2.1$ |
| $\ln L_2$ | $-389.17$ | |
| $\ln L_3$ | | $-379.95$ |

**Tab. 4.5.** The maximal likelihoods and corresponding probabilities that introducing of the third group is accidental for maximum likelihood fittings (one-dimensional) with two and three log-normal functions of ten different changes of durations by their uncertainties.

| $\ln L_2$ | $\ln L_3$ | Prob. (%) |
|---|---|---|
| $-388.12$ | $-378.86$ | $0.03$ |
| $-390.82$ | $-383.92$ | $0.32$ |
| $-391.90$ | $-380.97$ | $0.01$ |
| $-391.75$ | $-385.37$ | $0.52$ |
| $-392.25$ | $-384.24$ | $0.11$ |
| $-390.62$ | $-383.67$ | $0.30$ |
| $-386.26$ | $-375.54$ | $0.01$ |
| $-392.33$ | $-384.97$ | $0.21$ |
| $-389.16$ | $-380.93$ | $0.09$ |
| $-390.21$ | $-384.32$ | $0.82$ |

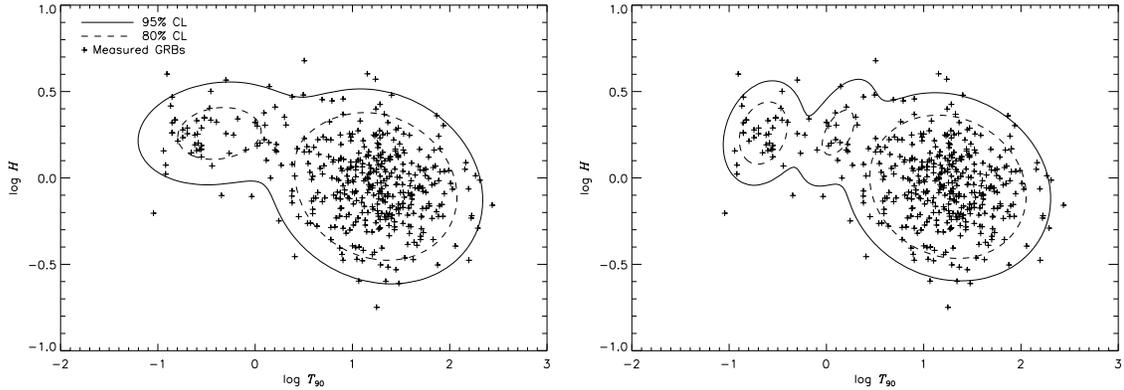

**Fig. 4.46.** Hardness ratio $H$ vs. duration $T_{90}$ of RHESSI GRBs with the best fit of two bivariate log-normal functions (*left panel*) and three bivariate log-normal functions (*right panel*).

In order to estimate how GRB duration and hardness ratio uncertainties effect our result, it was proceeded similarly to the $\chi^2$ fitting. Ten different data sets randomly changed in durations and hardness ratios by the full amount of their uncertainties were generated. The results are presented in Tab. 4.7. From this table, it is seen that all of the simulations infer probabilities lower than 5%, of introducing the third group being accidental. Thus, the hypothesis of introducing the third group is highly acceptable.

### 4.1.4   Discussion of the statistics

The analysis of the one-dimensional duration distribution, by $\chi^2$ fitting, showed a class of long-duration GRBs (about 83% of all RHESSI GRBs) with typical durations from 5 to 70 seconds, the most probable duration being $T_{90} \approx 19$ s. Another class are short GRBs (about 9% of all RHESSI GRBs) with typical durations from 0.1 to 0.4 seconds, the most probable duration being $T_{90} \approx 0.21$ s. By fitting 3 log-normal functions, a class of intermediate-duration GRBs (about 8% of all RHESSI GRBs) with typical length from 0.8 to 3 seconds with the most probable duration being $T_{90} \approx 1.5$ s was identified.

The existence of the intermediate class from the RHESSI $T_{90}$ distribution is not confirmed to a sufficiently high significance using only the $\chi^2$ fit. However, the maximum likelihood ratio test on the same data revealed that the introduction of a third class is statistically significant. The $\chi^2$ method might not be as sensitive and hence robust as the likelihood method, because of the low number of bursts in the data-sample (Horváth et al. (2008), 2nd section, 1st paragraph). The hardness ratio versus duration analysis demonstrates further the existence of a third class. The typical durations are similar to those obtained with the one-dimensional analysis, the percentages being slightly different ($\approx 86\,\%$ long, $\approx 9\,\%$ short, $\approx 5\,\%$ intermediate).

**Tab. 4.6.** Parameters of the best fit with two and three bivariate log-normal functions done by the maximum likelihood method on the RHESSI data. $\mu_x$ are the means on the $x$-axis ($x = \log T_{90}$), $\mu_y$ are the means on the $y$-axis ($y = \log H$), $\sigma_x$ are the dispersions on the $x$-axis, $\sigma_y$ are the dispersions on the $y$-axis, $r$ are the correlation coefficients, $w$ are the weights of the distribution and $L_2, L_3$ are the likelihoods. The given uncertainties are $(\text{Max}(p_i)\text{-Min}(p_i))/2$, where $p_i$ are the parameters obtained by fitting of data sets, where the durations and hardness ratios were randomly changed by their uncertainties.

| Parameter | 2 log-normal | 3 log-normal |
|---|---|---|
| $\mu_{x,short}$ | -0.38 ± 0.03 | -0.65 ± 0.03 |
| $\mu_{y,short}$ | 0.26 ± 0.02 | 0.26 ± 0.01 |
| $\sigma_{x,short}$ | 0.42 ± 0.02 | 0.20 ± 0.03 |
| $\sigma_{y,short}$ | 0.15 ± 0.01 | 0.15 ± 0.02 |
| $w_{short}$ (%) | 14.2 ± 0.5 | 9.2 ± 0.8 |
| $r_{short}$ | 0.14 ± 0.15 | 0.22 ± 0.17 |
| $\mu_{x,long}$ | 1.25 ± 0.01 | 1.25 ± 0.01 |
| $\mu_{y,long}$ | -0.05 ± 0.01 | -0.05 ± 0.01 |
| $\sigma_{x,long}$ | 0.42 ± 0.01 | 0.42 ± 0.01 |
| $\sigma_{y,long}$ | 0.22 ± 0.01 | 0.22 ± 0.005 |
| $w_{long}$ (%) | 85.8 ± 0.5 | 85.5 ± 1.2 |
| $r_{long}$ | -0.14 ± 0.03 | -0.13 ± 0.03 |
| $\mu_{x,middle}$ | | 0.11 ± 0.04 |
| $\mu_{y,middle}$ | | 0.27 ± 0.03 |
| $\sigma_{x,middle}$ | | 0.21 ± 0.10 |
| $\sigma_{y,middle}$ | | 0.17 ± 0.06 |
| $w_{middle}$ (%) | | 5.3 ± 1.9 |
| $r_{middle}$ | | 0.59 ± 0.29 |
| $\ln L_2$ | -323.91 | |
| $\ln L_3$ | | -313.00 |

**Tab. 4.7.** The maximal likelihoods and corresponding probabilities that introducing of the third group is accidental for maximum likelihood fittings (two-dimensional) with two and three bivariate log-normal functions of ten different changes of durations and hardness ratios by their uncertainties.

| $\ln L_2$ | $\ln L_3$ | Prob. (%) |
|---|---|---|
| -349.40 | -339.92 | 0.42 |
| -348.28 | -336.36 | 0.06 |
| -350.66 | -340.80 | 0.31 |
| -347.12 | -338.15 | 0.64 |
| -349.64 | -340.47 | 0.54 |
| -344.59 | -338.17 | 4.56 |
| -350.36 | -342.47 | 1.50 |
| -344.44 | -334.71 | 0.35 |
| -349.37 | -336.51 | 0.03 |
| -348.06 | -338.57 | 0.42 |

Three classes of GRBs have already been identified in the past in the BATSE data (Horváth et al. 2006) and in the Swift data (Horváth et al. 2008). For BATSE, $\approx 65\,\%$ of all GRBs are long, $\approx 24\,\%$ short, and $\approx 11\,\%$ intermediate (Horváth et al. 2006) (Table 2 of that article). The typical durations found for BATSE are roughly a factor of 2 longer than for RHESSI, but consistent for all three classes. For Swift, $\approx 58\,\%$ of all GRBs are long, $\approx 7\,\%$ short and $\approx 35\,\%$ intermediate (Horváth et al. 2008). The percentage of each group obviously depends on the used instrument. As known from BATSE, also in the RHESSI data set, the short GRBs are on average harder than the long GRBs. The most remarkable difference is the hardness of the intermediate class. In the BATSE data, the intermediate class has the lowest hardness ratio, which is anticorrelated with the duration (Horváth et al. 2006), whereas we find for the RHESSI data that its hardness is comparable with that of the short group and rather forward correlated with the duration (although this correlation is inconclusive because of its large error). The hardness of the intermediate class found for the RHESSI data is surprising since the intermediate class in the BATSE data was found to be the softest. It should be noted that for RHESSI the hardness $H$ is defined as $H = F_{120-1500}/F_{25-120}$, whereas for BATSE as $H = F_{100-320}/F_{50-100}$, where the numbers denote energy in keV (the BATSE fluences at higher energies than 320 keV are noisy (Bagoly et al. 1998)). This means that hardnesses do not measure the identical behaviour of bursts. The situation differs even more significantly if we compare hardnesses in the Swift and RHESSI databases, because the Swift's hardnesses are defined as $H = F_{100-150}/F_{50-100}$ and $H = F_{50-100}/F_{25-50}$ (Horváth et al. 2008; Sakamoto et al. 2008).

The shorter durations of the RHESSI GRBs compared to the BATSE GRBs can be understood in the following way. For RHESSI, which is practically unshielded, the background is high (minimum around 1000 counts per second in the $(25 - 1500)$ keV band) and varies by up to a factor 3. Additionally, RHESSI's sensitivity declines rapidly below $\approx 50$ keV. Weak GRBs (in the sense of counts per second) and soft GRBs are not so well observed by RHESSI. Since GRBs tend to be softer and weaker at later times, they should soon fall bellow RHESSI's detection limit, resulting in a shorter duration being inferred.

From the following papers it results that the mean duration for the intermediate group seems to be different for the Swift database on one side (Horváth et al. 2008; Huja et al. 2009), and for the RHESSI and BATSE database on the other side (Horváth 1998; Horváth et al. 2006; Mukherjee et al. 1998; Řípa et al. 2009); for the Swift data this mean duration is $\sim 12$ s resembling the durations of long GRBs, but for the remaining two databases, this mean is below 10 s (1.4 s for RHESSI and 4.3 s for BATSE). Hence, the third group in these three databases not necessarily defines the same GRB class.

### 4.1.5   Conclusions of the statistics

The RHESSI data confirm that GRBs can be separated into short and long classes, and that the short GRBs are on average harder than the long ones. A two-dimensional analysis of the hardness/duration plane as well as a maximum likelihood fit of the duration distribution, however, also indicate a third class with intermediate duration and similar hardness as the short class. From the two-dimensional analysis it follows that the introduction of the third group is statistically significant at the 0.13 % level (of being accidental). The ML analysis applied only on the durations gives that the introduction of the third group is even more statistically significant, i.e., significant on the 0.036 % level (of being accidental).

## 4.2   Effects of annealing on the RHESSI GRB measurements

This chapter discusses the effects of the annealing of the spectrometer aboard the RHESSI satellites on the GRB measurements. Results presented here were also published in papers Řípa et al. (2010) and Veres et al. (2009).

### 4.2.1   The annealing procedure

In November 4-27, 2007, the RHESSI spectrometer underwent the first annealing of its detectors. It was intended to restore its sensitivity that had been gradually deteriorating because of radiation damage (Bellm et al. 2008). It resided in heating up the germanium detectors to over $90°\,$C for one week (operating temperature is about $90\,$K)[1,2]. This procedure was, however, successful only partly because the low-energy response has not been improved as well as the high-energy one[3]. It results in a systematical shift of the measured hardness ratios (at lower energy bands) of observed GRBs after the date of annealing compared to those measured before.

### 4.2.2   The effect of the annealing on the measured spectral indices

An example of pre-annealing and post-annealing spectral fits of the RHESSI data are shown in Tab. 4.8, where the fits with the cutoff power-law model and Band function are performed. Best fitted low-energy and high-energy spectral indices are compared with that of obtained from the Konus and Fermi data. The fitted Band function is of the form:

$$\frac{dN}{dE} \sim \begin{cases} E^\alpha \exp\left(-\frac{E}{E_0}\right) & \text{if } E \leq E_{\text{break}} \\ E^\beta & \text{if } E \geq E_{\text{break}}. \end{cases} \tag{4.2}$$

The cutoff power-law model is of the form:

$$\frac{dN}{dE} \sim E^\alpha \exp\left(-\frac{E}{E_0}\right), \tag{4.3}$$

which is basically the low-energy part of the Band function with $\beta = -\infty$.

**Tab. 4.8.** Spectral fits of cutoff power-law (CPL) and Band function of some selected GRBs. Low-energy spectral indices $\alpha$, high-energy spectral indices $\beta$, peak energies $E_{\text{p}}$ and values of reduced chi squared of the best fits are listed. The RHESSI off-axis angle for all these GRBs is near $90°$. The RHESSI software gives absurd values of low-energy spectral indices while fitting post-annealing data with the simulations of the pre-annealing response matrix. If one fits the data with the Band function the $\alpha$ must be higher than -2 to ensure the integrated $\nu F_\nu$ over all energies be finite.

| GRB | model | data | $\alpha$ | $\beta$ | $E_{\text{p}}$ (keV) | $\chi^2_{\text{r}}$ |
|---|---|---|---|---|---|---|
| | | Swift | -1.37 ± 0.02 | | | 1.27 |
| 061121 | CPL | Konus* | -1.32 ± 0.05 | | 606 ± 80 | 1.01 |
| | | RHESSI | -1.37 ± 0.10 | | 532 ± 57 | 1.01 |
| 080607 | CPL | Swift | -1.15 ± 0.03 | | | 0.70 |
| | | RHESSI | **-2.33 ± 0.18** | | 432 ± 19 | 1.39 |
| 080825 | Band | Fermi | -0.54 ± 0.21 | -2.29 ± 0.35 | 180 ± 23 | 1.23 |
| | | RHESSI | **-5.36 ± 0.86** | -2.92 ± 0.57 | 256 ± 25 | 0.80 |

*GCN 5837

### 4.2.3   The effect of the annealing on the measured hardness ratios

The evolution of the average hardness ratios and their relation to the annealing is presented here. Figures 4.47 and 4.49 show the development of the average GRB RHESSI hardness ratios $H_{21}$ and $H_{32}$ over the years. $H_{21}$ is a low-energy ratio, i.e., the count ratio of a GRB signal over the $T_{90}$ in the energy ranges $(120 - 400)\,\text{keV}$ / $(25 - 120)\,\text{keV}$. $H_{32}$ is a higher energy hardness ratio and it was calculated in the same way like $H_{21}$, but in the ranges $(400 - 1500)\,\text{keV}$ / $(120 - 400)\,\text{keV}$. The development of the average hardness ratio $H$ (here denoted as $H_{231}$) used in the statistical analysis in Section 4.1.3 is presented in Fig. 4.48. The development of the average $T_{90}$ durations is presented in Fig. 4.50.

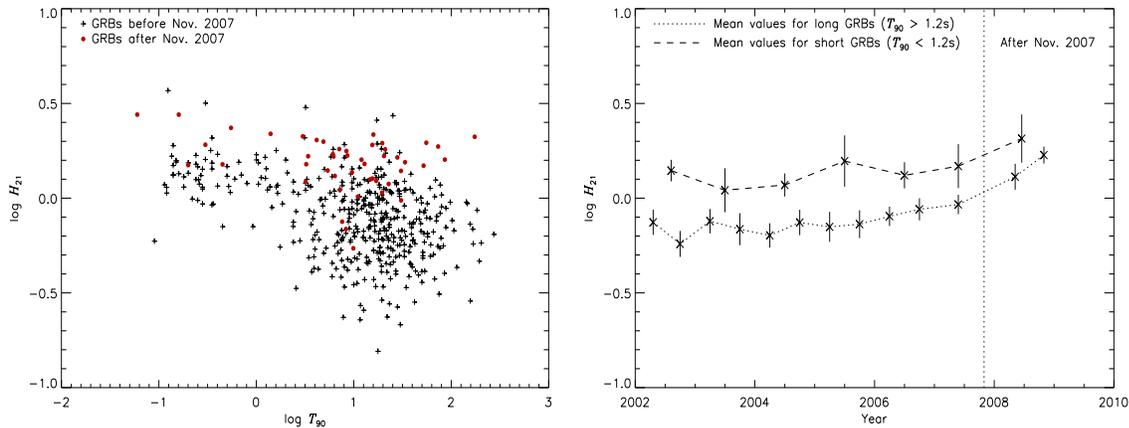

**Fig. 4.47.** The lower energy hardness ratio $H_{21}$ plotted as a function of duration $T_{90}$ (*right panel*) and mean values over different time periods (*left panel*). The error bars are of $2\,\sigma$. Data after the annealing are emphasized by full dots and their systematical shift to the higher values is seen.

### 4.2.4   Conclusions of the effects of the annealing

From Fig. 4.47 and Fig. 4.48 it is seen that the measurements of the low-energy hardness ratio $H_{21}$ and the ratio $H_{231}$, used in the statistical test in the previous chapter, were affected by the annealing procedure and their average values of the post-annealing period are systematically shifted upwards. Contrary to this, the hardness ratio $H_{32}$ of the higher energy channels remains, on average, the same. The $T_{90}$ durations seems not to be affected (Fig. 4.50).

Table 4.8 presents a comparison of the spectral parameters of one GRB from the pre- and two from the post-annealing time. Whereas the low-energy photon index for the pre-annealing burst GRB 061121, detected by Swift, Konus, and RHESSI, was found to be approximately the same, in case of the post-annealing bursts the situation is different. The RHESSI low-energy index for bursts GRB 080607 and GRB 080825 markedly differ from those obtained by the Swift or Fermi satellites.

All this indicates that the RHESSI low-energy sensitivity has not been recovered as well as the high-energy one by the annealing procedure. The detector response matrix has changed and thus using pre- and post-annealing count fluxes of an energy band $\lesssim 120\,\mathrm{keV}$ and ratios of fluxes including counts from these low energies might be problematic. Also any spectral analysis at these low energies, employing the simulations of the old response matrix and the post-annealing data, might be problematic.

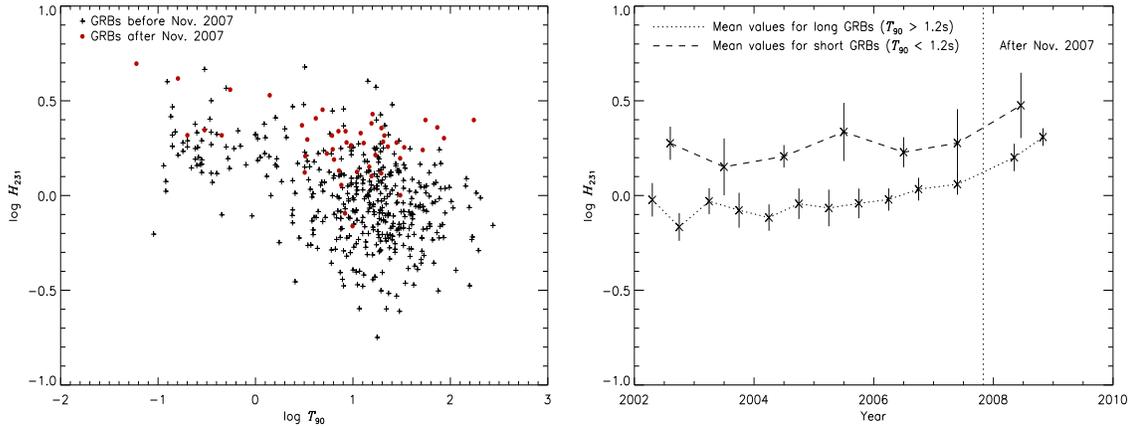

**Fig. 4.48.** The hardness ratio $H_{231}$ used throughout the previous chapter plotted as a function of duration $T_{90}$ (*right panel*) and its mean values over different time intervals (*left panel*). The error bars are of $2\sigma$. Data after the annealing are emphasized by full dots. Also this hardness ratio has been effected by the annealing and its systematical shift to the higher values in the post-annealing time period is seen.

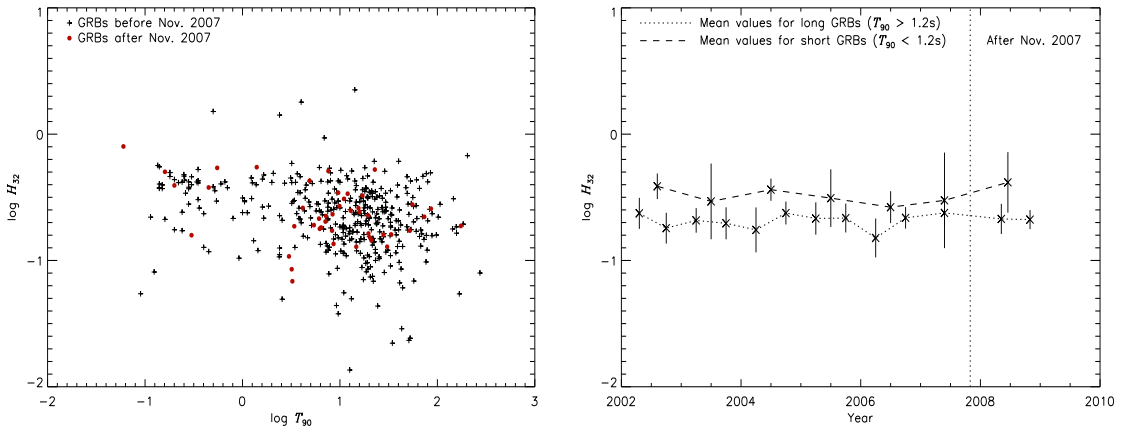

**Fig. 4.49.** The higher energy hardness ratio $H_{32}$ plotted as a function of duration $T_{90}$ (*right panel*) and mean values over different time intervals (*left panel*). The error bars are of $2\sigma$. Data after the annealing are emphasized by full dots. There is no significant difference between the pre- and the post-annealing time period.

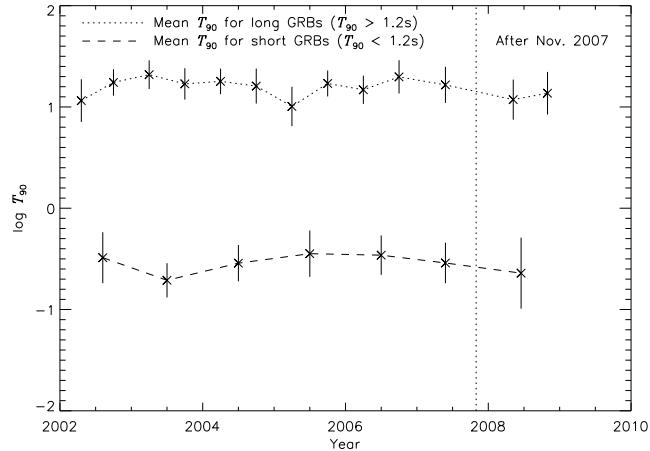

**Fig. 4.50.** Average $T_{90}$ for long- and short-duration GRBs over the time. The error bars are of $2\,\sigma$. There is not seen any significant difference between the pre- and the post-annealing time period.

## 4.3  Results and discussion of the statistics with the pre-annealing data only

In the statistical analysis presented in Section 4.1 and published in the paper Řípa et al. (2009) there were used also the measurements obtained after the annealing of the RHESSI detectors. Afterwards it was found that the measurements of the hardness ratio of the counts between $(25-120)\,\mathrm{keV}$ and $(120-1500)\,\mathrm{keV}$ have been effected by the annealing procedure. Measurements of the $T_{90}$ durations seems not to be affected. Therefore in this section the bivariate ML analysis is repeated, but only with the data from the pre-annealing period, in order to ascertain whether the found intermediate GRB group was not only an artefact. The logarithmic likelihoods for the best fit with the sum of two and three log-normal functions are $\ln L_2 = -305.8$ and $\ln L_3 = -295.2$. The difference is 10.6 and it should be equal to one half of the $\chi^2$ distribution for 6 degrees of freedom. And thus one can infer the probability of $0.16\,\%$ that the intermediate group is statistically accidental. The best fits are shown in Fig. 4.51 and the best fit parameters in Tab. 4.9.

This resultant probability is smaller than $5\,\%$ and similar to the probability obtained by the analysis including both the pre- and post-annealing period (Sect. 4.1.3). Therefore one can conclude that the intermediate group identified in Section 4.1.3 is not an artefact caused by using of the post-annealing data.

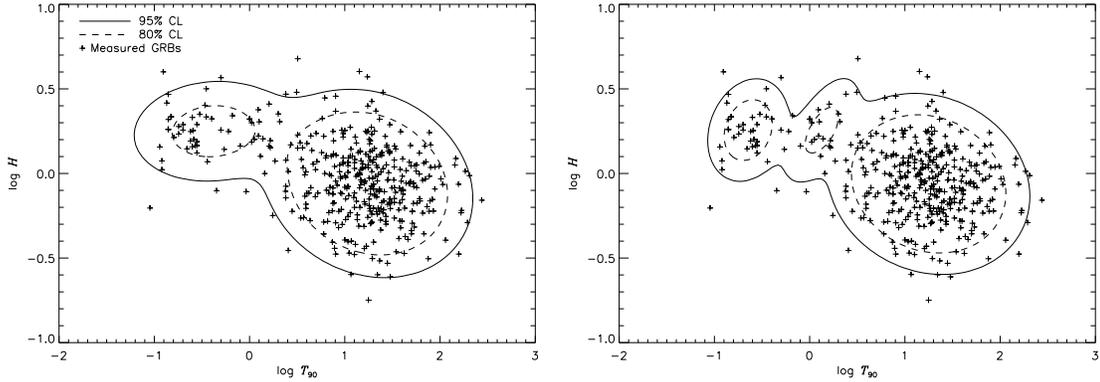

**Fig. 4.51.** Hardness ratio $H$ vs. duration $T_{90}$ of the RHESSI GRB data only from the pre-annealing period. The best fits of two bivariate log-normal functions (*left panel*) and three bivariate log-normal functions (*right panel*) are displayed.

**Tab. 4.9.** Parameters of the best fit with two and three bivariate log-normal functions done by the maximum likelihood method on the RHESSI data from the pre-annealing period only. The nomenclature is the same as in Tab. 4.6.

| Parameter | 2 log-normal | 3 log-normal |
|---|---|---|
| $\mu_{x,\text{short}}$ | -0.39 | -0.65 |
| $\mu_{y,\text{short}}$ | 0.25 | 0.26 |
| $\sigma_{x,\text{short}}$ | 0.42 | 0.21 |
| $\sigma_{y,\text{short}}$ | 0.15 | 0.15 |
| $w_{\text{short}}$ (%) | 14.2 | 9.5 |
| $r_{\text{short}}$ | 0.08 | 0.21 |
| $\mu_{x,\text{long}}$ | 1.24 | 1.25 |
| $\mu_{y,\text{long}}$ | -0.06 | -0.06 |
| $\sigma_{x,\text{long}}$ | 0.42 | 0.42 |
| $\sigma_{y,\text{long}}$ | 0.21 | 0.21 |
| $w_{\text{long}}$ (%) | 85.8 | 85.4 |
| $r_{\text{long}}$ | -0.16 | -0.14 |
| $\mu_{x,\text{middle}}$ | | 0.12 |
| $\mu_{y,\text{middle}}$ | | 0.26 |
| $\sigma_{x,\text{middle}}$ | | 0.20 |
| $\sigma_{y,\text{middle}}$ | | 0.17 |
| $w_{\text{middle}}$ (%) | | 5.1 |
| $r_{\text{middle}}$ | | 0.64 |
| $\ln L_2$ | -305.84 | |
| $\ln L_3$ | | -295.20 |

# 5  A Closer Look at the Intermediate Bursts

The group of intermediate GRBs found in the RHESSI data in Section 4.1.3 in the duration-hardness plane are in this chapter investigated in more detail with respect to their spectral lags, peak count rates, redshifts, supernova observations, and other properties. The Kolmogorov-Smirnov test is used to decide whether these bursts belong to the group of short or long GRBs, or if they significantly differ from both groups. The content of in this section draws from the paper Řípa et al. (2010b).

At first one has to decide which particular GRBs belong to the group of short, intermediate, and long bursts, respectively. It was proceeded in the following way. The probability density function used in the fits in Section 4.1.3 composes of the sum of three bivariate log-normal functions (see equation 8.10 in Appendix A) $f(x, y) = f_1(x, y) + f_2(x, y) + f_3(x, y)$, where $x$ are logarithms of durations and $y$ are logarithms of hardness ratios of different GRBs and $f_1$, $f_2$, and $f_3$ are components corresponding to the particular groups. Here a burst at a point $[x_0; y_0]$ is considered as short, intermediate or long depending on whether the $f_1(x_0, y_0)$, $f_2(x_0, y_0)$ or $f_3(x_0, y_0)$ is maximal. The number of GRBs in the short, intermediate, and long group is 40, 24, 363, respectively (see Fig. 5.52).

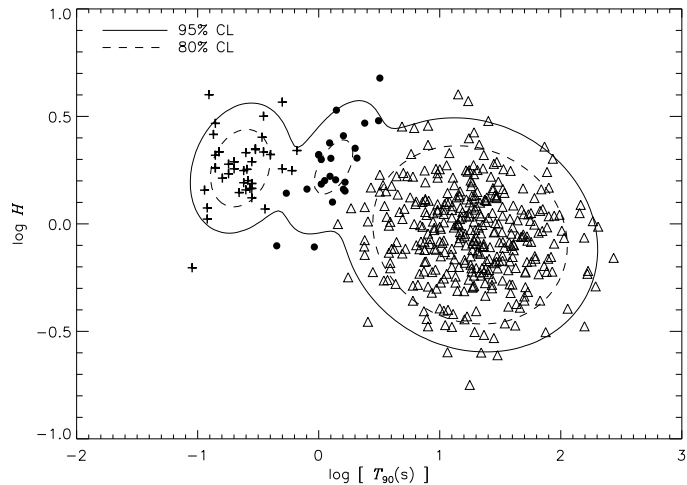

**Fig. 5.52.** The hardness ratio $H$ vs. $T_{90}$ duration plot with the best ML fit of three bivariate log-normal functions is shown. Different symbols correspond to different identified GRB groups. The crosses, full circles, and triangles mark GRBs belonging to the group of short-, intermediate-, or long-duration bursts, respectively.

## 5.1  Spectral lags

It has been found that spectral lags, i.e., the time delay between low and high energy photons, of gamma-ray bursts from short and long groups are different. For short bursts, an average lag is $\sim 20 - 40$ times shorter than for long bursts, and the lag distribution is close to be symmetric about zero - unlike long bursts (Norris et al. 2000a,b, 2001).

In this work the spectral lags of the RHESSI data were calculated similarly to works Norris (2002); Foley et al. (2009), i.e., by fitting of the peak of the cross-correlation function (CCF, see Appendix A) of the background-subtracted count light-curves at two channels, $400 - 1500$ keV and $25 - 120$ keV, by a third order polynomial. The position of the maximum of the polynomial fit is then the spectral lag. An example of such a fit is shown in Fig. 5.53.

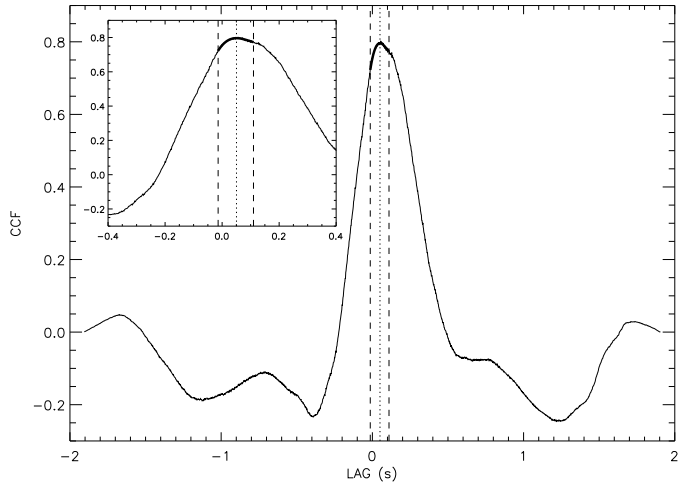

**Fig. 5.53.** An example of cross-correlation function of background-subtracted count light-curve (here very bright GRB 060306) at two different energy bands, $400 - 1500$ and $25 - 120$ keV. The inset presents the fit of the CCF peak by a third order polynomial (thick solid curve). The position of the maximum of the polynomial fit (dotted line) defines the spectral lag. The boundaries of the fit are marked with dashed lines.

To account for the statistical errors, the Monte Carlo method was used. 1001 synthetic count profiles for each GRB were prepared in this way: The measured count profiles were randomly influenced by the Poisson noise and then the background was subtracted. The RHESSI count rates are sometimes 'decimated', which means that, as the rate becomes too high or the onboard solid-state recorder becomes too full, a part of the recorded photon counts are removed. If a decimation occurs, the fraction $(F-1)/F$ of the counts bellow a decimation energy $E_0$ is removed. $F$ is a decimation factor (weight), usually equal to 4 or 6. All events above $E_0$ are downlinked[1]. To prepare the synthetic count profiles the number of counts in each bin were changed accordingly to the Poisson distribution with the average rate $C$, $F.C_{dc}$, or $C_1 + F.C_{2,dc}$ for non-decimated, fully decimated, or partially decimated data, respectively. $C$ is the measured count number in a bin for non-decimated data. $C_{dc}$ is the count number in a bin of fully decimated data and consequently corrected for this decimation. $C_1$ is the count number in the non-decimated portion and $C_{2,dc}$ is the corrected count number in the decimated portion of the measured rate in case of partially decimated data. A detailed explanation is written in Appendix C. The CCF was fitted for each of the 1001 synthetic profiles and for each burst in our sample. The median of such a distribution of 1001 maxima of polynomial fits was taken as the true lag $L$ for each burst. Results are presented in Fig. 5.54.

---

[1]http://sprg.ssl.berkeley.edu/ dsmith/hessi/decimationrecord.html

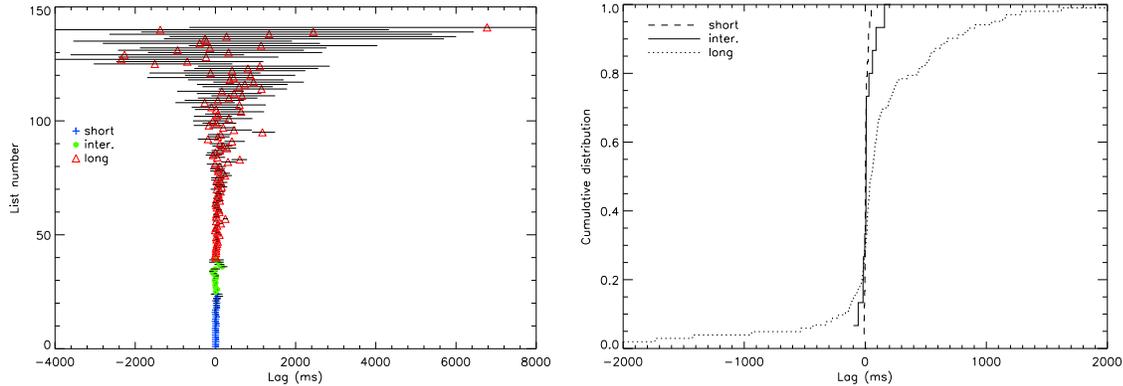

**Fig. 5.54.** *Left panel:* The spectral lags of RHESSI GRBs sorted along the *y*-axis with respect to the group type, and the +error and |-error| are shown. Crosses, full circles, and triangles denote median lags for short-, intermediate-, and long-duration GRBs, respectively. The median lags, for each GRB, were taken from the lags of 1001 synthetic background-subtracted count time profiles obtained by the Monte Carlo simulations of the measured profiles that were randomly influenced by the Poissonian noise. The error bars are of more than 95% CL and are composed of the statistical error (0.025 quantile of lag distribution from the 1001 simulations) and the profile time resolution. A positive lag means that low-energy counts are delayed. *Right panel:* The cumulative distributions of the obtained median lags for the three GRB groups. A difference between lags of long- and short- or long- and intermediate-duration bursts is clearly seen. The lag distributions of short and intermediate GRBs are similar.

This technique allows to calculate the statistical errors. The total uncertainties $\sigma_L$ of such calculated lags were estimated to be $\sigma_L = \sqrt{\sigma_s^2 + \delta_t^2}$, where $\sigma_s$ is the difference between the median lag $L$ and 0.025 quantile of the $L$ distribution and $\delta_t$ is the time resolution of the GRB time profile. Thus the total errors are of more than 95% CL. The time resolutions $\delta_t$ vary from 1 to 1000 ms depending on the intensity of the burst and it was kept as fine as possible, but to ensure the CCF be smooth enough to make the polynomial fitting possible. The evaluation of the spectral lags were performed only on the bursts with signal-to-noise ratio (defined in the same way as in Section 4.1) during $T_{90}$ duration higher than 3.5, in both channels. In Section 4.1 there are 427 GRBs studied. In 286 cases the signal-to-noise ratio, in at least one of the two channels, during $T_{90}$ and in one of the used channels was lower than 3.5 or the CCF did not have a clear peak and thus the evaluation of the lag was not possible. Therefore in what follows only the spectral lags of 141 GRBs are studied. The data can be found in Tab. 9.34−9.34 in Appendix B.

The Kolmogorov-Smirnov (K-S) test was used to compare the lags between different GRB groups. The results are summarized in Tab. 5.10.

As one can see in Fig. 5.54, in some cases the error bars are very large (in the order of seconds) with respect to the values of the median lags itself. This might make the results not so decisive. Therefore the K-S test was repeated for a sample from witch $N$ events with the highest values of (+error + |-error|) had been removed. $N$ has been selected to 5, 10, 15, and 20, respectively. The results are presented in Fig. 5.55 and Tab. 5.11.

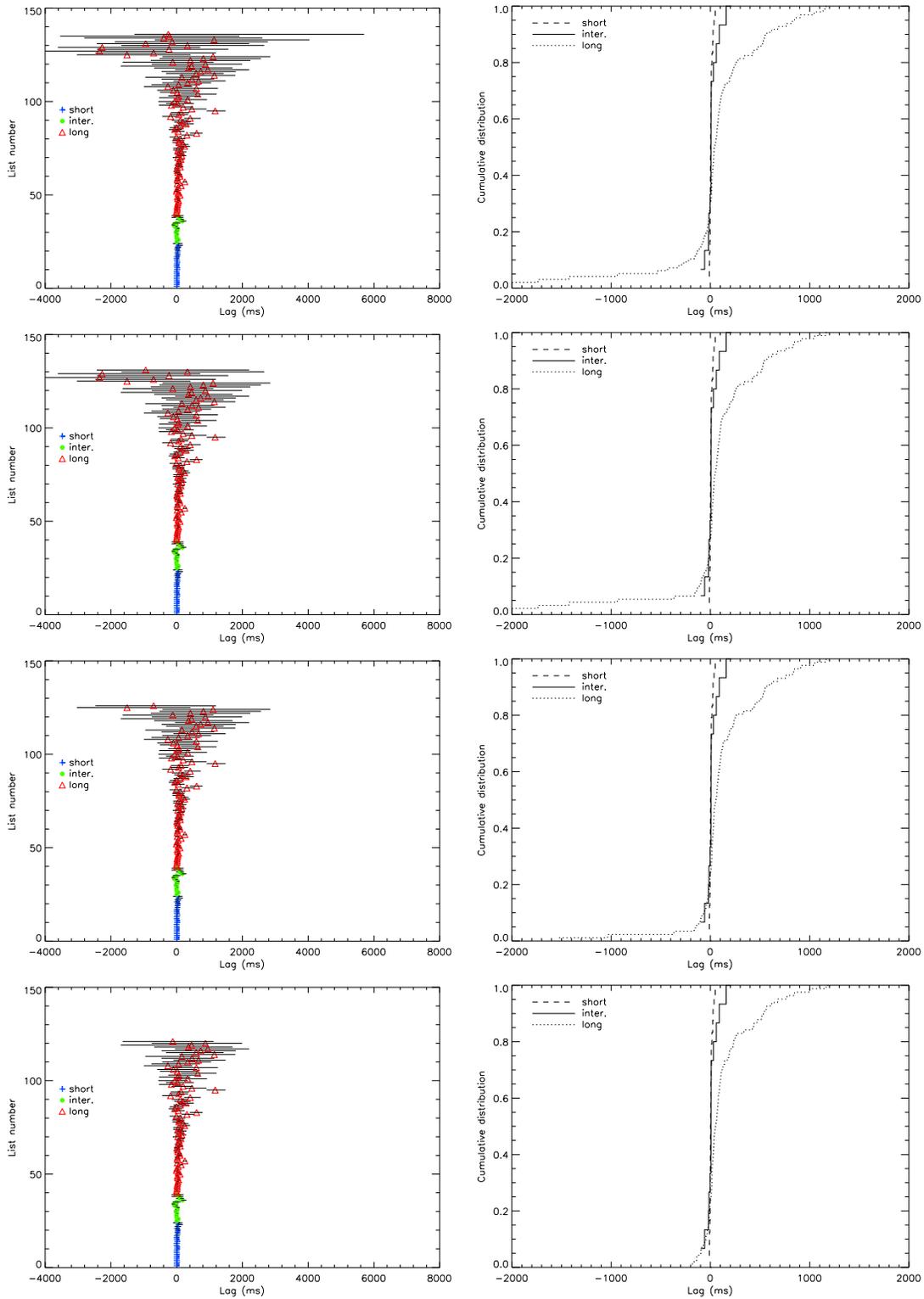

**Fig. 5.55.** Spectral lags of three different GRB groups and its cumulative distributions for the sample with $N$ highest-error events removed. $N$ is equal to 5, 10, 15, and 20 from top to down. Error bars are the same as in Figure 5.54.

| Groups | $D$ | K-S $P$ (%) | | Group | Mean $L$ (ms) | Median $L$ (ms) | $\sigma$ (ms) |
|---|---|---|---|---|---|---|---|
| inter.-short | 0.33 | 20.7 | | short | 7.7 | 2.9 | 16.2 |
| inter.-long | 0.42 | **1.36** | | inter. | 22.5 | 5.9 | 68.8 |
| short-long | 0.49 | **0.009** | | long | 177.0 | 52.8 | 884.7 |

**Tab. 5.10.** *Left part:* Results from the Kolmogorov-Smirnov test of equality of the spectral lag distributions are presented. The K-S distance, and the K-S probability is mentioned. The null hypothesis, that two samples are drawn from the same distribution, can be rejected only at a higher or equal significance level than the value of the K-S probability. Probabilities lower than 5 % are highlighted by boldface. *Right part:* The mean, median and standard deviations of the lags are shown.

| $N$ | Groups | $D$ | K-S $P$ (%) | Group | Mean $L$ (ms) | Median $L$ (ms) | $\sigma$ (ms) |
|---|---|---|---|---|---|---|---|
| 5 | inter.-long | 0.41 | **1.6** | long | 72.1 | 38.4 | 505.3 |
|   | short-long | 0.47 | **0.02** | | | | |
| 10 | inter.-long | 0.44 | **0.89** | long | 79.0 | 52.8 | 498.7 |
|    | short-long | 0.50 | **0.008** | | | | |
| 15 | inter.-long | 0.47 | **0.44** | long | 145.7 | 61.2 | 347.5 |
|    | short-long | 0.53 | **0.003** | | | | |
| 20 | inter.-long | 0.48 | **0.38** | long | 151.8 | 61.2 | 270.3 |
|    | short-long | 0.54 | **0.002** | | | | |

**Tab. 5.11.** *Left part:* Results of the K-S test of spectral lags for samples with $N$ highest-error events removed. *Right part:* Means, medians, and standard deviations of the lags for the sample of long GRBs with $N$ removed events are noted. The samples of short and intermediate bursts remain the same.

In order to test how robust are the results obtained by the K-S test on the sample of median lags, one can use the Bootstrap method (Davison & Hinkley 2004). We randomly resample (with replacements) the original data sets 10,000 times (for each GRB group) so that the number of data points in a bootstrap resample is equal to the number of data points in the original data set and a bootstrap resample may contain duplications. Then we recalculated the K-S test for these resampled data sets. It was obtained that, when comparing intermediate and short GRBs, the K-S probability is higher than 5 % in 6,538 cases out of 10,000. For intermediate vs. long GRBs the K-S probability is higher than 5 % in 1,629 cases. For short vs. long GRBs the K-S probability is never higher than 5 %. The result is shown in Fig. 5.56.

Again to avoid the inclusion of the events without accurate determined lags, the K-S test was repeated for the sample in which the $N$ events with the highest errors, i.e., with the value (+error + |-error|), had been removed. $N$ has been selected to 5, 10, 15, and 20, respectively. The results are presented in Fig. 5.57 and Tab. 5.12.

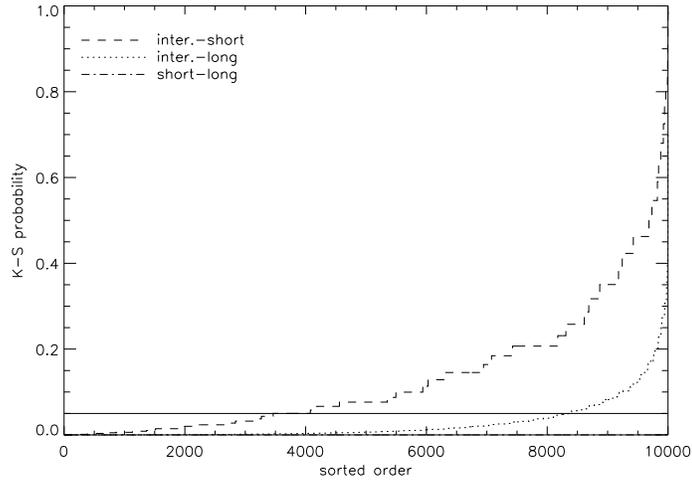

**Fig. 5.56.** K-S probabilities of the tests applied on the lag samples obtained from 10,000 random resamplings with replacements of the original datasets (Bootstrap method) for different GRB groups. The horizontal solid line is the 5 % threshold.

| Groups | $N = 5$ | $N = 10$ | $N = 15$ | $N = 20$ |
|---|---|---|---|---|
| inter.-short | 6647 | 6647 | 6626 | 6513 |
| inter.-long | 2094 | 1560 | 1025 | 1038 |
| short-long | 0 | 0 | 0 | 0 |

**Tab. 5.12.** The listed numbers mean how many times was the K-S probability higher than 5 % obtained from 10,000 tests of bootstrap-resampled data sets of median lags for different GRB groups. $N$ means how many events with the highest errors had been removed from the sample.

## 5.2   Peak count rates

Next we focused on the count peak rates. Figure 5.58 shows the RHESSI peak count rates vs. $T_{90}$ durations and the cumulative distributions of the three GRB groups identified by the ML analysis of the hardnesses and durations. The K-S probabilities were calculated to compare the peak rate distributions. The results are noted in Tab. 5.13.

## 5.3   Redshifts

Concerning redshifts, the number of redshift measurements for the short RHESSI GRBs is only two and their average value is $0.49 \pm 0.69$. For long bursts the number is 17 and the average redshift is $1.16 \pm 0.33$. Unfortunately no redshift measurements are available for the RHESSI intermediate bursts.

## 5.4   The sky distribution

Figure 5.59 displays the sky distribution of all localized bursts in our RHESSI sample. Unfortunately there are only two well localized intermediate events, therefore any statistical analysis is meaningless.

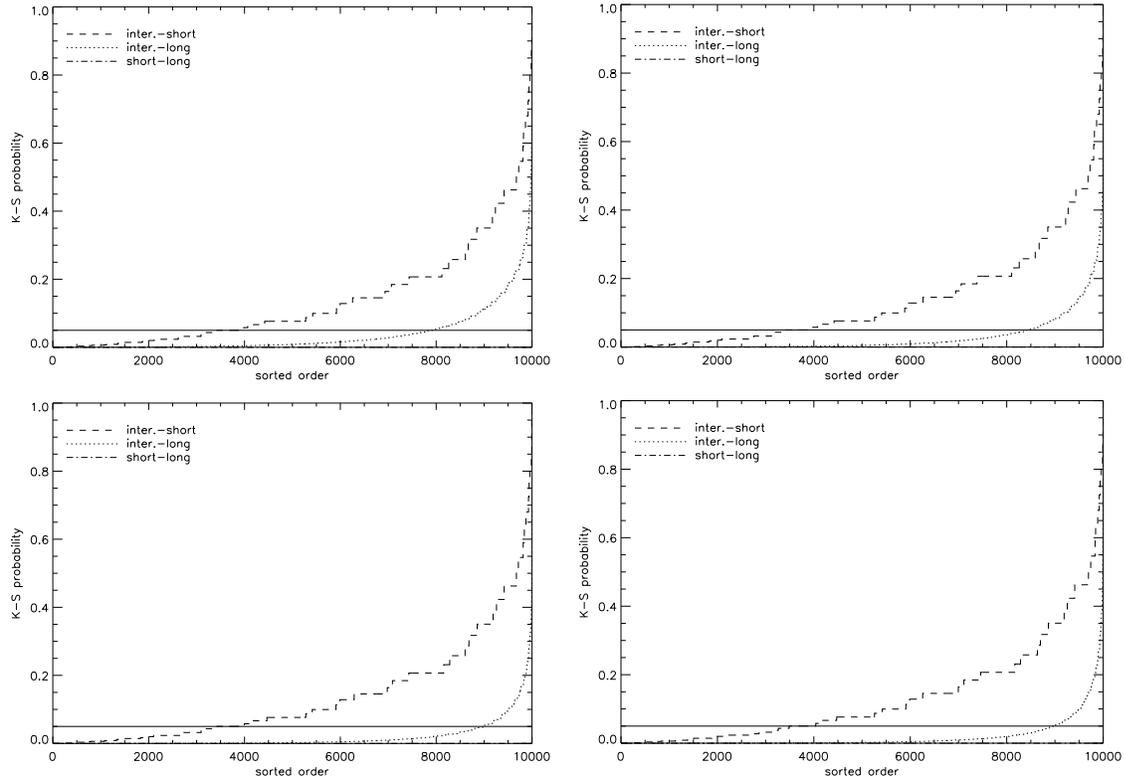

**Fig. 5.57.** The same matter like in Fig. 5.56, but for the samples with $N$ highest-error events removed. $N$ is equal to 5 (top left), 10 (top right), 15 (bottom left), and 20 (bottom right).

## 5.5   SN observations

Concerning supernova observations, it was found that all GRBs with underlying or possible underlying SN have, by RHESSI, long duration and there is no SN observation for any RHESSI intermediate-duration burst (Tab. 5.14). This finding is similar to the fact that, in general, there is no SN-GRB connection for short-duration bursts associated with the model of the merger of two compact astrophysical objects.

## 5.6   Star formation rates of the host galaxies

It was also tried to find the star formation rates (SFR) of the host galaxies of the intermediate bursts of the RHESSI sample, particularly in works Berger (2009) and Levesque et al. (2010), and compare them with the SFRs of the hosts of the short and long bursts. However, there are not such measurements for RHESSI intermediate-duration bursts.

## 5.7   Results and discussion of the intermediate bursts

The K-S test applied on the spectral lag distributions of the short- and intermediate-duration bursts gives the K-S probability 20.7 % and K-S distance 0.33. The tabular critical value of the K-S distance at the significance level $\alpha = 5\,\%$ and the given number of the elements (24 and 15) is 0.45. Therefore the null hypothesis that the lags of the short and intermediate bursts are drawn from the same distribution cannot be rejected at the 5 % (or lower ) significance level.

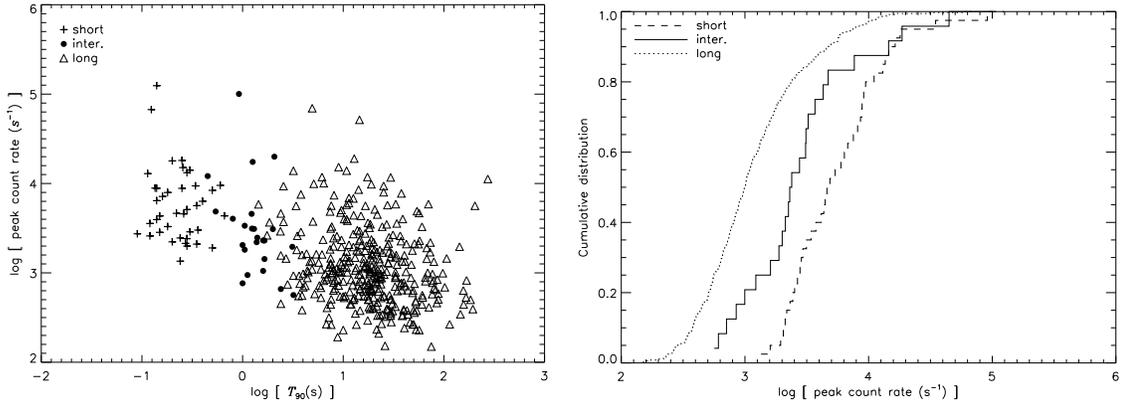

**Fig. 5.58.** *Left panel:* Logarithmic peak count rates as a function of logarithmic $T_{90}$ durations for the three GRB groups, identified by the analysis of the hardnesses and durations, are displayed. *Right panel:* Cumulative distributions of peak count rates for the short-, intermediate-, and long-duration RHESSI GRBs.

| Groups | $D$ | K-S $P$ (%) | | Group | Mean $\log F\,(\mathrm{s}^{-1})$ | Median $\log F\,(\mathrm{s}^{-1})$ | $\sigma$ |
|--------|-----|-------------|---|-------|------|--------|----------|
| inter.-short | 0.36 | **3.1** | | short | 3.76 | 3.71 | 0.41 |
| inter.-long | 0.48 | **3x10$^{-3}$** | | inter. | 3.48 | 3.40 | 0.51 |
| short-long | 0.72 | **2x10$^{-15}$** | | long | 3.06 | 3.00 | 0.43 |

**Tab. 5.13.** *Left part:* Results of the K-S test applied on the logarithmic peak count rates ($\log F$). The shortcuts have the same meaning as in the Tab. 5.10. *Right part:* The mean, median and standard deviations of the logarithmic peak count rates. Probabilities lower than 5 % are again emphasize by the boldface.

On the other hand, the same test applied on the lags of the intermediate vs. long (and short vs. long) bursts gives the K-S probability 1.36 % (0.01 %) and the K-S distance 0.42 (0.49). The tabular critical value of the K-S distance at the 5 % significance level for the number of the intermediate and long bursts (15 and 102) is 0.38. Therefore the hypothesis that the lags of the intermediate and long bursts are drawn from the same distribution can be rejected at this 5 % significance level. Similarly, it can be rejected, at the 5 % significance level, that the lags of short and long bursts are drawn from the same distribution. The critical K-S distance is in this case 0.31.

On the other side, concerning the results obtained in the part based on the K-S tests of 10,000 resampled sets of median lags (whole sample, i.e., $N=0$), the following was obtained: when comparing intermediate and short GRBs, the K-S probability is higher than 5 % in 6,538 cases out of 10,000, for intermediate vs. long GRBs the K-S probability is higher than 5 % in 1,629 cases, and for short vs. long GRBs the K-S probability is never higher than 5 %. These results indicate that it is not entirely sure that the intermediate bursts have the same lag distribution as the short bursts do and that its distribution differ from that of the long bursts.

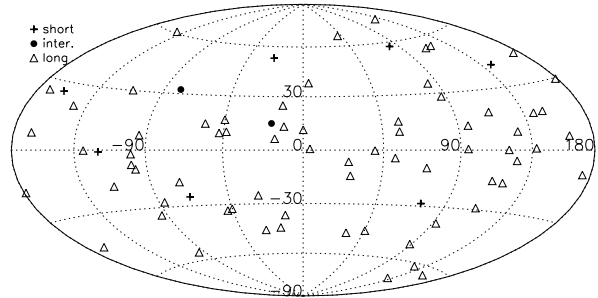

**Fig. 5.59.** Sky distribution of the three RHESSI GRB groups in the Hammer-Aitoff projection with Galactic coordinates (longitude is horizontal).

From the results of the K-S test applied on the peak count rates it follows that the distributions of the peak count rates are different over all three groups. Particularly the K-S probability for the intermediate vs. short busts is 3.1 %, for intermediate vs. long bursts it is $3 \times 10^{-3}$ %, and for short vs. long bursts $2 \times 10^{-15}$ %.

Figure 3.41 in Section 3 shows that initial hard peaks of the short bursts with extended emissions have longer duration than ordinary short bursts and simultaneously shorter duration than the long bursts. This invokes a question whether the intermediate bursts in the RHESSI sample are in real (or are contaminated with) SGBEE. We therefore looked at the RHESSI count light-curves of the intermediate bursts, but any soft extended emission cannot be seen for any of these bursts. Either these burst do not have any extended emission, or it cannot be simple seen by the RHESSI instrument, because its sensitivity rapidly drops bellow $\approx 50$ keV. One should explore the observations from other instruments with better low-energy sensitivity, e.g., with Konus-Wind, Super-AGILE, or HETE-2.

Veres et al. (2010) studied Swift GRBs and found a significant intermediate group. They concluded that these Swift intermediate bursts are closely related to XRFs. The RHESSI intermediate bursts have approximately the same average hardness ratio as the short bursts. Also the finding that spectral lags of the RHESSI intermediate bursts are similar to the lags of the short bursts is different from the result found by de Ugarte Postigo et al. (2010), i.e., that the intermediate bursts in the Swift sample have lags similar with the lags of the long-duration bursts. Therefore it seems improbable that the RHESSI intermediate bursts would be also related to XRFs.

| GRB | $T_{90}$ (s) | SN | Reference |
|---|---|---|---|
| 021211 | 4.3 | possible | Della Valle et al. (2003) |
| 030329A | 17.4 | 2003dh | Stanek et al. (2003) |
| 041006 | 11.1 | possible | Garg et al. (2004) |
| 050525A | 7.4 | 2005nc | Della Valle et al. (2006) |
| 060614 | 52.5 | possible | Brown & Holland (2006) |

**Tab. 5.14.** SN/GRB connections for the RHESSI long-duration GRBs.

# 6 Cosmological Effects on the Observed Fluxes

Next part of this thesis focuses on the cosmological effects on the observed flux and fluence distributions of gamma-ray bursts and tries to answer the simple question: "Are the most distant bursts in general the faintest ones"?

It draws from the article Mészáros et al. (2011), in which the author of this thesis is the second co-author (out of three). A big contribution to this part is thus thanks to the other two co-authors.

Several claims have been put forward that an essential fraction of long-duration BATSE gamma-ray bursts should lie at redshifts larger than 5. This point-of-view follows from the natural assumption that fainter objects should, on average, lie at larger redshifts. However, redshifts larger than 5 are rare for bursts observed by Swift, seemingly contradicting the BATSE estimates. The purpose of this work is to clarify this contradiction. In what follows the cosmological relationships between the observed and emitted quantities are derived, and it is arrived at a prediction that can be tested on the ensembles of bursts with determined redshifts. This analysis is independent on the assumed cosmology, on the observational biases, as well as on any gamma-ray burst model. Four different samples are studied: 8 BATSE bursts with redshifts, 13 bursts with derived pseudo-redshifts, 134 Swift bursts with redshifts, and 6 Fermi bursts with redshifts. The controversy can be explained by the fact that apparently fainter bursts need not, in general, lie at large redshifts. Such a behaviour is possible, when the luminosities (or emitted energies) in a sample of bursts increase more than the dimming of the observed values with redshift. In such a case $dP(z)/dz > 0$ can hold, where $P(z)$ is either the peak-flux or the fluence. All four different samples of the long bursts suggest that this is really the case. This also means that the hundreds of faint, long-duration BATSE bursts need not lie at high redshifts, and that the observed redshift distribution of long Swift bursts might actually represent the real distribution.

## 6.1 Introduction

Until the last years, the redshift distribution of gamma-ray bursts has only been poorly known. For example, the Burst and Transient Source Experiment instrument on Compton Gamma Ray Observatory detected around 2700 GRBs[1], but only a few of these have directly measured redshifts from the afterglow (AG) observations (Schaefer 2003; Piran 2004). During the last couple of years the situation has improved, mainly due to the observations made by the Swift satellite[2]. There are already dozens of directly determined redshifts[3] (Mészáros 2006). Nevertheless, this sample is only a small fraction of the, in total, thousands of detected bursts.

Beside the direct determination of redshifts from the AGs, there are several indirect methods, which utilize the gamma-ray data alone. In essence, there are two different methods which provide such determinations. The first one makes only statistical estimations; the fraction of the bursts in a given redshift interval is deduced. The second one determines an actual value of the redshift for a given GRB ("pseudo-redshift").

---

[1]http://gammaray.msfc.nasa.gov/batse/grb/catalog
[2]http://swift.gsfc.nasa.gov/docs/swift/swiftsc.html
[3]http://www.mpe.mpg.de/~jcg/grbgen.html

Already at the early 1990's, that is, far before the observation of any AG, and when even the cosmological origin was in doubt, there were estimations made in the sense of the first method (see, e.g., Paczyński (1992) and the references therein). In Mészáros & Mészáros (1996) a statistical study indicated that a fraction of GRBs should be at very large redshifts (up to $z \simeq 20$). In addition, there was no evidence for the termination of the increase of the numbers of GRBs for $z > 5$ (see also Mészáros & Mészáros (1995); Horváth et al. (1996), and Reichart & Mészáros (1997)). In other words, an essential fraction (a few or a few tens of %) of GRBs should be in the redshift interval $5 < z < 20$. Again using this type of estimation, Lin et al. (2004) claims that even the *majority* of bursts should be at these high redshifts.

The estimations of the pseudo-redshifts in the sense of the second method are more recent. Ramirez-Ruiz & Fenimore (2000); Reichart et al. (2001); Schaefer et al. (2001) and Lloyd-Ronning et al. (2002) developed a method allowing to obtain from the so-called variability the intrinsic luminosity of a GRB, and then from the measured flux its redshift. The redshifts of hundred of bursts were obtained by this method. Nevertheless, the obtained pseudo-redshifts are in doubt, because there is an $(1 + z)$ factor error in the cosmological formulas (Norris 2004; Band et al. 2004). Other authors also query these redshifts (e.g., Guidorzi et al. (2005)). Norris et al. (2000a) found another relation between the spectral lag and the luminosity. This method seems to be a better one (Schaefer et al. 2001; Norris 2002; Ryde et al. 2005), and led to the estimation of $\sim 1200$ burst redshifts. Remarkably, again, an essential fraction of long GRBs should have $z > 5$, and the redshift distribution should peak at $z \sim 10$. A continuation of this method (Band et al. 2004; Norris 2004), which corrected the $(1 + z)$ error in Norris (2002), gave smaller redshifts on average, but the percentage of long GRBs at $z > 5$ remains open. Bagoly et al. (2003) developed a different method allowing to obtain the redshifts from the gamma spectra for 343 bright GRBs. Due to the two-fold character of the estimation, the fraction of GRBs at $z > 5$ remains further open. No doubt has been yet emerged concerning this method. Atteia (2003) also proposed a method in a similar sense for bright bursts. Other methods of such estimations also exist (Amati et al. 2002; Ghirlanda et al. 2005). These pseudo-redshift estimations (for a survey see, e.g., Section 2.6 of Mészáros (2006)) give the results that even bright GRBs should be at redshifts $z \sim (1 - 5)$. For faint bursts in the BATSE data set (i.e., for GRBs with peak-fluxes smaller than $\simeq 1 \, \mathrm{photon}/(\mathrm{cm^2 s})$) one can hardly have good pseudo-redshift estimations, but on average they are expected to be at larger redshifts, using the natural assumption that these bursts are observationally fainter due to their larger distances. Hence, it is remarkable that all these pseudo-redshift estimations also supports the expectation, similarly to the results of the first method, that an essential fraction of GRBs is at very high redshifts.

Contrary to these estimations for the BATSE data set, only five long bursts with $z > 5$ bursts have yet been detected from direct redshift measurements from AGs using more recent satellites. In addition, the majority of measured $z$-s are around $z = 2 - 3$, and there is a clear decreasing tendency[1] towards the larger redshifts. In other words, the redshifts of GRBs detected by the Swift satellite do not support the BATSE redshift estimations; the redshifts of GRBs detected by the non-Swift satellites are on average even at smaller redshifts (Bagoly et al. 2006).

This can be interpreted in two essentially different ways. The first possibility is that a large fraction (a few tens of % or even the majority) of GRBs are at very high redshifts (at $z > 5$ or so). In such a case these bursts should mainly be seen only in the gamma-ray band due to some observational selection (Lamb & Reichart 2000). The second possibility is that the AG data reflect the true distribution of bursts at high redshifts, and bursts at $z > 5$ are really very rare. In this case, however, there has to be something wrong in the estimations of redshifts from the gamma-ray data alone. Since observational selections for AG detections of bursts at $z > 5$ cannot be excluded (Lamb & Reichart 2000), the first point-of-view could be also quite realistic.

The purpose of this work is to point out some statistical properties of the GRBs, which may support the second possibility. Section 6.2 discusses these properties theoretically. In Sections 6.3 and 6.4 we discuss the impact of such a behaviour on several observed burst samples. Section 6.5 summarizes the results of this part.

## 6.2   Theoretical considerations

### 6.2.1   The general consideration

Using the standard cosmological theory and some simple statistical considerations, it will be shown now that, under some conditions, apparently fainter bursts might very well be at smaller redshifts compared to the brighter ones.

As shown by Mészáros & Mészáros (1995), if there are given two photon-energies $E_1$ and $E_2$, where $E_1 < E_2$, then the flux $F$ (in units photons/(cm²s)) of the photons with energies $E_1 \leq E \leq E_2$ detected from a GRB having a redshift $z$ is given by

$$F_{\mathrm{ph}}(z) = \frac{(1+z)\tilde{L}_{\mathrm{ph}}(z)}{4\pi d_1(z)^2} = \frac{\tilde{L}_{\mathrm{ph}}(z)}{4\pi d_{\mathrm{M}}(z)^2(1+z)}, \qquad (6.4)$$

where $\tilde{L}_{\mathrm{ph}}(z)$ is the isotropic luminosity of a GRB (in units photons/s) between the energy range $E_1(1 + z) \leq E \leq E_2(1 + z)$, and $d_1(z)$ is the luminosity distance of the source. The reason of the notation $\tilde{L}(z)$, instead of the simple $L(z)$, is that $L(z)$ should mean the luminosity from $E_1 \leq E \leq E_2$ (Mészáros et al. 2006). One has $d_1(z) = (1 + z)d_{\mathrm{M}}(z)$, where $d_{\mathrm{M}}(z)$ is the proper motion distance of the GRB, given by the standard cosmological formula (Carroll et al. 1992), and depends on the cosmological parameters $H_0$ (Hubble-constant), $\Omega_{\mathrm{M}}$ (the ratio of the matter density and the critical density), and $\Omega_{\Lambda} = 3\lambda c^2/(3H_0^2)$ ($\lambda$ is the cosmological constant, $c$ is the speed of light). In energy units one may write $F_{\mathrm{en}}(z) = \overline{E}F_{\mathrm{ph}}(z)$ and $\tilde{L}_{\mathrm{en}}(z) = (1+z)\overline{E}\tilde{L}_{\mathrm{ph}}(z)$, where $\overline{E}$ ($E_1 < \overline{E} < E_2$) is a typical photon energy ensuring that the flux $F_{\mathrm{en}}(z)$ has the dimension erg/(cm²s). $\tilde{L}_{\mathrm{en}}(z)$ in erg/s unit gives the luminosity in the interval $E_1(1+z) \leq E \leq E_2(1+z)$. Except for an $(1+z)$ factor the situation is the same, when considering the fluence.

Hence, in the general case, we have

$$P(z) = \frac{(1+z)^N \tilde{L}(z)}{4\pi d_\mathrm{l}(z)^2}, \tag{6.5}$$

where $P(z)$ is either the fluence or the flux, and $\tilde{L}(z)$ is either the emitted isotropic total number of photons, or the isotropic total emitted energy or the luminosity. The following values of $N$ can be used: $N = 0$ if the flux is taken in energy units erg/(cm²s) and $\tilde{L}$ is the luminosity with dimension erg/s; $N = 1$ if either the flux and luminosity are in photon units, or the fluence in energy units and the total energy are taken; $N = 2$ if the total number of photons are considered. All this means that for a given GRB - for which redshift, flux and fluence are measured - Equation (6.5) defines immediately $\tilde{L}(z)$, which is then either the isotropic total emitted energy or the luminosity in the interval $E_1(1+z) \leq E \leq E_2(1+z)$. Hence, $\tilde{L}(z)$ is from the range $E_1(1+z) \leq E \leq E_2(1+z)$ and not from $E_1 \leq E \leq E_2$.

Let have a measured value of $P(z)$. Fixing this $P(z)$ Equation (6.5) defines a functional relationship between the redshift $z$ and $\tilde{L}(z)$. For its transformation into the real intrinsic luminosities $L(z)$ the beaming must be taken into account as well (Lee et al. 2000; Ryde et al. 2005; Bagoly & Veres 2009a; Petrosian et al. 2009; Bagoly & Veres 2009b; Butler et al. 2010). Additionally, we need to study the dependence of the obtained $\tilde{L}(z)$ on $z$, and to determine the real luminosities $L(z)$ by the K-corrections Mészáros (2006). It is not the aim of this work to solve the transformation of $\tilde{L}(z)$ into $L(z)$. The purpose of this work is to study the functional relationship among $P(z)$, $z$ and $\tilde{L}(z)$.

Using the proper motion distance $d_\mathrm{M}(z)$ Equation (6.5) can be reordered as

$$4\pi d_\mathrm{M}(z)^2 (1+z)^{2-N} P(z) = \tilde{L}(z). \tag{6.6}$$

The proper motion distance $d_\mathrm{M}(z)$ is bounded as $z \to \infty$ (Weinberg (1972), Chapter 14.4.). This finiteness of the proper motion distance is fulfilled for any $H_0$, $\Omega_\mathrm{M}$ and $\Omega_\Lambda$. Hence, $\tilde{L}(z)$ is a monotonically increasing function of the redshift along with $(1+z)^{2-N}$ for the fixed $P(z) = P_0$ and for the given value of $N \leq 1$. It means $z_1 < z_2$ implies $\tilde{L}(z_1) < \tilde{L}(z_2)$. Expressing this result in other words: the more distant and brighter sources may give the same observed value of $P_0$. Now, if a source at $z_2$ has a $\tilde{L} > \tilde{L}(z_2)$, its observed value $P'_{obs}$ will have $P'_{obs} > P_0$, i.e., it becomes apparently brighter despite its greater redshift than that of the source at $z_1$. The probability of the occurrence this kind of *inversion* depends on $f(\tilde{L}|z)$, on the conditional probability density of $\tilde{L}$ assuming $z$ is given, and on the spatial density of the objects.

It is obvious from Equation (6.6) that for the increasing $z$ the $\tilde{L}(z)$ is also increasing. This effect gives a bias (Lee et al. 2000; Bagoly & Veres 2009a; Petrosian et al. 2009; Bagoly & Veres 2009b; Butler et al. 2010) towards the higher $\tilde{L}(z)$ values among GRBs observed above a given detection threshold. These questions are discussed in detail mainly by Petrosian et al. (2009). There can be also that $\tilde{L}(z)$ is increasing with $z$ due to the metallicity evolution (Wolf & Podsiadlowski 2007). There can be also an intrinsic evolution of the $L(z)$ itself in the energy range $[E_1, E_2]$. Hence, keeping all this in mind, $\tilde{L}(z)$ can well be increasing on $z$, and the inverse behaviour can also occur.

### 6.2.2    A special assumption

Assume now that we have $\tilde{L}(z) \propto (1+z)^q$, where $q$ is a real number, and this relation holds for any $0 < z < \infty$. This means that it holds $\tilde{L}(z) = \tilde{L}_0(1+z)^q$, where $\tilde{L}(z=0) = \tilde{L}_0$. Of course, this special assumption is a loss of generality, because $\tilde{L}(z)$ can be really a function of $z$, but in general it need not have this special form. In addition, to calculate $\tilde{L}(z)$ the cosmological parameters must be specified, which is a further loss of generality. Nevertheless, if this special assumption is taken, the inverse behavior may well be demonstrated.

If $(N+q) > 2$, then one has a highly interesting mathematical behaviour of the function $P(z)$ (Equation 6.5). For $z \ll 0.1$, $P(z)$ decreases as $z^{-2}$, that is, larger redshifts gives smaller fluxes or fluences as expected. However, after some $z = z_{\text{turn}}$ this behavior must change, because as $z$ tends to $\infty$, the function $(1+z)^N \tilde{L}/d_l^2$ tends to infinity following $\propto (1+z)^{N+q-2}$. In other words, for $z > z_{\text{turn}}$ *as redshift increases, the measured $P(z)$ will also increase.* Equivalently stated, "fainter bursts are closer to us". The possibility of this "inverse" behaviour is quite remarkable. It is important to note that the existence of a $z_{\text{turn}}$ is exclusively determined by the value $q$, and the necessary and sufficient condition for it is given by the inequality $(N+q) > 2$. For the existence of a $z_{\text{turn}}$ the values $H_0$, $\Omega_{\text{M}}$ and $\Omega_{\Lambda}$ are unimportant. The value of $z_{\text{turn}}$ can vary depending on the choice of the $\Omega$ parameters, but, however, its existence or non-existence is unaffected.

Moreover, the value of $z_{\text{turn}}$ itself is independent on the Hubble-constant $H_0$. This can be seen as follows. To find $z_{\text{turn}}$ one must simply search for the minimum of the function $Q(z) = (1+z)^{N+q}/d_l(z)^2$, that is, when $dQ(z)/dz = 0$. But, trivially, $Q(z)$ and $Q(z)/H_0^2$ have the same minimum.

The solution of the equation $dQ(z)/dz = 0$ must be found numerically for the general case of Omega parameters. The left panel of Fig. 6.60 shows the function $Q(z)$ for $\Omega_{\text{M}} = 0.27$ and $\Omega_{\Lambda} = 0.73$. For $\Omega_{\Lambda} = 0$ it can be found analytically, because $d_{\text{M}}(z)$ is then an analytical function. For $\Omega_{\text{M}} = 1$ and $\Omega_{\Lambda} = 0$ the condition $dQ(z)/dz = 0$ is easily solvable. For this special case one has to search for the minimum of the function:

$$\frac{(1+z)^{N+q-2}}{(1-(1+z)^{-1/2})^2},\tag{6.7}$$

because here $d_{\text{M}} = (2c/H_0)/(1-(1+z)^{-1/2})$. Then one has

$$z_{\text{turn}} = \left(\frac{N+q-1}{N+q-2}\right)^2 - 1.\tag{6.8}$$

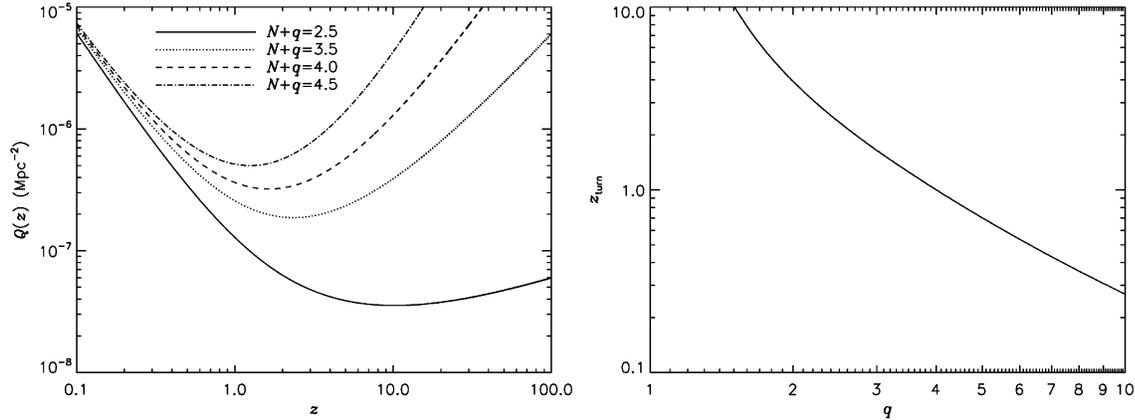

**Fig. 6.60.** *Left panel*: Function $Q(z)$ for $\Omega_M = 0.27$ and $\Omega_\Lambda = 0.73$.   *Right panel*: Dependence of $z_{\mathrm{turn}}$ on $q$ for $\Omega_M = 0.27$ and $\Omega_\Lambda = 0.73$.

## 6.3   The samples

### 6.3.1   The choice of burst samples

The frequency of the occurrence $\tilde{L}(z_1) < \tilde{L}(z_2)$ at $z_1 < z_2$, but for their observed values $P(z_1) < P(z_2)$, i.e., the more distant GRB is apparently brighter for the observer, can be verified on a sample for which there are well-defined redshifts, as well as measured peak-fluxes and/or fluences. At a given redshift the luminosity $\tilde{L}(z)$ is a stochastic variable and starting from Equation (6.6) one can get the probability for $P'_{obs} > P(z)$, assuming that $f(\tilde{L}|z)$ is given.

There are two different subgroups of GRBs, which can be denoted as "short-" and "long-"duration bursts (Norris et al. (2001); Balázs et al. (2003); Mészáros et al. (2006); Zhang et al. (2009)). In addition, the existence of additional subgroups cannot be excluded (Horváth (1998, 2002); Hakkila et al. (2003); Borgonovo (2004); Varga et al. (2005); Horváth et al. (2008); Vavrek et al. (2008); Horváth et al. (2009)). The first direct redshift for a long (short) GRBs was determined in 1997 (2005) (for a detailed survey see, e.g., Piran (2004) and Mészáros et al. (2006)). The few redshifts measured for short bursts are on average small (Mészáros et al. 2009), which motivates us to concentrate on long-duration bursts only in this study.

Since we try to obtain consequences of the GRBs' redshifts also in the BATSE Catalog, we should obviously study the BATSE sample. But only a few of these bursts have directly measured redshifts from afterglow data. Therefore an attempt to obtain conclusions from a sample that uses the so-called "pseudo-redshifts", i.e., the redshifts estimated exclusively from the gamma photon measurements alone, will be made. But for them one should keep in mind that they can be uncertain. Thus, the best is to compare the BATSE samples with other samples of long GRBs. The redshifts of GRBs detected by Swift satellite - obtained from afterglow data - can well serve for this comparison. On the other hand, the redshifts of GRBs - detected by other satellites (BeppoSAX, HETE-2, INTEGRAL) - are not good for our purpose, since they are strongly biased with selection effects (Lee et al. 2000; Bagoly et al. 2006; Butler et al. 2010), and represent only the brightest bursts. All this means that we will discuss four samples here: BATSE GRBs with known redshifts, BATSE

GRBs with pseudo-redshifts, the Swift sample and the Fermi sample. We will try to show the occurrence of the inverse behaviour, first, without the special assumptions of Section 6.2.2, and, second, using this section.

### 6.3.2 Swift GRBs and the inversion in this sample

In the period of 20 November 2004 − 30 April 2010 the Swift database gives a sample of 134 bursts with well determined redshifts from the afterglows together with BAT fluences and peak-fluxes in the energy range of $15 - 150$ keV. To abandon the short bursts only those with $T_{90}/(1 + z) > 2 s$ were taken. They are collected in Tab. $9.27 - 9.30$ in Appendix B.

The effect of inversion can be demonstrated by the scatter plots of the [log fluence; $z$] and [log peak-flux; $z$] planes as it can be seen in Fig. 6.61. To demonstrate the effect of the inversion the medians of the fluences and peak-fluxes are marked with the horizontal lines and the medians of the redshifts with the vertical dashed lines. The medians split the plotting area into four quadrants. It is easy to see that GRBs in the upper right quadrant are apparently brighter than those in the lower left one, although, their redshifts are larger. It is worth mentioning that the GRB having the greatest redshift in the sample has higher fluence than $50\%$ of all the items in the sample.

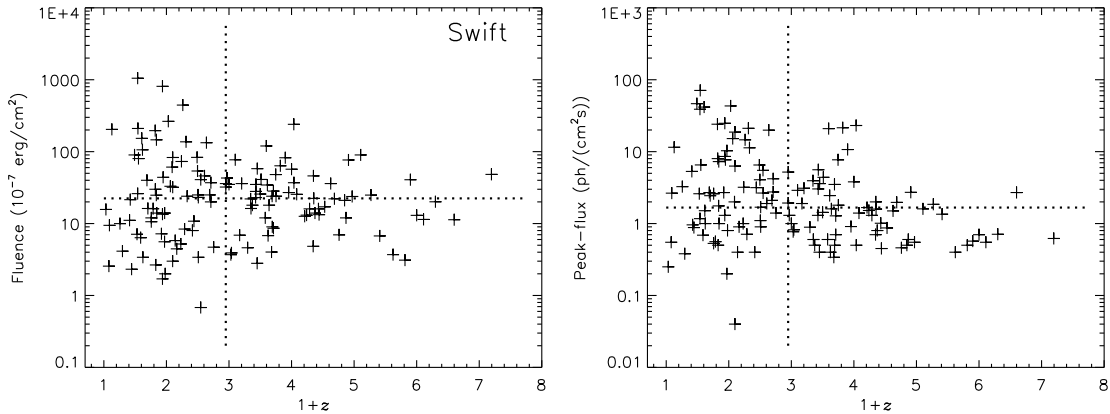

**Fig. 6.61.** Distribution of the fluences (left panel) and peak-fluxes (right panel) of the Swift GRBs with known redshifts. The medians separate the area into four quadrants. The objects in the upper right quadrant are brighter and have larger redshifts than the that of GRBs in the lower left quadrant.

Using the special assumption of Section 6.2.2 the effect of inversion may be illustrated in the Swift sample distinctly also as follows. In Fig. 6.62 the fluences and peak-fluxes are typified against the redshifts. Let be the luminosity distances calculated for $H_0 = 71$ km/(s Mpc), $\Omega_M = 0.27$ and $\Omega_\Lambda = 0.73$. Then the total emitted energy $\tilde{E}_{iso}$ and the peak-luminosity $\tilde{L}_{iso}$ can be calculated using Equation (6.5) with $N = 1$. In the figure the curves of fluences and peak-fluxes are shown after substituting $\tilde{L}_{iso} = \tilde{L}_0(1+z)^q$ and $\tilde{E}_{iso} = \tilde{E}_0(1+z)^q$ where $\tilde{L}_0$ and $\tilde{E}_0$ are constants, and $q = 0; 1; 2$. The inverse behaviour is quite obvious for $q > 1$ and roughly for $z > 2$.

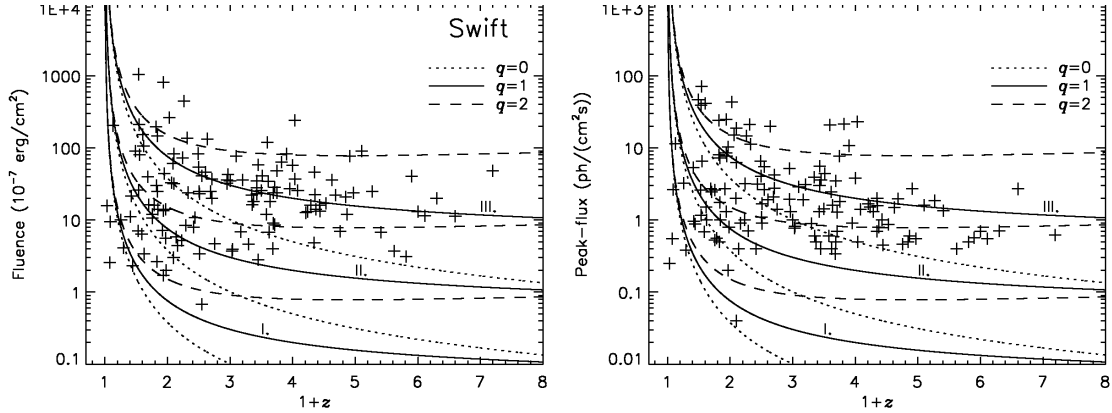

**Fig. 6.62.** Distribution of the fluences (left panel) and peak-fluxes (right panel) of the Swift GRBs with known redshifts. On the left panel the curves denote the values of fluences for $\tilde{E}_{iso} = \tilde{E}_0(1+z)^q$ (the three constant $\tilde{E}_0$ are in units $10^{51}$ erg: I. 0.1; II. 1.0; III. 10.0). On the right panel the curves denote the values of peak-fluxes for $\tilde{L}_{iso} = \tilde{L}_0(1+z)^q$ (the three constant $\tilde{L}_0$ are in units $10^{58}$ ph/s: I. 0.01; II. 0.1; III. 1.0).

The same effect can be similarly illustrated also in Fig. 6.63 showing the relation $\tilde{E}_{iso}$ vs. $(1+z)$, and the relation $\tilde{L}_{iso}$ vs. $(1+z)$. They were calculated again for $H_0 = 71$ km/(s Mpc), $\Omega_M = 0.27$ and $\Omega_\Lambda = 0.73$. In the figure the curves of constant observable peak-fluxes and fluences are also shown. These curves discriminate the bursts of lower/higher measured values. GRBs below a curve at smaller redshifts are representing the inverse behaviour with respect to those at higher redshifts and above the curve.

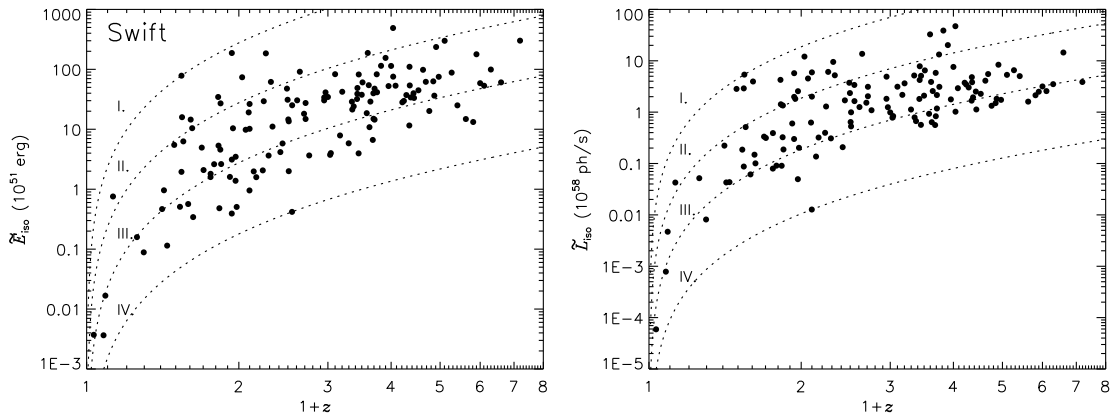

**Fig. 6.63.** *Left panel*: $\tilde{E}_{iso}$ vs. $(1+z)$ (dots). Dashed contours denote constant fluences (in units $10^{-7}$ erg/cm²): I. the maximal fluence, i.e., 1050; II. 105; III. 10.5; IV. the minimal fluence, i.e., 0.68. *Right panel*: $\tilde{L}_{iso}$ vs. $(1+z)$ (dots). Dashed contours denote constant peak-fluxes (in units ph cm⁻² s⁻¹): I. the maximal peak-flux, i.e., 71; II. 7.1; III. 0.71; IV. the minimal peak-flux, i.e., 0.04. The objects below a curve at smaller redshifts together with those at higher redshifts and above the curve illustrate the inverse behaviour.

### 6.3.3 BATSE sample with known redshifts

There are 11 bursts, which were observed by BATSE during the period $1997 - 2000$ and for which there are observed redshifts from the afterglow data. For one of them, GRB 980329 (BATSE trigger 6665), the redshift has only an upper limit ($z < 3.5$), and hence will not be used here. Two cases (GRB 970828 [6350] $z = 0.9578$ and GRB 000131 [7975] $z = 4.5$) have determined redshifts, but they have no measured fluences and peak-fluxes, hence they are also excluded. There are remaining 8 GRBs, which are collected in Tab. 9.31 in Appendix B (see also Bagoly et al. (2003) and Borgonovo (2004)). For the definition of fluence we have chosen the fluence from the third BATSE channel between 100 and 300 keV ($F_3$). This choice is motivated by the observational fact that these fluences in the BATSE Catalog are usually well measured and correlate with other fluences (Bagoly et al. 1998; Veres et al. 2005).

### 6.3.4 BATSE pseudo-redshifts

In the choice of a BATSE sample with estimated pseudo-redshifts one has to take care, since these redshifts are less reliable than the direct redshifts from AGs. The pseudo-redshifts based on the luminosity-lag relation, restricted to the sample in Ryde et al. (2005), where also the spectroscopic studies support the estimations, will be used here. In Table 9.32, in Appendix B, 13 GRBs, using Table 3 of Ryde et al. (2005), are collected. We do not consider two GRBs (BATSE triggers 973 and 3648) from Ryde et al. (2005), since their estimated pseudo-redshifts are ambiguous.

### 6.3.5 Fermi sample

The Fermi sample contains only 6 GRBs with known redshifts together with peak-fluxes and fluences (Bissaldi et al. 2009a,b; van der Horst et al. 2009; Rau et al. 2009a,b; Foley et al. 2010). They are collected in Tab. 9.33, Appendix B. The peak-fluxes and fluences were measured over the energy range $50 \, \text{keV} - 10 \, \text{MeV}$ for GRB 090902B and in the range $8 \, \text{keV} - 1 \, \text{MeV}$ for the remaining five objects.

### 6.3.6 Inversion in the BATSE and Fermi samples

The fluence (peak-flux) vs. redshift relationship of the Fermi and of the two BATSE samples are summarized in Fig. 6.64. For demonstrating the inversion effect - similarly to the case of the Swift sample - the medians also marked with dashed lines. Here it is quite evident that some of the distant bursts exceed in their observed fluence and peak-fluxes those of having smaller redshifts. Here again the GRBs in the upper right quadrants are apparently brighter than those in the lower left one, although their redshifts are larger. Note that the upper right quadrants are even more populated than the lower right quadrants. In other words, a trend of the increasing of peak-flux (fluence) with redshift is evident, and the assumption of the Section 6.2.2 need not be used.

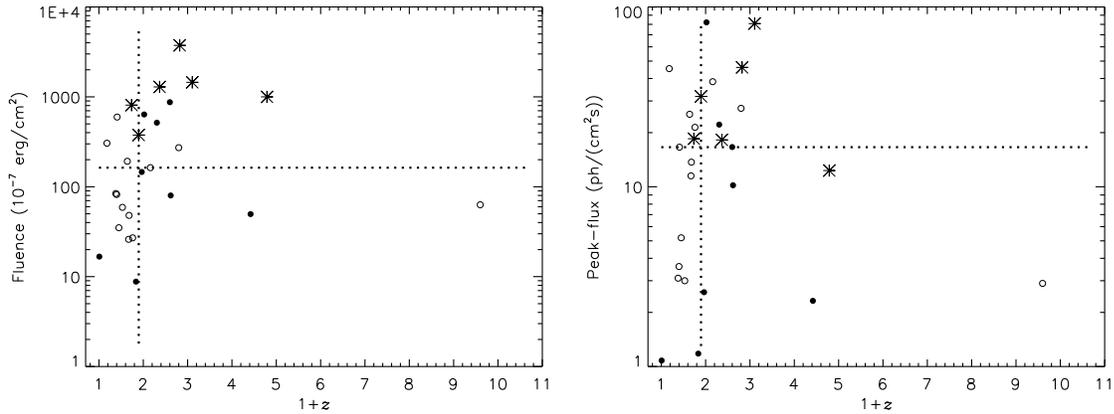

**Fig. 6.64.** Distribution of the fluences (left panel) and peak-fluxes (right panel) of the GRBs with known redshifts, where the Fermi GRBs are denoted by asterisks, BATSE GRBs with determined redshifts (pseudo-redshifts) are denoted by dots (circles). The medians separate the area into four quadrants. The objects in the upper right quadrant are brighter and have larger redshifts than that of GRBs in the lower left quadrant.

## 6.4   Results and discussion of the cosmological effects

It follows from the previous section that in all samples both for the fluences and for the peak-fluxes the "inverse" behaviour, discussed in Section 6.2, can happen. The answer on the question at the beginning of this part is therefore that "this does not need to be the case." Simply, the apparently faintest GRBs need not be also the most distant ones. This is in essence the key result of this section.

It is essential to note that no assumptions were made about the models of the long GRBs. Also the cosmological parameters did not need to be specified.

All this means that faint bursts in the BATSE Catalog simply need not be at larger redshifts, because the key "natural" assumption - apparently fainter GRBs should on average be at higher redshifts - does not hold. All this also means that the controversy about the fraction of GRBs at very high redshifts in BATSE Catalog may well be overcame: No large fraction of GRBs needs to be at very high redshifts, and the redshift distribution of long GRBs - coming from the Swift sample - may well occur also for the BATSE sample. Of course, this does not mean that no GRBs can be at, say, $8.3 < z < 20$. As proposed first by Mészáros & Mészáros (1996), at these very high redshifts GRBs may well occur, but should give a minority (say 10 % or so) in the BATSE Catalog similarly to the Swift sample. This point of view is supported by newer studies, as well (Bromm & Loeb 2006; Jakobsson et al. 2006; Gehrels et al. 2009; Butler et al. 2010).

At the end it is useful to raise again that the purpose of this part was not to study the intrinsic evolution of the luminosities $L(z)$ from the energy range $E_1 \leq E \leq E_2$. To carry out such study, one should consider three additional reasons that may be responsible for the growth of the average value of $\tilde{L}_{\mathrm{iso}}$ with $(1+z)$: 1. K-correction; 2. the beaming; 3. selection biases due to the instrument's threshold and other instrumental effects. For example, in Fig. 6.63 the main parts in the right-bottom sections below the curves IV - corresponding to the values below the instrumental thresholds in fluence/peak-flux - are not observable. Nonetheless, even using these biased data, the theoretical considerations stated in the Section 6.2 and conclusions of the following sections are valid.

## 6.5   Conclusions of the observed flux/fluence distributions

The results of this part can be summarized as follows:

1. The theoretical study of the $z$-dependence of the observed fluences and peak-fluxes of GRBs have shown that fainter bursts could well have smaller redshifts.

2. This is really fulfilled for the four different samples of long GRBs.

3. These results do not depend on the cosmological parameters and on the GRB models.

4. All this suggests that the estimations, leading to a large fraction of BATSE bursts at $z > 5$, need not be correct.

# 7   Conclusions

In Section 1 a brief overview of gamma-ray bursts was given. Section 2 introduced the RHESSI satellite. In Section 3 an insight into the issue of GRB groups was presented.

Section 4.1 dealt with the RHESSI data in order to search of the in duration intermediate class of bursts claimed by different authors in the past. These RHESSI data confirm that GRBs can be separated into short and long classes, and that the short GRBs are on average harder than the long ones. A two-dimensional analysis of the hardness/duration plane as well as a maximum likelihood fit of the duration distribution, however, also indicate a third class with intermediate duration and similar hardness as the short class. From the two-dimensional analysis it follows that the introduction of the third group is statistically significant at the 0.13 % level (of being accidental). The ML analysis applied only on the durations gives that the introduction of the third group is even more statistically significant, i.e., significant on the 0.036 % level (of being accidental).

Section 4.2 then discussed an instrumental issue of the RHESSI satellite, particulary the effect of the first annealing of the detectors, on the GRB measurements. It was found that the detector response matrix has changed and thus using pre- and post-annealing count fluxes of an energy band $\lesssim 120\,\mathrm{keV}$ and ratios of the fluxes including counts from these low energies might be problematic. However, it is concluded in the following discussion that, although the average hardness ratios $H_{231}$, used in the statistical test in the previous chapter, were systematically shifted to higher values in the period after the annealing, the intermediate group found in Section 4.1.3 is not an artefact caused by employing the pre- and post-annealing data together.

Section 5 investigated in more detail the intermediate-duration bursts with respect to their spectral lags, peak count rates, redshifts, supernova observations, and other properties. The K-S test applied on the spectral lag distributions of the short- and intermediate-duration bursts gives a K-S probability of 20.7 % and a K-S distance of 0.33. The tabular critical value of the K-S distance at the significance level $\alpha = 5\,\%$ and for the given number of elements (24 and 15) is 0.45. Therefore the null hypothesis that the lags of the short and intermediate bursts are drawn from the same distribution cannot be rejected at the 5 % (or lower ) significance level. On the other hand, the same test applied on the lags of the intermediate vs. long (and short vs. long) bursts gives the K-S probability 1.36 % (0.01 %) and the K-S distance 0.42 (0.49). The tabular critical value of the K-S distance at the 5 % significance level for the number of the intermediate and long bursts (15 and 102) is 0.38. Therefore the hypothesis that the lags of the intermediate and long bursts are drawn from the same distribution can be rejected at this 5 % significance level. Similarly, it can be rejected, at the 5 % significance level, that the lags of short and long bursts are drawn from the same distribution. The critical K-S distance is in this case 0.31. The results obtained in the part based on K-S tests of 10,000 resampled sets of median lags (Bootstrap method), however, indicate that it is not entirely sure that the intermediate bursts have the same lag distribution as the short bursts do and that its distribution differ from that of the long bursts.

From the results of the K-S test applied on the peak count rates it follows that the distributions of the peak count rates are different over all three groups. Particularly the K-S probability for intermediate vs. short busts is 3.1 %, for intermediate vs. long bursts it is $3 \times 10^{-3}$ %, and for short vs. long bursts $2 \times 10^{-15}$ %.

The origin of the intermediate bursts found in the RHESSI sample remains unclear. Veres et al. (2010) studied Swift GRBs and found a significant intermediate group. They concluded that these Swift intermediate bursts are closely related to XRFs. The RHESSI intermediate bursts have approximately the same average hardness ratio as the short bursts. Also the finding that spectral lags of the RHESSI intermediate bursts are similar to the lags of the short bursts is different from the result found by de Ugarte Postigo et al. (2010), i.e., that the intermediate bursts in the Swift sample have lags similar with the lags of the long-duration bursts. Therefore it seems improbable that the RHESSI intermediate bursts would be also related to XRFs.

Section 6 focused on the cosmological effects on the observed flux and fluence distributions of GRBs and tried to answer the question if the most distant bursts are, in general, the faintest ones. Four different samples were studied, BATSE bursts with redshifts, BATSE bursts with derived pseudo-redshifts, Swift and Fermi bursts with known redshifts. The results of that part can be summarized as follows: 1. The theoretical study of the $z$-dependence of the observed fluences and peak-fluxes of GRBs have shown that fainter bursts could well have smaller redshifts; 2. This is really fulfilled for the four different samples of long GRBs; 3. These results do not depend on the cosmological parameters and on the GRB models; 4. All this suggests that the estimations, leading to a large fraction of BATSE bursts at $z > 5$, need not be correct.

# 8    Appendix A - Used Statistical Methods

In this section all statistical tests used throughout this work are briefly described. More and detailed information can be found, e.g., in works Band et al. (1997), their Appendix A; Anděl (2003); Press et al. (1992); Rektorys (2000); Trumpler & Weaver (1953); Zey et al. (2006); Zvára & Štěpán (2002).

## 8.1    Log-normal functions

The sum of $m$ fitted univariate log-normal functions is defined as:

$$f(x) = \sum_{k=1}^{m} \frac{w_k}{\sigma_k \sqrt{2\pi}} \exp\left(-\frac{(x - \mu_k)^2}{2\sigma_k^2}\right), \tag{8.9}$$

where $\sigma_k$ are the standard deviations, $\mu_k$ are the means, and $w_k$ are the weights ($\sum_{k=1}^{m} w_k = 1$).

The sum of $m$ bivariate log-normal functions is defined as:

$$f(x,y) = \sum_{k=1}^{m} f_k(x,y) = \tag{8.10}$$

$$= \sum_{k=1}^{m} \frac{w_k}{2\pi \sigma_{k,x} \sigma_{k,y} \sqrt{1 - r_k^2}} \exp\left[-\frac{1}{2(1-r_k^2)} \left(\frac{(x - \mu_{k,x})^2}{\sigma_{k,x}^2} + \frac{(y - \mu_{k,y})^2}{\sigma_{k,y}^2} - \frac{C_k}{\sigma_{k,x}\sigma_{k,y}}\right)\right],$$

where $C_k = 2r_k(x - \mu_{k,x})(y - \mu_{k,x})$, and $\sigma_k$, $\mu_k$, and $w_k$ are again the standard deviations, the means and the weights, respectively ($\sum_{k=1}^{m} w_k = 1$).

## 8.2    $\chi^2$ test

Let us have binned data. Then $\chi^2$ is defined as:

$$\chi^2 = \sum_{i=1}^{N} \frac{(O_i - E_i)^2}{E_i}, \tag{8.11}$$

where $N$ is the number of bins, $O_i$ is the observed frequency for bin $i$, and $E_i$ is the expected frequency for bin $i$ given by the fitted model with $p$ free parameters. The null hypothesis, that the observed values are equal to the expected values, is rejected at a significance level $\alpha$ when $\chi^2 \geq \chi^2_{N-1-p}(\alpha)$.

## 8.3    F-test

To evaluate the significance of an intermediate GRB group, the F-test, which compares fits with nested models (i.e., one model is a subset of the other) has been used. The quantity

$$F_0 = \frac{\dfrac{\chi_1^2 - \chi_2^2}{\Delta\nu}}{\dfrac{\chi_2^2}{\nu_2}} \tag{8.12}$$

follows Fisher-Snedecor distribution as $F(\Delta\nu, \nu_2)$. $\chi_1^2$ is the minimal $\chi^2$ resulting from fitting of the data divided into $N$ bins by a model with $r_1$ free parameters (thus $\nu_1 = N - r_1$ degrees of freedom). $\chi_2^2$ is the minimal $\chi^2$ resulting from fitting of the data divided into $N$ bins by a model with $r_2$ free parameters, where $r_2 > r_1$ ($\nu_2 = N - r_2$ degrees of freedom). The probability that the improvement in $\chi^2$ by adding the additional $\Delta\nu = r_2 - r_1$ parameters is only an accidental fluctuation is the probability $P(F \geq F_0)$ of finding $F$ greater than or equal to $F_0$.

## 8.4 Maximum likelihood method and the ratio test

The maximum likelihood estimation is based on searching for the maximum of the likelihood $L$, which is defined for an univariate function $f(x)$ as:

$$L = \sum_{i=1}^{N} \ln f(x_i) \tag{8.13}$$

and for a bivariate function $f(x, y)$ as:

$$L = \sum_{i=1}^{N} \ln f(x_i, y_i) \tag{8.14}$$

where $x_i$ and $y_i$ are the data points of the sample with $N$ elements.

To compare two fit models with number of parameters $r_1$ and $r_2$ ($r_2 > r_1$), respectively, the maximum likelihood ratio test can be used. Twice the difference of the likelihoods follows $\chi^2$ distribution with $\Delta\nu = r_2 - r_1$ degrees of freedom:

$$2(L_{r_2} - L_{r_1}) \simeq \chi^2(\Delta\nu) \tag{8.15}$$

## 8.5 Cross correlation function

The cross correlation function $P_{xy}(L)$ of two sample populations $x$ and $y$ with lengths $N$ as a function of the lag $L$ is defined as follows:

$$P_{xy}(L) = \begin{cases} \dfrac{\displaystyle\sum_{i=1}^{N-|L|}(x_{i+|L|}-\overline{x})(y_i-\overline{y})}{\sqrt{\left[\displaystyle\sum_{i=1}^{N}(x_i-\overline{x})^2\right]\left[\displaystyle\sum_{i=1}^{N}(y_i-\overline{y})^2\right]}} & \text{for } L < 0 \\[4ex] \dfrac{\displaystyle\sum_{i=1}^{N-L}(x_i-\overline{x})(y_{i+L}-\overline{y})}{\sqrt{\left[\displaystyle\sum_{i=1}^{N}(x_i-\overline{x})^2\right]\left[\displaystyle\sum_{i=1}^{N}(y_i-\overline{y})^2\right]}} & \text{for } L \geq 0, \end{cases} \tag{8.16}$$

where $\overline{x}$ and $\overline{y}$ are the means of the samples.

## 8.6   Two-sample Kolmogorov-Smirnov test

Let $S(x)$ have a continuous distribution function and $S_M(x)$ and $S_N(x)$ are its two independent random selections of sizes M and N, respectively. We test the null hypothesis that these two selections have the same distribution function. Let quantity $D$ (distance) be defined as

$$D = \sup_x |S_M(x) - S_N(x)|, \tag{8.17}$$

and $K(\lambda)$ as

$$K(\lambda) = 2\sum_{j=1}^{\infty}(-1)^{j-1}e^{-2j^2\lambda^2}. \tag{8.18}$$

Then for any $\lambda$ it holds that

$$\lim_{m,n\to\infty} P\left(\sqrt{\frac{MN}{M+N}}D < \lambda\right) = K(\lambda). \tag{8.19}$$

For $m, n \to \infty$ the null hypothesis is rejected at a significance level $\alpha$, if $K(\lambda_0) \geq 1 - \alpha$, where $\lambda_0 = \sqrt{\frac{MN}{M+N}}D$.

# 9    Appendix B - Data Samples

The RHESSI data obtained by the author other collected data samples used in Sections 4.1, 5, and 6 are listed.

## 9.1    RHESSI data of GRB durations and hardness ratios

**Tab. 9.15.** The RHESSI GRB data set including: I. GRB name which corresponds to the observation date (letters after the GRB names are internal and do not have to be in accordance with, e.g., GRB labeling in GCN), II. peak time in the light curve, III. $T_{90}$ duration, IV. light curve time resolution $\delta t_{res}$, and V. logarithmic hardness ratio log $H$. These data are explained in Section 4.1.

| GRB | Peak time UTC | $T_{90}$ (s) | $\delta t_{res}$ (s) | Hardness ratio $H$ |
|---|---|---|---|---|
| 020214 | 18:49:47.700 | $(1.42\pm0.03)$E+1 | 2.0E$-$1 | $(6.03\pm0.16)$E$-$1 |
| 020218 | 19:49:41.750 | $(3.40\pm0.06)$E+1 | 5.0E$-$1 | $-(1.12\pm0.11)$E$-$1 |
| 020302 | 12:23:54.400 | $(4.88\pm0.32)$E+1 | 8.0E$-$1 | $(1.65\pm0.53)$E$-$1 |
| 020306 | 18:58:02.893 | $(1.35\pm0.16)$E$-$1 | 1.5E$-$2 | $(4.16\pm0.40)$E$-$1 |
| 020311 | 01:21:31.550 | $(1.08\pm0.06)$E+1 | 3.0E$-$1 | $(6.55\pm3.88)$E$-$2 |
| 020313 | 01:17:53.400 | $(2.32\pm0.10)$E+1 | 4.0E$-$1 | $-(5.17\pm0.50)$E$-$1 |
| 020315 | 15:42:47.550 | $(1.26\pm0.11)$E+1 | 3.0E$-$1 | $(2.97\pm6.94)$E$-$2 |
| 020331 | 10:23:26.750 | $(3.90\pm0.53)$E+1 | 5.0E$-$1 | $-(0.46\pm1.05)$E$-$1 |
| 020407 | 04:14:44.300 | $(2.10\pm0.09)$E+1 | 6.0E$-$1 | $-(1.72\pm2.76)$E$-$2 |
| 020409 | 09:27:23.500 | $(1.46\pm0.14)$E+2 | 1.0E+0 | $(8.84\pm7.37)$E$-$2 |
| 020413 | 16:20:15.500 | $(9.00\pm0.64)$E+0 | 2.0E$-$1 | $-(2.89\pm5.57)$E$-$2 |
| 020417 | 05:36:26.250 | $(7.85\pm0.52)$E+1 | 5.0E$-$1 | $-(1.62\pm0.58)$E$-$1 |
| 020418 | 17:43:08.850 | $(3.40\pm0.13)$E+0 | 1.0E$-$1 | $(1.81\pm0.20)$E$-$1 |
| 020426 | 23:56:14.795 | $(1.20\pm0.32)$E$-$1 | 3.0E$-$2 | $(7.40\pm7.13)$E$-$2 |
| 020430 | 21:22:01.650 | $(1.35\pm0.05)$E+1 | 3.0E$-$1 | $-(2.73\pm0.32)$E$-$1 |
| 020509 | 00:01:17.675 | $(3.50\pm0.38)$E+0 | 5.0E$-$2 | $(8.32\pm8.52)$E$-$2 |
| 020524 | 02:12:47.300 | $(1.48\pm0.08)$E+1 | 2.0E$-$1 | $(7.46\pm4.26)$E$-$2 |
| 020525A | 03:47:53.650 | $(7.60\pm0.61)$E+0 | 1.0E$-$1 | $(1.10\pm0.65)$E$-$1 |
| 020525B | 04:26:53.210 | $(1.80\pm0.29)$E$-$1 | 2.0E$-$2 | $(2.78\pm0.96)$E$-$1 |
| 020527 | 05:17:20.525 | $(1.65\pm0.18)$E+0 | 5.0E$-$2 | $(1.94\pm0.88)$E$-$1 |
| 020602 | 17:30:28.085 | $(1.75\pm0.19)$E+0 | 7.0E$-$2 | $-(2.49\pm0.91)$E$-$1 |
| 020603 | 17:50:34.905 | $(1.44\pm0.06)$E+0 | 3.0E$-$2 | $-(0.07\pm3.08)$E$-$2 |
| 020604 | 14:13:42.850 | $(7.20\pm0.41)$E+0 | 3.0E$-$1 | $-(1.37\pm3.19)$E$-$2 |
| 020620 | 12:58:06.875 | $(3.45\pm0.22)$E+0 | 5.0E$-$2 | $-(2.63\pm0.56)$E$-$1 |
| 020623 | 04:23:07.350 | $(6.60\pm0.46)$E+0 | 1.0E$-$1 | $-(2.84\pm0.66)$E$-$1 |
| 020630 | 07:58:52.650 | $(3.30\pm0.26)$E+1 | 3.0E$-$1 | $-(1.48\pm0.66)$E$-$1 |
| 020702 | 15:54:16.250 | $(3.69\pm0.24)$E+1 | 3.0E$-$1 | $-(3.85\pm0.71)$E$-$1 |
| 020708 | 04:34:12.500 | $(2.70\pm0.40)$E+1 | 1.0E+0 | $-(3.00\pm1.29)$E$-$1 |
| 020712 | 06:09:44.100 | $(1.80\pm0.20)$E+1 | 2.0E$-$1 | $-(1.51\pm0.93)$E$-$1 |

**Tab. 9.16.** The RHESSI GRB data set continuation.

| GRB | Peak time UTC | $T_{90}$ (s) | $\delta t_{res}$ (s) | Hardness ratio $H$ |
|-----|---------------|--------------|----------------------|--------------------|
| 020715A | 15:14:26.735 | $(2.40\pm0.40)$E$-1$ | 3.0E$-2$ | $(1.92\pm0.88)$E$-1$ |
| 020715B | 19:21:09.100 | $(6.40\pm0.21)$E$+0$ | 2.0E$-1$ | $(1.05\pm0.10)$E$-1$ |
| 020725 | 16:25:41.150 | $(5.00\pm0.31)$E$+0$ | 1.0E$-1$ | $(2.37\pm0.51)$E$-1$ |
| 020801 | 11:52:41.550 | $(1.43\pm0.09)$E$+1$ | 1.0E$-1$ | $-(8.54\pm5.30)$E$-2$ |
| 020819A | 07:56:39.455 | $(1.05\pm0.08)$E$+0$ | 3.0E$-2$ | $(1.85\pm0.60)$E$-1$ |
| 020819B | 14:57:38.600 | $(9.20\pm0.75)$E$+0$ | 4.0E$-1$ | $-(2.13\pm0.64)$E$-1$ |
| 020828 | 05:45:37.925 | $(6.60\pm0.45)$E$-1$ | 3.0E$-2$ | $(3.41\pm0.48)$E$-1$ |
| 020910 | 19:57:43.150 | $(2.85\pm0.10)$E$+1$ | 3.0E$-1$ | $(1.33\pm0.29)$E$-1$ |
| 020914 | 21:53:20.100 | $(1.08\pm0.09)$E$+1$ | 2.0E$-1$ | $-(2.54\pm0.73)$E$-1$ |
| 020926 | 04:52:54.200 | $(2.24\pm0.15)$E$+1$ | 8.0E$-1$ | $-(2.05\pm0.53)$E$-1$ |
| 021008A | 07:01:03.550 | $(1.45\pm0.01)$E$+1$ | 1.0E$-1$ | $(1.34\pm0.03)$E$-1$ |
| 021008B | 14:30:04.500 | $(1.42\pm0.12)$E$+1$ | 2.0E$-1$ | $(0.09\pm6.92)$E$-2$ |
| 021011 | 04:38:11.250 | $(5.10\pm0.32)$E$+1$ | 5.0E$-1$ | $-(2.94\pm0.61)$E$-1$ |
| 021016 | 10:29:43.500 | $(8.30\pm0.54)$E$+1$ | 1.0E$+0$ | $-(1.54\pm0.55)$E$-1$ |
| 021020 | 20:12:57.350 | $(1.44\pm0.04)$E$+1$ | 3.0E$-1$ | $-(0.64\pm1.86)$E$-2$ |
| 021023 | 02:53:47.300 | $(1.36\pm0.05)$E$+1$ | 2.0E$-1$ | $-(0.60\pm2.85)$E$-2$ |
| 021025 | 20:18:30.150 | $(1.86\pm0.22)$E$+1$ | 3.0E$-1$ | $-(1.22\pm0.95)$E$-1$ |
| 021102 | 15:58:31.850 | $(1.08\pm0.05)$E$+1$ | 3.0E$-1$ | $-(4.04\pm0.41)$E$-1$ |
| 021105 | 05:27:18.650 | $(6.40\pm0.80)$E$+0$ | 1.0E$-1$ | $-(9.92\pm9.89)$E$-2$ |
| 021108 | 05:39:55.800 | $(2.20\pm0.17)$E$+1$ | 4.0E$-1$ | $-(5.99\pm1.07)$E$-1$ |
| 021109 | 08:42:49.000 | $(1.96\pm0.13)$E$+1$ | 4.0E$-1$ | $-(2.87\pm5.23)$E$-2$ |
| 021113 | 13:37:37.625 | $(3.98\pm0.48)$E$+1$ | 2.5E$-1$ | $-(2.82\pm1.08)$E$-1$ |
| 021115 | 13:33:04.250 | $(1.26\pm0.15)$E$+1$ | 3.0E$-1$ | $-(4.79\pm1.31)$E$-1$ |
| 021119 | 12:54:07.700 | $(2.82\pm0.08)$E$+1$ | 2.0E$-1$ | $-(1.83\pm0.26)$E$-1$ |
| 021125 | 17:58:31.250 | $(6.75\pm0.12)$E$+1$ | 5.0E$-1$ | $-(3.57\pm0.18)$E$-1$ |
| 021201 | 05:30:04.175 | $(3.40\pm0.16)$E$-1$ | 1.0E$-2$ | $(4.03\pm0.36)$E$-1$ |
| 021205 | 03:18:29.750 | $(7.65\pm0.33)$E$+1$ | 1.5E$+0$ | $-(2.23\pm0.36)$E$-1$ |
| 021206 | 22:49:16.650 | $(4.92\pm0.02)$E$+0$ | 2.0E$-2$ | $(2.26\pm0.22)$E$-2$ |
| 021211 | 11:18:35.040 | $(4.32\pm0.32)$E$+0$ | 8.0E$-2$ | $-(2.79\pm0.68)$E$-1$ |
| 021214 | 03:27:25.500 | $(3.00\pm0.28)$E$+1$ | 1.0E$+0$ | $-(2.07\pm0.77)$E$-1$ |
| 021223 | 01:10:00.250 | $(8.00\pm0.64)$E$+0$ | 1.0E$-1$ | $-(2.92\pm0.77)$E$-1$ |
| 021226 | 14:53:39.675 | $(3.50\pm0.53)$E$-1$ | 5.0E$-2$ | $(3.34\pm0.43)$E$-1$ |
| 030102 | 23:18:59.350 | $(1.32\pm0.09)$E$+1$ | 3.0E$-1$ | $-(1.65\pm0.56)$E$-1$ |
| 030103 | 21:46:46.750 | $(1.00\pm0.12)$E$+1$ | 5.0E$-1$ | $-(3.93\pm1.14)$E$-1$ |
| 030105 | 14:34:11.995 | $(1.23\pm0.06)$E$+0$ | 3.0E$-2$ | $(3.77\pm0.40)$E$-1$ |
| 030110 | 09:39:30.205 | $(9.00\pm3.14)$E$-2$ | 3.0E$-2$ | $-(2.04\pm0.88)$E$-1$ |
| 030115A | 06:25:37.250 | $(7.95\pm0.14)$E$+1$ | 5.0E$-1$ | $(3.54\pm1.40)$E$-2$ |
| 030115B | 08:15:48.500 | $(2.70\pm0.10)$E$+1$ | 2.0E$-1$ | $-(2.83\pm0.36)$E$-1$ |
| 030127 | 12:32:42.750 | $(3.80\pm0.40)$E$+1$ | 5.0E$-1$ | $(2.34\pm8.30)$E$-2$ |
| 030204 | 12:45:36.500 | $(5.60\pm0.13)$E$+1$ | 1.0E$+0$ | $(1.39\pm1.34)$E$-2$ |
| 030206 | 11:00:32.010 | $(1.40\pm0.24)$E$-1$ | 2.0E$-2$ | $(2.62\pm0.83)$E$-1$ |

**Tab. 9.17.** The RHESSI GRB data set continuation.

| GRB | Peak time UTC | $T_{90}$ (s) | $\delta t_{res}$ (s) | Hardness ratio $H$ |
|---|---|---|---|---|
| 030212 | 22:17:18.450 | $(2.31\pm0.26)$E+1 | 3.0E−1 | $-(2.25\pm0.97)$E−1 |
| 030214 | 14:48:21.325 | $(1.83\pm0.10)$E+1 | 1.5E−1 | $-(2.07\pm0.49)$E−1 |
| 030216 | 16:13:44.150 | $(9.30\pm1.40)$E+0 | 3.0E−1 | $-(1.20\pm1.16)$E−1 |
| 030217 | 02:45:42.600 | $(5.40\pm0.08)$E+1 | 8.0E−2 | $(4.46\pm1.28)$E−2 |
| 030222 | 16:32:53.250 | $(1.44\pm0.18)$E+1 | 3.0E−1 | $-(2.01\pm1.02)$E−1 |
| 030223 | 09:45:15.250 | $(2.05\pm0.11)$E+1 | 5.0E−1 | $(2.50\pm0.42)$E−1 |
| 030225 | 15:02:53.750 | $(2.00\pm0.11)$E+1 | 5.0E−1 | $-(9.23\pm4.20)$E−2 |
| 030227 | 13:12:05.250 | $(7.10\pm1.03)$E+1 | 5.0E−1 | $(0.68\pm1.11)$E−1 |
| 030228 | 20:26:47.250 | $(2.80\pm0.18)$E+1 | 5.0E−1 | $-(3.96\pm4.98)$E−2 |
| 030301 | 20:27:21.500 | $(4.30\pm0.35)$E+1 | 1.0E+0 | $-(3.73\pm0.85)$E−1 |
| 030306 | 03:38:16.150 | $(1.26\pm0.04)$E+1 | 3.0E−1 | $-(1.16\pm0.22)$E−1 |
| 030307 | 14:31:58.950 | $(3.80\pm0.11)$E+0 | 1.0E−1 | $-(1.34\pm0.13)$E−1 |
| 030320A | 10:12:01.150 | $(4.86\pm0.15)$E+1 | 3.0E−1 | $-(6.10\pm2.51)$E−2 |
| 030320B | 18:49:35.500 | $(1.56\pm0.06)$E+2 | 3.0E+0 | $-(6.38\pm2.83)$E−2 |
| 030326 | 10:43:44.150 | $(1.26\pm0.06)$E+1 | 3.0E−1 | $-(2.01\pm0.34)$E−1 |
| 030328 | 07:28:48.250 | $(6.55\pm0.26)$E+1 | 5.0E−1 | $-(4.09\pm3.25)$E−2 |
| 030329A | 11:37:40.850 | $(1.74\pm0.03)$E+1 | 3.0E−1 | $-(2.60\pm0.06)$E−1 |
| 030329B | 15:34:18.500 | $(6.10\pm0.29)$E+1 | 1.0E+0 | $-(1.50\pm0.39)$E−1 |
| 030331 | 05:38:49.500 | $(2.38\pm0.11)$E+1 | 2.0E−1 | $-(3.34\pm0.47)$E−1 |
| 030406 | 22:42:57.450 | $(7.02\pm0.13)$E+1 | 3.0E−1 | $-(3.30\pm1.54)$E−2 |
| 030410 | 11:23:42.275 | $(1.60\pm0.16)$E+0 | 5.0E−2 | $(4.10\pm1.01)$E−1 |
| 030413 | 07:34:45.550 | $(2.04\pm0.09)$E+1 | 3.0E−1 | $(1.48\pm0.35)$E−1 |
| 030414 | 13:48:28.250 | $(2.85\pm0.08)$E+1 | 5.0E−1 | $-(3.66\pm1.80)$E−2 |
| 030419 | 01:12:14.300 | $(3.78\pm0.06)$E+1 | 2.0E−1 | $-(2.81\pm0.14)$E−1 |
| 030421 | 00:36:32.350 | $(1.27\pm0.04)$E+1 | 1.0E−1 | $(8.23\pm2.40)$E−2 |
| 030422 | 09:01:30.850 | $(1.47\pm0.11)$E+1 | 3.0E−1 | $-(3.83\pm6.08)$E−2 |
| 030428 | 22:31:23.300 | $(9.60\pm0.26)$E+0 | 2.0E−1 | $(5.70\pm1.42)$E−2 |
| 030501A | 01:17:22.300 | $(7.40\pm0.49)$E+0 | 2.0E−1 | $-(2.21\pm0.55)$E−1 |
| 030501B | 03:10:23.850 | $(1.17\pm0.16)$E+1 | 3.0E−1 | $-(4.00\pm1.43)$E−1 |
| 030501C | 20:44:49.520 | $(1.24\pm0.12)$E+0 | 4.0E−2 | $(2.21\pm0.74)$E−1 |
| 030505A | 07:39:13.100 | $(8.82\pm0.43)$E+1 | 6.0E−1 | $(3.77\pm3.96)$E−2 |
| 030505B | 09:03:26.750 | $(1.07\pm0.06)$E+2 | 1.5E+0 | $-(6.53\pm4.77)$E−2 |
| 030506 | 02:04:27.350 | $(2.31\pm0.07)$E+1 | 3.0E−1 | $-(9.31\pm2.34)$E−2 |
| 030518A | 01:23:49.650 | $(2.52\pm0.02)$E+1 | 1.0E−1 | $(4.79\pm0.07)$E−1 |
| 030518B | 03:12:23.050 | $(1.74\pm0.12)$E+1 | 3.0E−1 | $(3.98\pm0.54)$E−1 |
| 030519A | 09:32:22.700 | $(3.20\pm0.39)$E+0 | 2.0E−1 | $(6.78\pm1.13)$E−1 |
| 030519B | 14:05:00.100 | $(9.00\pm0.21)$E+0 | 2.0E−1 | $-(1.81\pm0.47)$E−2 |
| 030523 | 14:10:53.190 | $(1.20\pm0.26)$E−1 | 2.0E−2 | $(0.22\pm1.04)$E−1 |
| 030528 | 13:03:07.750 | $(2.15\pm0.17)$E+1 | 5.0E−1 | $-(2.26\pm0.69)$E−1 |
| 030601 | 22:12:06.050 | $(1.83\pm0.08)$E+1 | 3.0E−1 | $(7.59\pm3.59)$E−2 |
| 030614 | 01:30:42.500 | $(1.68\pm0.05)$E+2 | 3.0E+0 | $-(2.19\pm0.21)$E−1 |

**Tab. 9.18.** The RHESSI GRB data set continuation.

| GRB | Peak time UTC | $T_{90}$ (s) | $\delta t_{res}$ (s) | Hardness ratio $H$ |
|---|---|---|---|---|
| 030626 | 01:46:54.950 | $(3.85\pm0.10)$E+1 | 7.0E−1 | $(2.46\pm0.18)$E−1 |
| 030703 | 19:14:02.400 | $(5.68\pm0.74)$E+1 | 8.0E−1 | $-(0.84\pm1.03)$E−1 |
| 030706 | 00:02:17.750 | $(4.50\pm0.45)$E+0 | 3.0E−1 | $(1.36\pm0.62)$E−1 |
| 030710 | 23:05:02.350 | $(1.77\pm0.07)$E+1 | 3.0E−1 | $-(7.49\pm0.71)$E−1 |
| 030714 | 22:14:50.300 | $(1.02\pm0.08)$E+1 | 2.0E−1 | $-(2.73\pm0.73)$E−1 |
| 030716 | 11:57:19.750 | $(5.35\pm0.40)$E+1 | 5.0E−1 | $-(3.16\pm0.75)$E−1 |
| 030721 | 23:41:10.100 | $(2.96\pm0.05)$E+1 | 2.0E−1 | $(2.88\pm0.14)$E−1 |
| 030725 | 11:46:27.800 | $(1.84\pm0.07)$E+1 | 4.0E−1 | $-(2.22\pm0.32)$E−1 |
| 030726A | 06:38:41.750 | $(2.75\pm0.06)$E+1 | 5.0E−1 | $(1.56\pm0.11)$E−1 |
| 030726B | 09:51:37.500 | $(1.94\pm0.12)$E+2 | 1.0E+0 | $-(2.91\pm0.63)$E−1 |
| 030728 | 09:05:55.100 | $(1.92\pm0.28)$E+1 | 2.0E−1 | $(4.26\pm1.50)$E−1 |
| 030824 | 06:31:10.850 | $(3.00\pm0.20)$E+1 | 3.0E−1 | $-(6.11\pm0.95)$E−1 |
| 030827 | 16:08:40.325 | $(1.85\pm0.08)$E+0 | 5.0E−2 | $(7.30\pm2.77)$E−2 |
| 030830 | 18:37:46.250 | $(2.05\pm0.09)$E+1 | 5.0E−1 | $-(1.27\pm0.34)$E−1 |
| 030831 | 15:07:22.450 | $(2.31\pm0.09)$E+1 | 3.0E−1 | $-(1.04\pm0.33)$E−1 |
| 030919 | 21:10:36.750 | $(2.00\pm0.14)$E+1 | 5.0E−1 | $-(2.05\pm0.59)$E−1 |
| 030921 | 08:38:23.650 | $(1.47\pm0.06)$E+1 | 3.0E−1 | $(2.23\pm3.15)$E−2 |
| 030922A | 08:43:35.950 | $(2.28\pm0.08)$E+1 | 3.0E−1 | $-(1.07\pm0.30)$E−1 |
| 030922B | 18:30:56.900 | $(1.08\pm0.03)$E+1 | 2.0E−1 | $(4.63\pm1.41)$E−2 |
| 030926 | 16:52:28.290 | $(2.80\pm0.31)$E−1 | 2.0E−2 | $(2.89\pm0.77)$E−1 |
| 031005 | 08:21:50.250 | $(1.60\pm0.18)$E+1 | 5.0E−1 | $-(4.44\pm1.19)$E−1 |
| 031019 | 22:00:57.175 | $(7.50\pm0.71)$E+0 | 1.5E−1 | $(2.48\pm0.79)$E−1 |
| 031024 | 09:24:14.350 | $(4.00\pm0.32)$E+0 | 1.0E−1 | $(1.77\pm0.58)$E−1 |
| 031027 | 17:07:50.250 | $(3.90\pm0.07)$E+1 | 5.0E−1 | $-(1.03\pm0.10)$E−1 |
| 031107 | 18:24:07.400 | $(1.92\pm0.06)$E+1 | 4.0E−1 | $(1.02\pm2.13)$E−2 |
| 031108 | 14:11:19.250 | $(2.75\pm0.07)$E+1 | 5.0E−1 | $(1.67\pm0.16)$E−1 |
| 031111 | 16:45:20.920 | $(3.20\pm0.08)$E+0 | 8.0E−2 | $(1.50\pm0.08)$E−1 |
| 031118 | 06:26:16.410 | $(2.60\pm0.27)$E−1 | 2.0E−2 | $(2.01\pm0.57)$E−1 |
| 031120 | 05:52:38.250 | $(9.75\pm0.30)$E+1 | 1.5E+0 | $(4.24\pm2.27)$E−2 |
| 031127 | 18:58:55.050 | $(1.92\pm0.17)$E+1 | 3.0E−1 | $-(2.88\pm0.83)$E−1 |
| 031130 | 02:04:53.775 | $(4.65\pm0.29)$E+0 | 1.5E−1 | $(5.52\pm4.44)$E−2 |
| 031214 | 22:50:44.150 | $(8.30\pm1.01)$E+0 | 1.0E−1 | $-(6.98\pm9.55)$E−2 |
| 031218 | 06:28:09.370 | $(2.20\pm0.24)$E−1 | 2.0E−2 | $(1.46\pm0.51)$E−1 |
| 031219 | 05:39:05.850 | $(7.50\pm0.21)$E+0 | 1.0E−1 | $-(1.21\pm0.21)$E−1 |
| 031226A | 15:43:08.750 | $(3.45\pm0.29)$E+1 | 5.0E−1 | $-(4.63\pm1.03)$E−1 |
| 031226B | 17:51:29.750 | $(4.40\pm0.23)$E+1 | 5.0E−1 | $-(5.74\pm4.22)$E−2 |
| 040102 | 19:35:26.975 | $(7.80\pm0.83)$E+0 | 1.5E−1 | $-(4.43\pm1.15)$E−1 |
| 040108 | 07:46:39.900 | $(3.30\pm0.37)$E+1 | 6.0E−1 | $-(2.64\pm1.03)$E−1 |
| 040113 | 01:36:46.350 | $(1.62\pm0.10)$E+1 | 3.0E−1 | $-(4.60\pm5.01)$E−2 |

**Tab. 9.19.** The RHESSI GRB data set continuation.

| GRB | Peak time UTC | $T_{90}$ (s) | $\delta t_{res}$ (s) | Hardness ratio $H$ |
|---|---|---|---|---|
| 040115 | 18:30:11.550 | (2.19±0.24)E+1 | 3.0E−1 | −(2.40±0.94)E−1 |
| 040125 | 22:14:47.100 | (1.60±0.23)E+1 | 2.0E−1 | −(1.67±1.16)E−1 |
| 040205A | 05:19:47.350 | (3.90±0.62)E+0 | 1.0E−1 | −(2.06±1.29)E−1 |
| 040205B | 09:27:45.750 | (2.65±0.33)E+1 | 5.0E−1 | (2.19±0.98)E−1 |
| 040207 | 22:12:22.600 | (2.44±0.06)E+1 | 4.0E−1 | (1.07±0.17)E−1 |
| 040211 | 15:02:08.050 | (3.70±0.52)E+0 | 1.0E−1 | −(2.61±1.25)E−1 |
| 040215 | 00:28:02.500 | (5.00±0.29)E+1 | 1.0E+0 | −(3.57±0.57)E−1 |
| 040220 | 00:55:15.800 | (1.72±0.09)E+1 | 4.0E−1 | (5.71±0.46)E−1 |
| 040225A | 05:30:53.850 | (1.05±0.11)E+1 | 3.0E−1 | −(8.84±8.12)E−2 |
| 040225B | 10:02:12.200 | (1.40±0.16)E+1 | 4.0E−1 | −(3.12±1.10)E−1 |
| 040228 | 00:09:08.500 | (2.73±0.02)E+2 | 2.0E−1 | −(1.58±0.07)E−1 |
| 040302A | 04:14:35.250 | (2.37±0.27)E+1 | 3.0E−1 | −(2.76±1.05)E−1 |
| 040302B | 12:24:03.900 | (1.08±0.02)E+1 | 2.0E−1 | −(7.71±0.64)E−2 |
| 040303 | 15:32:37.750 | (2.97±0.27)E+1 | 3.0E−1 | −(1.52±0.76)E−1 |
| 040312 | 00:02:36.550 | (1.60±0.22)E−1 | 2.0E−2 | (2.13±0.48)E−1 |
| 040316 | 18:16:14.350 | (1.50±0.02)E+1 | 1.0E−1 | −(4.87±1.30)E−2 |
| 040323 | 13:03:04.750 | (1.89±0.19)E+1 | 3.0E−1 | −(2.92±0.94)E−1 |
| 040324 | 10:21:12.908 | (2.55±0.17)E−1 | 1.5E−2 | (2.54±0.26)E−1 |
| 040327 | 16:19:27.500 | (2.00±0.32)E+1 | 1.0E+0 | −(4.06±1.54)E−1 |
| 040329 | 11:10:51.965 | (2.07±0.04)E+0 | 3.0E−2 | (3.06±0.12)E−1 |
| 040330 | 13:14:42.150 | (2.88±0.41)E+1 | 3.0E−1 | (0.04±1.09)E−1 |
| 040404 | 10:58:52.150 | (4.90±0.35)E+0 | 1.0E−1 | −(1.38±0.60)E−1 |
| 040413 | 13:09:56.490 | (2.80±0.31)E−1 | 2.0E−2 | (1.21±0.68)E−1 |
| 040414 | 11:09:22.500 | (7.70±0.24)E+1 | 1.0E+0 | −(1.63±2.36)E−2 |
| 040421 | 02:30:27.300 | (1.14±0.02)E+1 | 2.0E−1 | (2.73±0.08)E−1 |
| 040423 | 02:23:30.350 | (4.20±0.53)E+0 | 3.0E−1 | −(1.16±0.87)E−1 |
| 040425 | 16:23:34.275 | (8.10±0.19)E+0 | 5.0E−2 | (1.08±0.20)E−1 |
| 040427 | 20:12:37.700 | (8.00±0.77)E+0 | 2.0E−1 | −(4.55±7.43)E−2 |
| 040429 | 10:53:04.400 | (2.52±0.19)E+1 | 4.0E−1 | (2.85±6.00)E−2 |
| 040502A | 06:37:06.900 | (1.86±0.04)E+1 | 2.0E−1 | −(2.97±0.22)E−1 |
| 040502B | 13:30:02.500 | (1.66±0.07)E+2 | 1.0E+0 | −(2.32±0.39)E−1 |
| 040506 | 23:45:18.800 | (6.56±0.42)E+1 | 8.0E−1 | −(2.34±0.58)E−1 |
| 040508 | 10:15:44.750 | (4.05±0.56)E+1 | 5.0E−1 | −(0.95±1.08)E−1 |
| 040510 | 09:59:37.300 | (6.00±0.57)E+0 | 2.0E−1 | −(2.87±0.83)E−1 |
| 040513 | 03:02:17.500 | (5.30±0.83)E+1 | 1.0E+0 | −(0.78±1.20)E−1 |
| 040526 | 20:21:13.350 | (1.29±0.19)E+1 | 3.0E−1 | −(4.19±1.53)E−1 |
| 040528 | 16:55:58.150 | (2.16±0.07)E+1 | 3.0E−1 | −(1.06±0.26)E−1 |
| 040531 | 23:15:04.750 | (4.20±0.33)E+1 | 5.0E−1 | −(4.73±0.91)E−1 |
| 040601 | 06:33:24.625 | (2.73±0.29)E+1 | 2.5E−1 | (5.10±8.45)E−2 |

**Tab. 9.20.** The RHESSI GRB data set continuation.

| GRB | Peak time UTC | $T_{90}$ (s) | $\delta t_{res}$ (s) | Hardness ratio $H$ |
|-----|-----|-----|-----|-----|
| 040603A | 15:40:58.700 | $(1.24\pm0.13)$E+1 | 2.0E$-$1 | $-(1.42\pm0.84)$E$-$1 |
| 040603B | 19:15:38.500 | $(7.50\pm1.00)$E+1 | 3.0E+0 | $-(5.04\pm1.49)$E$-$1 |
| 040605A | 04:31:44.500 | $(5.80\pm0.73)$E+0 | 2.0E$-$1 | $(3.25\pm1.04)$E$-$1 |
| 040605B | 18:46:16.390 | $(1.80\pm0.22)$E$-$1 | 2.0E$-$2 | $(2.33\pm0.48)$E$-$1 |
| 040605C | 23:58:45.800 | $(1.68\pm0.08)$E+1 | 4.0E$-$1 | $-(4.30\pm0.50)$E$-$1 |
| 040611 | 13:36:00.600 | $(1.80\pm0.10)$E+1 | 4.0E$-$1 | $(2.02\pm0.44)$E$-$1 |
| 040619 | 15:15:52.250 | $(7.70\pm0.27)$E+0 | 1.0E$-$1 | $-(1.74\pm0.28)$E$-$1 |
| 040701 | 22:46:45.250 | $(9.60\pm0.69)$E+0 | 3.0E$-$1 | $(9.16\pm5.34)$E$-$2 |
| 040719 | 01:16:31.250 | $(7.50\pm0.65)$E+0 | 1.0E$-$1 | $-(1.77\pm0.73)$E$-$1 |
| 040723 | 04:06:38.150 | $(1.02\pm0.04)$E+1 | 1.0E$-$1 | $-(1.51\pm0.27)$E$-$1 |
| 040731 | 10:24:42.750 | $(2.45\pm0.07)$E+1 | 5.0E$-$1 | $(1.23\pm0.18)$E$-$1 |
| 040803 | 15:08:57.750 | $(1.14\pm0.11)$E+2 | 1.5E+0 | $-(3.95\pm1.05)$E$-$1 |
| 040810 | 14:15:42.650 | $(1.92\pm0.05)$E+1 | 3.0E$-$1 | $-(7.96\pm1.96)$E$-$2 |
| 040818 | 01:29:04.250 | $(8.00\pm0.33)$E+0 | 1.0E$-$1 | $(4.57\pm0.42)$E$-$1 |
| 040822 | 21:21:55.125 | $(1.38\pm0.14)$E+0 | 3.0E$-$2 | $(2.05\pm0.83)$E$-$1 |
| 040824 | 05:16:07.500 | $(4.90\pm0.50)$E+1 | 1.0E+0 | $-(4.90\pm7.92)$E$-$2 |
| 040921 | 16:06:20.265 | $(2.80\pm0.73)$E$-$1 | 7.0E$-$2 | $(1.88\pm0.64)$E$-$1 |
| 040925 | 22:28:59.750 | $(4.75\pm0.16)$E+1 | 5.0E$-$1 | $(6.66\pm2.68)$E$-$2 |
| 040926 | 04:03:18.100 | $(7.60\pm0.22)$E+0 | 2.0E$-$1 | $(1.93\pm0.11)$E$-$1 |
| 041003 | 09:17:56.250 | $(9.00\pm0.74)$E+0 | 5.0E$-$1 | $-(1.16\pm0.52)$E$-$1 |
| 041006 | 12:18:40.350 | $(1.11\pm0.08)$E+1 | 3.0E$-$1 | $-(4.71\pm0.80)$E$-$1 |
| 041007 | 02:02:08.350 | $(2.00\pm0.13)$E+0 | 1.0E$-$1 | $(3.52\pm0.37)$E$-$1 |
| 041009 | 06:38:21.250 | $(7.50\pm0.76)$E+0 | 3.0E$-$1 | $-(8.60\pm7.65)$E$-$2 |
| 041010 | 00:14:57.555 | $(2.50\pm0.22)$E$-$1 | 1.0E$-$2 | $(1.58\pm0.64)$E$-$1 |
| 041012 | 12:40:51.500 | $(4.60\pm0.53)$E+1 | 1.0E+0 | $-(3.90\pm1.16)$E$-$1 |
| 041013A | 02:35:25.000 | $(1.84\pm0.09)$E+2 | 2.0E+0 | $(1.27\pm3.84)$E$-$2 |
| 041013B | 22:56:27.865 | $(3.60\pm0.46)$E$-$1 | 3.0E$-$2 | $(6.92\pm7.69)$E$-$2 |
| 041015 | 10:22:18.950 | $(3.90\pm0.29)$E+0 | 1.0E$-$1 | $-(1.82\pm0.60)$E$-$1 |
| 041016 | 04:39:37.000 | $(1.80\pm0.17)$E+1 | 4.0E$-$1 | $-(1.90\pm0.82)$E$-$1 |
| 041018 | 13:08:19.500 | $(1.02\pm0.04)$E+2 | 1.0E+0 | $-(3.19\pm3.19)$E$-$2 |
| 041101 | 01:49:36.150 | $(3.20\pm0.34)$E+0 | 1.0E$-$1 | $(2.28\pm0.85)$E$-$1 |
| 041102 | 11:12:23.750 | $(2.70\pm0.22)$E+0 | 1.0E$-$1 | $(1.69\pm0.60)$E$-$1 |
| 041107 | 15:49:29.250 | $(4.70\pm0.45)$E+1 | 5.0E$-$1 | $-(6.04\pm7.68)$E$-$2 |
| 041116 | 05:34:56.500 | $(5.00\pm0.53)$E+1 | 1.0E+0 | $-(3.38\pm1.06)$E$-$1 |
| 041117 | 15:18:00.950 | $(1.62\pm0.05)$E+1 | 3.0E$-$1 | $-(1.81\pm0.24)$E$-$1 |
| 041120 | 19:23:41.300 | $(8.60\pm0.42)$E+0 | 2.0E$-$1 | $-(3.20\pm0.44)$E$-$1 |
| 041125 | 16:07:27.400 | $(2.52\pm0.04)$E+1 | 4.0E$-$1 | $-(2.34\pm0.06)$E$-$1 |
| 041202 | 02:30:57.300 | $(1.66\pm0.03)$E+1 | 2.0E$-$1 | $(1.14\pm0.12)$E$-$1 |
| 041211A | 07:49:56.450 | $(2.06\pm0.05)$E+1 | 1.0E$-$1 | $-(1.33\pm0.23)$E$-$1 |

**Tab. 9.21.** The RHESSI GRB data set continuation.

| GRB | Peak time UTC | $T_{90}$ (s) | $\delta t_{res}$ (s) | Hardness ratio $H$ |
|-----|-----|-----|-----|-----|
| 041211B | 11:31:53.400 | $(2.00\pm0.09)\text{E}+1$ | $4.0\text{E}-1$ | $(8.20\pm3.24)\text{E}-2$ |
| 041211C | 23:57:42.925 | $(6.15\pm0.16)\text{E}+0$ | $1.5\text{E}-1$ | $(4.46\pm0.07)\text{E}-1$ |
| 041213 | 06:59:36.330 | $(1.40\pm0.24)\text{E}-1$ | $2.0\text{E}-2$ | $(3.19\pm0.82)\text{E}-1$ |
| 041218 | 15:45:50.500 | $(5.10\pm0.57)\text{E}+1$ | $1.0\text{E}+0$ | $-(2.22\pm0.98)\text{E}-1$ |
| 041219 | 01:42:19.400 | $(1.00\pm0.11)\text{E}+1$ | $4.0\text{E}-1$ | $-(1.32\pm7.84)\text{E}-2$ |
| 041223 | 14:06:42.250 | $(4.05\pm0.11)\text{E}+1$ | $5.0\text{E}-1$ | $(8.55\pm2.11)\text{E}-2$ |
| 041224 | 20:20:58.250 | $(3.85\pm0.58)\text{E}+1$ | $5.0\text{E}-1$ | $-(3.21\pm1.42)\text{E}-1$ |
| 041231 | 21:50:48.050 | $(1.00\pm0.12)\text{E}+0$ | $1.0\text{E}-1$ | $(3.20\pm0.56)\text{E}-1$ |
| 050124 | 11:30:03.250 | $(2.80\pm0.38)\text{E}+0$ | $1.0\text{E}-1$ | $-(2.24\pm1.12)\text{E}-1$ |
| 050126 | 21:07:37.700 | $(3.30\pm0.10)\text{E}+1$ | $6.0\text{E}-1$ | $-(1.71\pm0.22)\text{E}-1$ |
| 050203 | 17:22:00.850 | $(3.80\pm0.24)\text{E}+0$ | $1.0\text{E}-1$ | $-(0.85\pm4.62)\text{E}-2$ |
| 050213 | 19:24:04.750 | $(1.70\pm0.09)\text{E}+1$ | $5.0\text{E}-1$ | $(2.43\pm0.37)\text{E}-1$ |
| 050214 | 11:38:33.500 | $(4.00\pm0.66)\text{E}+1$ | $1.0\text{E}+0$ | $-(0.16\pm1.23)\text{E}-1$ |
| 050216 | 07:26:34.275 | $(5.00\pm0.72)\text{E}-1$ | $5.0\text{E}-2$ | $(5.66\pm1.00)\text{E}-1$ |
| 050219 | 21:05:51.650 | $(9.40\pm0.17)\text{E}+0$ | $1.0\text{E}-1$ | $-(3.01\pm0.15)\text{E}-1$ |
| 050311 | 17:06:58.850 | $(2.40\pm0.44)\text{E}+0$ | $3.0\text{E}-1$ | $-(1.44\pm1.09)\text{E}-1$ |
| 050312 | 05:40:13.575 | $(1.50\pm0.51)\text{E}+1$ | $5.0\text{E}-2$ | $(3.35\pm0.82)\text{E}-1$ |
| 050314 | 08:33:08.900 | $(8.00\pm0.35)\text{E}+0$ | $2.0\text{E}-1$ | $-(4.77\pm0.45)\text{E}-1$ |
| 050320 | 08:04:26.900 | $(1.54\pm0.13)\text{E}+1$ | $2.0\text{E}-1$ | $(5.51\pm6.52)\text{E}-2$ |
| 050321 | 22:11:51.500 | $(7.00\pm0.75)\text{E}+0$ | $2.0\text{E}-1$ | $-(8.21\pm8.41)\text{E}-2$ |
| 050326 | 09:53:56.500 | $(2.70\pm0.11)\text{E}+1$ | $2.0\text{E}-1$ | $-(2.54\pm0.36)\text{E}-1$ |
| 050328 | 03:25:14.875 | $(4.50\pm0.51)\text{E}-1$ | $5.0\text{E}-2$ | $-(1.02\pm0.20)\text{E}-1$ |
| 050404 | 17:27:48.500 | $(1.14\pm0.03)\text{E}+1$ | $2.0\text{E}-1$ | $(1.03\pm0.18)\text{E}-1$ |
| 050409 | 01:18:36.050 | $(1.26\pm0.05)\text{E}+0$ | $2.0\text{E}-2$ | $(3.05\pm0.30)\text{E}-1$ |
| 050411 | 21:51:09.500 | $(6.80\pm0.69)\text{E}+0$ | $2.0\text{E}-1$ | $-(1.18\pm0.80)\text{E}-1$ |
| 050412 | 18:58:45.900 | $(1.94\pm0.08)\text{E}+1$ | $2.0\text{E}-1$ | $-(5.03\pm0.51)\text{E}-1$ |
| 050429 | 14:09:50.900 | $(1.94\pm0.15)\text{E}+1$ | $2.0\text{E}-1$ | $(1.35\pm0.63)\text{E}-1$ |
| 050430 | 09:13:09.100 | $(1.42\pm0.17)\text{E}+1$ | $2.0\text{E}-1$ | $(0.20\pm9.23)\text{E}-2$ |
| 050501 | 08:19:38.825 | $(2.40\pm0.19)\text{E}+0$ | $1.5\text{E}-1$ | $(7.81\pm4.03)\text{E}-2$ |
| 050502 | 19:56:57.575 | $(1.60\pm0.20)\text{E}+0$ | $5.0\text{E}-2$ | $(1.61\pm0.97)\text{E}-1$ |
| 050509 | 09:31:26.700 | $(2.00\pm0.04)\text{E}+1$ | $2.0\text{E}-1$ | $-(6.35\pm1.57)\text{E}-2$ |
| 050516 | 12:58:05.200 | $(7.60\pm1.16)\text{E}+0$ | $4.0\text{E}-1$ | $(1.16\pm1.11)\text{E}-1$ |
| 050525A | 00:02:54.450 | $(7.40\pm0.19)\text{E}+0$ | $1.0\text{E}-1$ | $-(4.06\pm0.25)\text{E}-1$ |
| 050525B | 00:50:00.500 | $(1.26\pm0.03)\text{E}+1$ | $2.0\text{E}-1$ | $(6.42\pm1.15)\text{E}-2$ |
| 050528 | 07:05:23.500 | $(1.57\pm0.09)\text{E}+2$ | $1.0\text{E}+0$ | $-(4.77\pm0.69)\text{E}-1$ |
| 050530 | 04:44:44.900 | $(2.40\pm0.34)\text{E}+0$ | $2.0\text{E}-1$ | $(4.69\pm1.10)\text{E}-1$ |
| 050531 | 04:27:26.700 | $(3.32\pm0.05)\text{E}+1$ | $2.0\text{E}-1$ | $-(8.44\pm1.34)\text{E}-2$ |
| 050614 | 12:02:00.500 | $(3.10\pm0.45)\text{E}+1$ | $1.0\text{E}+0$ | $-(1.73\pm1.20)\text{E}-1$ |
| 050701 | 14:22:00.950 | $(1.00\pm0.08)\text{E}+1$ | $1.0\text{E}-1$ | $(2.54\pm0.69)\text{E}-1$ |

**Tab. 9.22.** The RHESSI GRB data set continuation.

| GRB | Peak time UTC | $T_{90}$ (s) | $\delta t_{res}$ (s) | Hardness ratio $H$ |
|---|---|---|---|---|
| 050702 | 07:47:44.160 | (1.12±0.16)E+0 | 8.0E−2 | (2.01±1.01)E−1 |
| 050703 | 05:31:50.825 | (8.65±0.44)E+0 | 5.0E−2 | −(2.24±0.47)E−1 |
| 050706 | 17:11:37.875 | (6.15±0.42)E+0 | 1.5E−1 | −(0.07±5.20)E−2 |
| 050713A | 04:29:11.750 | (1.86±0.07)E+1 | 3.0E−1 | −(9.16±2.81)E−2 |
| 050713B | 12:07:28.750 | (6.05±0.39)E+1 | 5.0E−1 | −(0.35±5.23)E−2 |
| 050715 | 01:15:49.250 | (1.14±0.03)E+1 | 1.0E−1 | −(2.32±0.20)E−1 |
| 050717 | 10:30:55.100 | (9.60±0.32)E+0 | 2.0E−1 | (2.36±0.23)E−1 |
| 050726 | 20:22:19.800 | (1.60±0.06)E+1 | 4.0E−1 | −(5.78±2.30)E−2 |
| 050729 | 01:09:41.120 | (4.40±0.47)E+0 | 8.0E−2 | −(8.49±8.47)E−2 |
| 050802 | 10:08:02.850 | (2.94±0.31)E+1 | 3.0E−1 | −(1.76±0.90)E−1 |
| 050805 | 13:29:47.625 | (1.05±0.08)E+0 | 5.0E−2 | (2.99±0.53)E−1 |
| 050809 | 20:15:26.720 | (2.40±0.14)E+0 | 8.0E−2 | −(7.56±3.92)E−2 |
| 050813 | 21:13:43.900 | (8.00±0.87)E+0 | 2.0E−1 | (3.68±0.99)E−1 |
| 050814 | 04:35:19.451 | (1.24±0.08)E−1 | 2.0E−3 | (6.01±0.64)E−1 |
| 050817 | 10:43:18.900 | (2.12±0.20)E+1 | 2.0E−1 | (3.69±0.86)E−1 |
| 050820 | 23:50:36.050 | (6.00±0.58)E+0 | 3.0E−1 | −(3.57±0.87)E−1 |
| 050824 | 11:57:42.535 | (2.50±0.14)E−1 | 1.0E−2 | (3.31±0.35)E−1 |
| 050825 | 03:34:27.850 | (8.00±0.44)E−1 | 2.0E−2 | (1.62±0.42)E−1 |
| 050902 | 12:24:30.650 | (1.16±0.10)E+1 | 1.0E−1 | −(5.97±1.18)E−1 |
| 050923 | 01:37:44.850 | (9.60±0.56)E+0 | 1.0E−1 | (6.55±4.65)E−2 |
| 051009 | 10:49:02.750 | (8.50±0.52)E+1 | 5.0E−1 | −(2.22±0.57)E−1 |
| 051012 | 12:00:11.925 | (2.55±0.28)E+0 | 5.0E−2 | (1.14±1.98)E−2 |
| 051021 | 14:01:14.625 | (5.10±0.10)E+1 | 2.5E−1 | −(2.26±0.17)E−1 |
| 051031 | 22:01:05.750 | (4.95±0.28)E+1 | 1.5E+0 | (4.49±3.88)E−2 |
| 051101 | 01:13:12.625 | (1.10±0.14)E+1 | 1.5E−1 | (9.19±9.99)E−2 |
| 051103 | 09:25:42.192 | (1.40±0.05)E−1 | 5.0E−3 | (4.68±0.15)E−1 |
| 051109 | 16:42:00.750 | (2.55±0.16)E+1 | 5.0E−1 | −(2.15±0.58)E−1 |
| 051111 | 05:59:40.100 | (2.38±0.15)E+1 | 2.0E−1 | (1.10±0.53)E−1 |
| 051117 | 12:34:25.300 | (3.32±0.13)E+1 | 2.0E−1 | (1.62±0.34)E−1 |
| 051119 | 13:10:59.900 | (3.78±0.31)E+1 | 2.0E−1 | −(1.74±0.73)E−1 |
| 051124A | 08:16:59.450 | (1.92±0.15)E+1 | 1.0E−1 | (2.31±0.65)E−1 |
| 051124B | 14:20:11.100 | (3.68±0.19)E+1 | 2.0E−1 | −(2.16±0.48)E−1 |
| 051201A | 18:31:37.250 | (2.25±0.29)E+1 | 1.5E+0 | −(0.50±8.84)E−2 |
| 051201B | 22:35:30.350 | (1.84±0.13)E+1 | 1.0E−1 | (1.29±0.59)E−1 |
| 051207 | 19:04:09.350 | (5.73±0.17)E+1 | 3.0E−1 | −(1.75±0.27)E−1 |
| 051211 | 05:28:13.250 | (2.94±0.09)E+1 | 7.0E−1 | (1.24±0.17)E−1 |
| 051217 | 09:54:08.650 | (1.80±0.27)E+1 | 3.0E−1 | −(0.28±1.14)E−1 |
| 051220A | 13:04:17.525 | (1.29±0.01)E+1 | 5.0E−2 | (2.15±0.04)E−1 |
| 051220B | 21:34:36.750 | (2.80±0.46)E+1 | 5.0E−1 | −(5.31±2.02)E−1 |

**Tab. 9.23.** The RHESSI GRB data set continuation.

| GRB | Peak time UTC | $T_{90}$ (s) | $\delta t_{res}$ (s) | Hardness ratio $H$ |
|---|---|---|---|---|
| 051221 | 01:51:15.975 | $(2.80\pm0.17)$E$-1$ | 1.0E$-2$ | $(1.65\pm0.41)$E$-1$ |
| 051222 | 15:07:35.600 | $(2.44\pm0.40)$E$+1$ | 4.0E$-1$ | $-(1.73\pm1.35)$E$-1$ |
| 060101 | 00:34:11.800 | $(1.96\pm0.09)$E$+1$ | 4.0E$-1$ | $(6.61\pm3.35)$E$-2$ |
| 060110 | 08:01:18.900 | $(1.20\pm0.16)$E$+1$ | 2.0E$-1$ | $(1.03\pm1.00)$E$-1$ |
| 060111 | 08:49:00.250 | $(6.90\pm0.25)$E$+1$ | 1.5E$+0$ | $(1.61\pm0.24)$E$-1$ |
| 060117 | 06:50:13.850 | $(1.59\pm0.10)$E$+1$ | 3.0E$-1$ | $-(3.13\pm0.63)$E$-1$ |
| 060121A | 04:12:56.500 | $(4.40\pm0.63)$E$+1$ | 1.0E$+0$ | $(0.17\pm1.08)$E$-1$ |
| 060121B | 22:24:56.755 | $(2.38\pm0.12)$E$+0$ | 7.0E$-2$ | $-(1.05\pm0.36)$E$-1$ |
| 060123 | 05:05:24.900 | $(1.10\pm0.03)$E$+1$ | 2.0E$-1$ | $-(1.31\pm0.17)$E$-1$ |
| 060124 | 16:04:22.400 | $(1.76\pm0.13)$E$+1$ | 4.0E$-1$ | $(4.99\pm5.58)$E$-2$ |
| 060130 | 13:48:31.100 | $(1.52\pm0.08)$E$+1$ | 2.0E$-1$ | $(2.98\pm4.17)$E$-2$ |
| 060203 | 07:28:58.535 | $(5.40\pm0.45)$E$-1$ | 1.0E$-2$ | $(1.43\pm0.66)$E$-1$ |
| 060217 | 09:47:43.325 | $(1.91\pm0.08)$E$+1$ | 1.5E$-1$ | $-(7.34\pm3.62)$E$-2$ |
| 060224 | 02:31:11.500 | $(1.00\pm0.06)$E$+1$ | 2.0E$-1$ | $-(8.14\pm4.54)$E$-2$ |
| 060228 | 03:17:33.850 | $(3.27\pm0.21)$E$+1$ | 3.0E$-1$ | $-(3.88\pm5.28)$E$-2$ |
| 060303 | 22:42:47.525 | $(5.00\pm0.53)$E$-1$ | 5.0E$-2$ | $(2.56\pm0.29)$E$-1$ |
| 060306 | 15:22:38.485 | $(9.20\pm0.11)$E$-1$ | 1.0E$-2$ | $-(1.07\pm0.04)$E$-1$ |
| 060309 | 14:38:54.250 | $(2.95\pm0.30)$E$+1$ | 5.0E$-1$ | $(1.32\pm0.81)$E$-1$ |
| 060312A | 06:17:21.115 | $(2.40\pm0.47)$E$-1$ | 3.0E$-2$ | $(2.48\pm1.16)$E$-1$ |
| 060312B | 16:44:53.400 | $(6.80\pm1.18)$E$+0$ | 4.0E$-1$ | $-(2.32\pm1.40)$E$-1$ |
| 060313 | 20:11:32.900 | $(3.60\pm0.51)$E$+0$ | 2.0E$-1$ | $(1.52\pm1.03)$E$-1$ |
| 060323 | 07:04:30.100 | $(1.76\pm0.40)$E$+1$ | 2.0E$-1$ | $(3.37\pm1.52)$E$-2$ |
| 060325 | 12:02:20.350 | $(1.08\pm0.02)$E$+1$ | 1.0E$-1$ | $(0.70\pm1.11)$E$-2$ |
| 060401 | 05:40:18.750 | $(6.30\pm0.17)$E$+0$ | 1.0E$-1$ | $(9.21\pm1.86)$E$-2$ |
| 060408 | 13:11:39.150 | $(6.90\pm0.95)$E$+0$ | 3.0E$-1$ | $-(0.06\pm1.01)$E$-1$ |
| 060415 | 05:31:00.050 | $(1.32\pm0.20)$E$+1$ | 3.0E$-1$ | $(1.80\pm1.14)$E$-1$ |
| 060418 | 03:06:35.800 | $(4.08\pm0.21)$E$+1$ | 4.0E$-1$ | $-(1.23\pm0.43)$E$-1$ |
| 060421A | 11:03:49.000 | $(3.12\pm0.11)$E$+1$ | 4.0E$-1$ | $-(8.58\pm2.76)$E$-2$ |
| 060421B | 20:36:38.200 | $(2.16\pm0.19)$E$+1$ | 4.0E$-1$ | $-(1.39\pm0.74)$E$-1$ |
| 060425 | 16:57:38.705 | $(1.40\pm0.12)$E$-1$ | 1.0E$-2$ | $(2.59\pm0.42)$E$-1$ |
| 060428 | 02:30:41.750 | $(1.35\pm0.23)$E$+1$ | 5.0E$-1$ | $-(2.45\pm1.44)$E$-1$ |
| 060429 | 12:19:51.250 | $(2.00\pm0.21)$E$-1$ | 2.0E$-2$ | $(2.89\pm0.33)$E$-1$ |
| 060505 | 23:32:01.050 | $(9.60\pm0.46)$E$+0$ | 3.0E$-1$ | $(2.15\pm0.30)$E$-1$ |
| 060528 | 22:53:05.750 | $(7.70\pm0.20)$E$+1$ | 5.0E$-1$ | $(2.41\pm0.21)$E$-1$ |
| 060530 | 19:19:11.300 | $(4.00\pm0.35)$E$+0$ | 2.0E$-1$ | $-(1.16\pm0.63)$E$-1$ |
| 060610 | 11:22:24.070 | $(6.00\pm0.33)$E$-1$ | 2.0E$-2$ | $(2.48\pm0.38)$E$-1$ |
| 060614 | 12:43:48.250 | $(5.25\pm0.45)$E$+1$ | 1.5E$+0$ | $-(1.78\pm0.72)$E$-1$ |
| 060622 | 17:19:48.750 | $(2.35\pm0.09)$E$+1$ | 5.0E$-1$ | $(1.12\pm0.27)$E$-1$ |
| 060624 | 13:46:56.255 | $(2.55\pm0.07)$E$+0$ | 3.0E$-2$ | $-(4.55\pm0.29)$E$-1$ |

**Tab. 9.24.** The RHESSI GRB data set continuation.

| GRB | Peak time UTC | $T_{90}$ (s) | $\delta t_{res}$ (s) | Hardness ratio $H$ |
|---|---|---|---|---|
| 060625 | 07:33:27.100 | (4.40±0.37)E+0 | 2.0E−1 | (2.21±0.58)E−1 |
| 060630 | 00:06:41.250 | (4.10±0.19)E+1 | 5.0E−1 | −(9.96±4.00)E−2 |
| 060708 | 04:30:38.485 | (1.14±0.08)E−1 | 6.0E−3 | (1.57±0.36)E−1 |
| 060729 | 04:07:38.600 | (5.20±0.53)E+0 | 2.0E−1 | −(1.16±0.79)E−1 |
| 060805 | 14:27:17.450 | (5.10±0.07)E+0 | 6.0E−2 | (6.98±0.69)E−2 |
| 060811 | 16:56:43.950 | (7.29±0.27)E+1 | 3.0E−1 | −(1.96±0.35)E−1 |
| 060819 | 18:28:20.700 | (2.40±0.11)E+1 | 6.0E−1 | (0.84±3.26)E−2 |
| 060823 | 08:05:33.750 | (1.00±0.18)E+0 | 1.0E−1 | (3.23±1.26)E−1 |
| 060919 | 21:52:12.750 | (2.95±0.37)E+1 | 5.0E−1 | (1.20±0.95)E−1 |
| 060920 | 15:32:38.700 | (2.18±0.03)E+1 | 2.0E−1 | (3.59±1.01)E−2 |
| 060925 | 20:14:35.375 | (1.65±0.08)E+1 | 2.5E−1 | −(1.00±0.38)E−1 |
| 060928 | 01:20:24.150 | (2.03±0.03)E+2 | 3.0E−1 | −(1.28±1.19)E−2 |
| 061005 | 13:38:01.800 | (4.24±0.12)E+1 | 4.0E−1 | −(2.05±0.25)E−1 |
| 061006A | 08:43:39.225 | (1.65±0.10)E+0 | 5.0E−2 | (1.53±0.43)E−1 |
| 061006B | 16:45:27.875 | (4.00±0.53)E−1 | 5.0E−2 | (3.23±0.39)E−1 |
| 061007 | 10:08:54.150 | (6.06±0.07)E+1 | 3.0E−1 | (1.20±0.08)E−1 |
| 061012 | 11:51:57.850 | (9.30±0.48)E+0 | 1.0E−1 | (2.56±0.44)E−1 |
| 061013 | 18:06:28.200 | (4.00±0.22)E+1 | 8.0E−1 | −(1.54±0.45)E−1 |
| 061014 | 06:17:02.375 | (2.00±0.55)E−1 | 5.0E−2 | (2.56±0.94)E−1 |
| 061022 | 12:23:42.850 | (2.19±0.29)E+1 | 3.0E−1 | −(0.44±1.02)E−1 |
| 061031 | 12:19:51.500 | (3.30±0.12)E+1 | 2.0E−1 | (7.92±3.06)E−2 |
| 061101 | 21:26:51.550 | (1.92±0.30)E+1 | 3.0E−1 | −(2.41±1.40)E−1 |
| 061108 | 01:09:54.850 | (3.75±0.09)E+1 | 3.0E−1 | −(8.68±1.90)E−2 |
| 061113 | 13:43:36.050 | (1.82±0.03)E+1 | 1.0E−1 | (1.76±0.13)E−1 |
| 061117 | 06:00:11.250 | (2.15±0.27)E+1 | 5.0E−1 | (9.86±9.49)E−2 |
| 061121 | 15:23:44.275 | (1.46±0.03)E+1 | 1.5E−1 | (1.27±0.17)E−1 |
| 061123 | 16:33:28.650 | (5.70±0.32)E+0 | 1.0E−1 | (1.23±0.43)E−1 |
| 061126 | 08:48:03.150 | (1.65±0.07)E+1 | 3.0E−1 | (2.49±0.31)E−1 |
| 061128 | 20:01:11.805 | (3.00±0.31)E−1 | 3.0E−2 | (3.50±0.27)E−1 |
| 061205 | 05:22:15.450 | (7.50±0.91)E+0 | 3.0E−1 | −(1.22±0.96)E−1 |
| 061212 | 05:31:30.970 | (1.90±0.01)E+1 | 6.0E−2 | (2.73±0.06)E−1 |
| 061222 | 03:30:19.300 | (1.16±0.05)E+1 | 2.0E−1 | (1.44±0.31)E−1 |
| 061229 | 22:25:44.250 | (7.95±0.97)E+1 | 5.0E−1 | (1.69±0.93)E−1 |
| 061230 | 23:09:31.000 | (2.56±0.19)E+1 | 8.0E−1 | −(1.63±0.60)E−1 |
| 070113 | 11:56:23.815 | (2.70±0.48)E−1 | 3.0E−2 | (1.65±1.04)E−1 |
| 070116 | 14:32:16.125 | (1.65±0.11)E+1 | 2.5E−1 | (3.79±5.24)E−2 |
| 070120 | 10:48:36.250 | (1.85±0.27)E+1 | 5.0E−1 | (1.94±1.07)E−1 |
| 070121 | 10:12:17.000 | (8.80±1.60)E+0 | 8.0E−1 | (0.11±1.19)E−1 |
| 070125 | 07:21:27.250 | (5.67±0.08)E+1 | 3.0E−1 | (5.03±1.12)E−2 |

**Tab. 9.25.** The RHESSI GRB data set continuation.

| GRB | Peak time UTC | $T_{90}$ (s) | $\delta t_{res}$ (s) | Hardness ratio $H$ |
|---|---|---|---|---|
| 070214 | 22:39:20.850 | $(1.77\pm0.27)$E+1 | 3.0E−1 | $(1.00\pm1.11)$E−1 |
| 070220 | 04:44:45.300 | $(2.14\pm0.11)$E+1 | 2.0E−1 | $(4.86\pm4.03)$E−2 |
| 070221 | 21:06:46.500 | $(1.02\pm0.13)$E+1 | 2.0E−1 | $-(2.28\pm9.64)$E−2 |
| 070307 | 21:15:43.250 | $(5.25\pm0.27)$E+1 | 5.0E−1 | $-(5.80\pm4.35)$E−2 |
| 070402 | 15:48:39.475 | $(8.85\pm0.56)$E+0 | 1.5E−1 | $(4.96\pm4.92)$E−2 |
| 070420 | 06:18:18.400 | $(6.16\pm0.36)$E+1 | 8.0E−1 | $-(1.63\pm0.51)$E−1 |
| 070508 | 04:18:25.050 | $(1.31\pm0.04)$E+1 | 1.0E−1 | $-(1.71\pm2.74)$E−2 |
| 070516 | 20:41:24.725 | $(3.50\pm0.56)$E−1 | 5.0E−2 | $(5.01\pm0.73)$E−1 |
| 070531 | 11:45:43.500 | $(2.60\pm0.34)$E+1 | 1.0E+0 | $-(1.08\pm1.03)$E−1 |
| 070614 | 05:05:09.425 | $(1.50\pm0.51)$E−1 | 5.0E−2 | $(3.35\pm0.64)$E−1 |
| 070622 | 02:25:17.850 | $(1.38\pm0.03)$E+1 | 1.0E−1 | $(1.23\pm0.18)$E−1 |
| 070626 | 04:08:44.500 | $(1.43\pm0.02)$E+2 | 1.0E+0 | $(5.00\pm1.07)$E−2 |
| 070710 | 08:22:07.850 | $(3.90\pm0.58)$E+0 | 3.0E−1 | $(0.94\pm9.91)$E−2 |
| 070717 | 21:50:38.750 | $(1.75\pm0.20)$E+1 | 5.0E−1 | $(2.50\pm0.87)$E−1 |
| 070722 | 06:00:31.500 | $(7.20\pm0.96)$E+0 | 2.0E−1 | $(1.02\pm0.99)$E−1 |
| 070724 | 23:25:46.500 | $(2.40\pm0.25)$E+1 | 1.0E+0 | $-(2.08\pm0.89)$E−1 |
| 070802 | 06:16:19.390 | $(3.12\pm0.25)$E+0 | 6.0E−2 | $(4.80\pm0.78)$E−1 |
| 070817 | 14:43:42.250 | $(8.80\pm0.71)$E+1 | 5.0E−1 | $-(9.18\pm6.84)$E−2 |
| 070819 | 10:17:04.750 | $(3.30\pm0.26)$E+1 | 5.0E−1 | $(3.83\pm6.10)$E−2 |
| 070821 | 12:51:31.750 | $(6.75\pm0.21)$E+1 | 5.0E−1 | $(6.03\pm2.47)$E−2 |
| 070824 | 20:50:10.425 | $(1.30\pm0.07)$E+0 | 5.0E−2 | $(1.01\pm0.29)$E−1 |
| 070825 | 01:55:54.250 | $(3.42\pm0.09)$E+1 | 3.0E−1 | $(1.44\pm0.19)$E−1 |
| 070917 | 09:40:31.250 | $(2.31\pm0.24)$E+1 | 3.0E−1 | $(3.20\pm0.85)$E−1 |
| 071013 | 08:53:39.475 | $(3.65\pm0.43)$E+0 | 5.0E−2 | $(3.27\pm8.94)$E−2 |
| 071014 | 03:19:52.450 | $(7.80\pm0.37)$E+0 | 1.0E−1 | $(4.13\pm3.71)$E−2 |
| 071030 | 08:52:41.900 | $(6.00\pm0.91)$E+0 | 2.0E−1 | $(2.20\pm1.14)$E−1 |
| 071104 | 11:41:09.525 | $(1.47\pm0.13)$E+1 | 1.5E−1 | $-(6.66\pm7.12)$E−2 |
| 071204 | 05:58:29.475 | $(3.00\pm0.56)$E−1 | 5.0E−2 | $(3.46\pm0.72)$E−1 |
| 071217 | 17:03:27.950 | $(8.30\pm0.48)$E+0 | 1.0E−1 | $-(9.40\pm4.89)$E−2 |
| 080114 | 16:10:22.300 | $(7.34\pm0.07)$E+1 | 2.0E−1 | $(3.60\pm0.09)$E−1 |
| 080202 | 13:04:37.250 | $(3.06\pm0.39)$E+1 | 3.0E−1 | $(0.15\pm9.87)$E−2 |
| 080204 | 13:56:34.760 | $(4.88\pm0.23)$E+0 | 8.0E−2 | $(4.53\pm0.42)$E−1 |
| 080211 | 07:23:46.250 | $(2.80\pm0.08)$E+1 | 5.0E−1 | $(2.79\pm0.19)$E−1 |
| 080218 | 05:57:28.375 | $(1.95\pm0.15)$E+1 | 2.5E−1 | $(1.18\pm0.61)$E−1 |
| 080224 | 16:58:51.050 | $(5.40\pm0.29)$E+0 | 1.0E−1 | $(2.22\pm0.41)$E−1 |
| 080318 | 08:31:45.050 | $(1.47\pm0.16)$E+1 | 3.0E−1 | $(1.51\pm0.79)$E−1 |
| 080319 | 12:25:56.900 | $(1.20\pm0.06)$E+1 | 2.0E−1 | $(3.29\pm0.42)$E−1 |
| 080320 | 11:52:02.625 | $(3.05\pm0.08)$E+1 | 2.5E−1 | $(1.96\pm0.22)$E−1 |
| 080328 | 08:03:14.500 | $(8.60\pm0.62)$E+1 | 1.0E+0 | $(3.03\pm0.58)$E−1 |

**Tab. 9.26.** The RHESSI GRB data set continuation.

| GRB | Peak time UTC | $T_{90}$ (s) | $\delta t_{res}$ (s) | Hardness ratio $H$ |
|-----|---------------|--------------|----------------------|---------------------|
| 080330 | 11:04:33.450 | (3.36±0.09)E+1 | 3.0E−1 | (2.54±0.20)E−1 |
| 080408 | 03:36:23.050 | (1.40±0.11)E+0 | 1.0E−1 | (5.29±0.37)E−1 |
| 080413 | 08:51:12.250 | (7.60±1.20)E+0 | 1.0E−1 | (0.55±1.17)E−1 |
| 080425 | 20:21:47.775 | (2.28±0.32)E+1 | 1.5E−1 | (2.58±1.05)E−1 |

## 9.2 Swift, BATSE and Fermi data of fluences and peak-fluxes

**Tab. 9.27.** Swift sample of GRBs with known redshifts. Fluences, peak-fluxes, redshifts $z$, luminosity distances $d_l$ ($H_0 = 71 \, \mathrm{km.s^{-1}.Mpc^{-1}}$, $\Omega_M = 0.27$, $\Omega_\Lambda = 0.73$), isotropic total emitted energies $\tilde{E}_{\mathrm{iso}}$, and isotropic luminosities $\tilde{L}_{\mathrm{iso}}$ are listed. These data are explained in Section 6.

| GRB | Fluence $10^{-7}$ erg/cm$^2$ | Peak-flux ph/(cm$^2$s) | $z$ | $d_l$ Gpc | $\tilde{E}_{\mathrm{iso}}$ $10^{51}$ erg | $\tilde{L}_{\mathrm{iso}}$ $10^{58}$ ph/s |
|-----|------------------------------|------------------------|-----|-----------|-------------------------------------------|--------------------------------------------|
| 050126 | 8.38 | 0.71 | 1.29 | 9.12 | 3.6E+0 | 3.1E−1 |
| 050223 | 6.36 | 0.69 | 0.588 | 3.44 | 5.7E−1 | 6.1E−2 |
| 050315 | 32.2 | 1.93 | 1.949 | 15.24 | 3.0E+1 | 1.8E+0 |
| 050318 | 10.8 | 3.16 | 1.44 | 10.46 | 5.8E+0 | 1.7E+0 |
| 050319 | 13.1 | 1.52 | 3.24 | 28.37 | 3.0E+1 | 3.5E+0 |
| 050401 | 82.2 | 10.7 | 2.9 | 24.81 | 1.6E+2 | 2.0E+1 |
| 050505 | 24.9 | 1.85 | 4.27 | 39.49 | 8.8E+1 | 6.6E+0 |
| 050525A | 153 | 41.7 | 0.606 | 3.57 | 1.5E+1 | 4.0E+0 |
| 050603 | 63.6 | 21.5 | 2.821 | 23.99 | 1.1E+2 | 3.9E+1 |
| 050724 | 9.98 | 3.26 | 0.258 | 1.29 | 1.6E−1 | 5.2E−2 |
| 050730 | 23.8 | 0.55 | 3.969 | 36.20 | 7.5E+1 | 1.7E+0 |
| 050802 | 20 | 2.75 | 1.71 | 12.96 | 1.5E+1 | 2.0E+0 |
| 050803 | 21.5 | 0.96 | 0.422 | 2.30 | 9.6E−1 | 4.3E−2 |
| 050814 | 20.1 | 0.71 | 5.3 | 50.99 | 9.9E+1 | 3.5E+0 |
| 050820A | 34.4 | 2.45 | 2.613 | 21.85 | 5.4E+1 | 3.9E+0 |
| 050824 | 2.66 | 0.5 | 0.83 | 5.26 | 4.8E−1 | 9.0E−2 |
| 050826 | 4.13 | 0.38 | 0.297 | 1.52 | 8.8E−2 | 8.1E−3 |
| 050904 | 48.3 | 0.62 | 6.195 | 61.22 | 3.0E+2 | 3.9E+0 |
| 050908 | 4.83 | 0.7 | 3.346 | 29.49 | 1.2E+1 | 1.7E+0 |
| 051016B | 1.7 | 1.3 | 0.936 | 6.11 | 3.9E−1 | 3.0E−1 |
| 051109A | 22 | 3.94 | 2.346 | 19.15 | 2.9E+1 | 5.2E+0 |
| 051109B | 2.56 | 0.55 | 0.08 | 0.36 | 3.7E−3 | 7.8E−4 |

**Tab. 9.28.** Swift sample of GRBs with known redshifts continuation.

| GRB | Fluence $10^{-7}$ erg/cm$^2$ | Peak-flux ph/(cm$^2$s) | $z$ | $d_l$ Gpc | $\tilde{E}_{iso}$ $10^{51}$ erg | $\tilde{L}_{iso}$ $10^{58}$ ph/s |
|---|---|---|---|---|---|---|
| 051111 | 40.8 | 2.66 | 1.549 | 11.46 | 2.5E+1 | 1.6E+0 |
| 060108 | 3.69 | 0.77 | 2.03 | 16.03 | 3.7E+0 | 7.8E−1 |
| 060115 | 17.1 | 0.87 | 3.53 | 31.45 | 4.5E+1 | 2.3E+0 |
| 060123 | 3 | 0.04 | 1.099 | 7.47 | 9.5E−1 | 1.3E−2 |
| 060124 | 4.61 | 0.89 | 2.298 | 18.67 | 5.8E+0 | 1.1E+0 |
| 060210 | 76.6 | 2.72 | 3.91 | 35.55 | 2.4E+2 | 8.4E+0 |
| 060218 | 15.7 | 0.25 | 0.033 | 0.14 | 3.7E−3 | 5.9E−5 |
| 060223A | 6.73 | 1.35 | 4.41 | 41.03 | 2.5E+1 | 5.0E+0 |
| 060418 | 83.3 | 6.52 | 1.49 | 10.91 | 4.8E+1 | 3.7E+0 |
| 060502A | 23.1 | 1.69 | 1.51 | 11.10 | 1.4E+1 | 9.9E−1 |
| 060505 | 9.44 | 2.65 | 0.089 | 0.40 | 1.7E−2 | 4.7E−3 |
| 060510B | 40.7 | 0.57 | 4.9 | 46.49 | 1.8E+2 | 2.5E+0 |
| 060512 | 2.32 | 0.88 | 0.443 | 2.44 | 1.1E−1 | 4.3E−2 |
| 060522 | 11.4 | 0.55 | 5.11 | 48.85 | 5.3E+1 | 2.6E+0 |
| 060526 | 12.6 | 1.67 | 3.21 | 28.05 | 2.8E+1 | 3.7E+0 |
| 060602A | 16.1 | 0.56 | 0.787 | 4.92 | 2.6E+0 | 9.1E−2 |
| 060604 | 4.02 | 0.34 | 2.68 | 22.54 | 6.6E+0 | 5.6E−1 |
| 060605 | 6.97 | 0.46 | 3.76 | 33.93 | 2.0E+1 | 1.3E+0 |
| 060607A | 25.5 | 1.4 | 3.082 | 26.71 | 5.3E+1 | 2.9E+0 |
| 060614 | 204 | 11.5 | 0.128 | 0.59 | 7.6E−1 | 4.3E−2 |
| 060707 | 16 | 1.01 | 3.43 | 30.39 | 4.0E+1 | 2.5E+0 |
| 060714 | 28.3 | 1.28 | 2.71 | 22.84 | 4.8E+1 | 2.2E+0 |
| 060729 | 26.1 | 1.17 | 0.54 | 3.10 | 1.9E+0 | 8.7E−2 |
| 060814 | 146 | 7.27 | 0.84 | 5.34 | 2.7E+1 | 1.3E+0 |
| 060904B | 16.2 | 2.44 | 0.703 | 4.28 | 2.1E+0 | 3.1E−1 |
| 060906 | 22.1 | 1.97 | 3.685 | 33.12 | 6.2E+1 | 5.5E+0 |
| 060908 | 28 | 3.03 | 2.43 | 20.00 | 3.9E+1 | 4.2E+0 |
| 060912 | 13.5 | 8.58 | 0.937 | 6.12 | 3.1E+0 | 2.0E+0 |
| 060927 | 11.3 | 2.7 | 5.6 | 54.40 | 6.1E+1 | 1.4E+1 |
| 061007 | 444 | 14.6 | 1.262 | 8.87 | 1.8E+2 | 6.1E+0 |
| 061110A | 10.6 | 0.53 | 0.758 | 4.70 | 1.6E+0 | 8.0E−2 |
| 061110B | 13.3 | 0.45 | 3.44 | 30.49 | 3.3E+1 | 1.1E+0 |
| 061121 | 137 | 21.1 | 1.314 | 9.33 | 6.2E+1 | 9.5E+0 |
| 061210 | 11.1 | 5.31 | 0.41 | 2.22 | 4.7E−1 | 2.2E−1 |
| 061222B | 22.4 | 1.59 | 3.355 | 29.59 | 5.4E+1 | 3.8E+0 |
| 070110 | 16.2 | 0.6 | 2.352 | 19.21 | 2.1E+1 | 7.9E−1 |
| 070208 | 4.45 | 0.9 | 1.165 | 8.03 | 1.6E+0 | 3.2E−1 |
| 070306 | 53.8 | 4.07 | 1.497 | 10.98 | 3.1E+1 | 2.4E+0 |
| 070318 | 24.8 | 1.76 | 0.838 | 5.32 | 4.6E+0 | 3.2E−1 |

**Tab. 9.29.** Swift sample of GRBs with known redshifts continuation.

| GRB | Fluence $10^{-7}$ erg/cm² | Peak-flux ph/(cm²s) | $z$ | $d_l$ Gpc | $\tilde{E}_{\rm iso}$ $10^{51}$ erg | $\tilde{L}_{\rm iso}$ $10^{58}$ ph/s |
|-----|-----|-----|-----|-----|-----|-----|
| 070411  | 27   | 0.91 | 2.954 | 25.37 | 5.3E+1 | 1.8E+0 |
| 070419A | 5.58 | 0.2  | 0.97  | 6.39  | 1.4E+0 | 5.0E−2 |
| 070508  | 196  | 24.1 | 0.82  | 5.18  | 3.5E+1 | 4.3E+0 |
| 070521  | 80.1 | 6.53 | 0.553 | 3.19  | 6.3E+0 | 5.1E−1 |
| 070529  | 25.7 | 1.43 | 2.5   | 20.70 | 3.8E+1 | 2.1E+0 |
| 070611  | 3.91 | 0.82 | 2.04  | 16.13 | 4.0E+0 | 8.4E−1 |
| 070612A | 106  | 1.51 | 0.617 | 3.65  | 1.0E+1 | 1.5E−1 |
| 070714B | 7.2  | 2.7  | 0.92  | 5.98  | 1.6E+0 | 6.0E−1 |
| 070721B | 36   | 1.5  | 3.626 | 32.48 | 9.8E+1 | 4.1E+0 |
| 070802  | 2.8  | 0.4  | 2.45  | 20.20 | 4.0E+0 | 5.7E−1 |
| 070810A | 6.9  | 1.9  | 2.17  | 17.40 | 7.9E+0 | 2.2E+0 |
| 071003  | 83   | 6.3  | 1.1   | 7.47  | 2.6E+1 | 2.0E+0 |
| 071010A | 2    | 0.8  | 0.98  | 6.47  | 5.1E−1 | 2.0E−1 |
| 071010B | 44   | 7.7  | 0.947 | 6.20  | 1.0E+1 | 1.8E+0 |
| 071021  | 13   | 0.7  | 5.0   | 47.61 | 5.9E+1 | 3.2E+0 |
| 071031  | 9    | 0.5  | 2.692 | 22.66 | 1.5E+1 | 8.3E−1 |
| 071112C | 30   | 8    | 0.823 | 5.20  | 5.3E+0 | 1.4E+0 |
| 071117  | 24   | 11.3 | 1.331 | 9.48  | 1.1E+1 | 5.2E+0 |
| 071122  | 5.8  | 0.4  | 1.14  | 7.82  | 2.0E+0 | 1.4E−1 |
| 080210  | 18   | 1.6  | 2.641 | 22.13 | 2.9E+1 | 2.6E+0 |
| 080310  | 23   | 1.3  | 2.426 | 19.95 | 3.2E+1 | 1.8E+0 |
| 080319B | 810  | 24.8 | 0.937 | 6.12  | 1.9E+2 | 5.7E+0 |
| 080319C | 36   | 5.2  | 1.95  | 15.25 | 3.4E+1 | 4.9E+0 |
| 080330  | 3.4  | 0.9  | 1.51  | 11.10 | 2.0E+0 | 5.3E−1 |
| 080411  | 264  | 43.2 | 1.03  | 6.88  | 7.4E+1 | 1.2E+1 |
| 080413A | 35   | 5.6  | 2.432 | 20.01 | 4.9E+1 | 7.8E+0 |
| 080413B | 32   | 18.7 | 1.1   | 7.47  | 1.0E+1 | 6.0E+0 |
| 080430  | 12   | 2.6  | 0.759 | 4.70  | 1.8E+0 | 3.9E−1 |
| 080603B | 24   | 3.5  | 2.69  | 22.64 | 4.0E+1 | 5.8E+0 |
| 080604  | 8    | 0.4  | 1.416 | 10.25 | 4.2E+0 | 2.1E−1 |
| 080605  | 133  | 19.9 | 1.64  | 12.30 | 9.1E+1 | 1.4E+1 |
| 080607  | 240  | 23.1 | 3.036 | 26.22 | 4.9E+2 | 4.7E+1 |
| 080707  | 5.2  | 1    | 1.23  | 8.59  | 2.1E+0 | 4.0E−1 |
| 080710  | 14   | 1    | 0.845 | 5.38  | 2.6E+0 | 1.9E−1 |
| 080721  | 120  | 20.9 | 2.597 | 21.68 | 1.9E+2 | 3.3E+1 |
| 080804  | 36   | 3.1  | 2.202 | 17.72 | 4.2E+1 | 3.6E+0 |
| 080805  | 25   | 1.1  | 1.505 | 11.06 | 1.5E+1 | 6.4E−1 |
| 080810  | 46   | 2    | 3.35  | 29.53 | 1.1E+2 | 4.8E+0 |
| 080905B | 18   | 0.5  | 2.374 | 19.43 | 2.4E+1 | 6.7E−1 |

**Tab. 9.30.** Swift sample of GRBs with known redshifts continuation.

| GRB | Fluence $10^{-7}$ erg/cm$^2$ | Peak-flux ph/(cm$^2$s) | $z$ | $d_{\mathrm{l}}$ Gpc | $\tilde{E}_{\mathrm{iso}}$ $10^{51}$ erg | $\tilde{L}_{\mathrm{iso}}$ $10^{58}$ ph/s |
|---|---|---|---|---|---|---|
| 080906 | 35 | 1 | 2 | 15.74 | 3.5E+1 | 9.9E−1 |
| 080916A | 40 | 2.7 | 0.689 | 4.18 | 4.9E+0 | 3.3E−1 |
| 080928 | 25 | 2.1 | 1.691 | 12.78 | 1.8E+1 | 1.5E+0 |
| 081007 | 7.1 | 2.6 | 0.53 | 3.03 | 5.1E−1 | 1.9E−1 |
| 081008 | 43 | 1.3 | 1.968 | 15.42 | 4.1E+1 | 1.2E+0 |
| 081028A | 37 | 0.5 | 3.038 | 26.24 | 7.6E+1 | 1.0E+0 |
| 081029 | 21 | 0.5 | 3.847 | 34.88 | 6.3E+1 | 1.5E+0 |
| 081118 | 12 | 0.6 | 2.58 | 21.51 | 1.9E+1 | 9.3E−1 |
| 081121 | 41 | 4.4 | 2.512 | 20.82 | 6.1E+1 | 6.5E+0 |
| 081203A | 77 | 2.9 | 2.1 | 16.71 | 8.3E+1 | 3.1E+0 |
| 081222 | 48 | 7.7 | 2.747 | 23.22 | 8.3E+1 | 1.3E+1 |
| 090102 | 0.68 | 5.5 | 1.548 | 11.45 | 4.2E−1 | 3.4E+0 |
| 090313 | 14 | 0.8 | 3.374 | 29.79 | 3.4E+1 | 1.9E+0 |
| 090418A | 46 | 1.9 | 1.608 | 12.01 | 3.0E+1 | 1.3E+0 |
| 090424 | 210 | 71 | 0.544 | 3.13 | 1.6E+1 | 5.4E+0 |
| 090516A | 90 | 1.6 | 4.105 | 37.68 | 3.0E+2 | 5.3E+0 |
| 090519 | 12 | 0.6 | 3.87 | 35.12 | 3.6E+1 | 1.8E+0 |
| 090529 | 6.8 | 0.4 | 2.62 | 21.97 | 1.1E+1 | 6.4E−1 |
| 090618 | 1050 | 38.9 | 0.54 | 3.10 | 7.8E+1 | 2.9E+0 |
| 090715B | 57 | 3.8 | 3 | 25.85 | 1.1E+2 | 7.6E+0 |
| 090726 | 8.6 | 0.7 | 2.71 | 22.84 | 1.4E+1 | 1.2E+0 |
| 090812 | 58 | 3.6 | 2.452 | 20.22 | 8.2E+1 | 5.1E+0 |
| 090926B | 73 | 3.2 | 1.24 | 8.68 | 2.9E+1 | 1.3E+0 |
| 091018 | 14 | 10.3 | 0.971 | 6.40 | 3.5E+0 | 2.6E+0 |
| 091020 | 37 | 4.2 | 1.71 | 12.96 | 2.7E+1 | 3.1E+0 |
| 091024 | 61 | 2 | 1.092 | 7.40 | 1.9E+1 | 6.3E−1 |
| 091029 | 24 | 1.8 | 2.752 | 23.28 | 4.1E+1 | 3.1E+0 |
| 091109A | 16 | 1.3 | 3.288 | 28.88 | 3.7E+1 | 3.0E+0 |
| 091127 | 90 | 46.5 | 0.49 | 2.75 | 5.5E+0 | 2.8E+0 |
| 091208B | 33 | 15.2 | 1.063 | 7.16 | 9.8E+0 | 4.5E+0 |
| 100219A | 3.7 | 0.4 | 4.622 | 43.38 | 1.5E+1 | 1.6E+0 |
| 100302A | 3.1 | 0.5 | 4.813 | 45.51 | 1.3E+1 | 2.1E+0 |
| 100418A | 3.4 | 1 | 0.624 | 3.70 | 3.4E−1 | 1.0E−1 |
| 100425A | 4.7 | 1.4 | 1.755 | 13.38 | 3.7E+0 | 1.1E+0 |

**Tab. 9.31.** BATSE sample of GRBs with known redshifts. GRB names, BATSE trigger numbers, redshifts $z$, fluences $F_3$, peak-fluxes $P_{256}$, luminosity distances $d_l$, isotropic total emitted energies $\tilde{E}_{iso}$, and isotropic luminosities $\tilde{L}_{iso}$ are listed. These data are explained in Section 6.

| GRB | BATSE trigger | $z$ | $F_3$ $10^{-6}$ erg/(cm$^2$s) | $P_{256}$ ph/(cm$^2$s) | $d_l$ Gpc | $\tilde{E}_{iso}$ $10^{51}$ erg | $\tilde{L}_{iso}$ $10^{58}$ ph/s |
|-----|-----|-----|-----|-----|-----|-----|-----|
| 970508 | 6225 | 0.835 | 0.88 | 1.18 | 5.3 | 1.6E+0 | 2.2E−1 |
| 971214 | 6533 | 3.42 | 4.96 | 2.32 | 30.3 | 1.2E+2 | 5.8E+0 |
| 980425 | 6707 | 0.0085 | 1.67 | 1.08 | 0.036 | 2.6E−4 | 1.7E−5 |
| 980703 | 6891 | 0.966 | 14.6 | 2.59 | 6.35 | 3.6E+1 | 6.4E−1 |
| 990123 | 7343 | 1.600 | 87.2 | 16.63 | 11.9 | 5.7E+2 | 1.1E+1 |
| 990506 | 7549 | 1.307 | 51.6 | 22.16 | 9.3 | 2.3E+2 | 9.9E+0 |
| 990510 | 7560 | 1.619 | 8.0 | 10.19 | 12.1 | 5.4E+1 | 6.8E+0 |
| 991216 | 7906 | 1.02 | 63.7 | 82.10 | 6.8 | 1.7E+2 | 2.3E+1 |

**Tab. 9.32.** BATSE sample of GRBs with pseudo-redshifts. GRB names, BATSE trigger numbers, pseudo-redshifts $z$, fluences $F_3$, peak-fluxes $P_{256}$, luminosity distances $d_l$, isotropic total emitted energies $\tilde{E}_{iso}$, and isotropic luminosities $\tilde{L}_{iso}$ are listed. These data are explained in Section 6.

| GRB | BATSE trigger | $z$ | $F_3$ $10^{-6}$ erg/(cm$^2$s) | $P_{256}$ ph/(cm$^2$s) | $d_l$ Gpc | $\tilde{E}_{iso}$ $10^{51}$ erg | $\tilde{L}_{iso}$ $10^{58}$ ph/s |
|-----|-----|-----|-----|-----|-----|-----|-----|
| 911016 | 907 | 0.40 | 8.2 | 3.6 | 2.2 | 3.4E+0 | 1.5E−1 |
| 911104 | 999 | 0.67 | 2.6 | 11.5 | 4.1 | 3.1E+0 | 1.4E+0 |
| 920525 | 1625 | 1.8 | 27.2 | 27.3 | 13.8 | 2.2E+2 | 2.3E+1 |
| 920830 | 1883 | 0.45 | 3.5 | 5.2 | 2.5 | 1.8E+0 | 2.7E−1 |
| 921207 | 2083 | 0.18 | 30.5 | 45.4 | 0.86 | 2.3E+0 | 3.4E−1 |
| 930201 | 2156 | 0.41 | 59.6 | 16.6 | 2.2 | 2.5E+1 | 6.6E−1 |
| 941026 | 3257 | 0.38 | 8.4 | 3.1 | 2.1 | 3.2E+0 | 1.2E−1 |
| 950818 | 3765 | 0.64 | 19.2 | 25.3 | 3.8 | 2.0E+1 | 2.7E+0 |
| 951102 | 3891 | 0.68 | 4.8 | 13.7 | 4.1 | 5.8E+0 | 1.6E+0 |
| 960530 | 5478 | 0.53 | 5.9 | 3.0 | 3.0 | 4.2E+0 | 2.1E−1 |
| 960804 | 5563 | 0.76 | 2.7 | 21.4 | 4.7 | 4.1E+0 | 3.2E+0 |
| 980125 | 6581 | 1.16 | 16.3 | 38.4 | 8.0 | 5.8E+1 | 1.4E+1 |
| 990102 | 7293 | 8.6 | 6.3 | 2.9 | 89.4 | 6.3E+2 | 2.8E+1 |

**Tab. 9.33.** Fermi sample of GRBs with known redshifts. GRB names, fluences, peak-fluxes, redshifts $z$, luminosity distances $d_l$, isotropic total emitted energies $\tilde{E}_{\mathrm{iso}}$, and isotropic luminosities $\tilde{L}_{\mathrm{iso}}$ are listed. These data are explained in Section 6.

| GRB | Fluence | Peak-flux | $z$ | $d_l$ | $\tilde{E}_{\mathrm{iso}}$ | $\tilde{L}_{\mathrm{iso}}$ |
|-----|---------|-----------|-----|-------|------------|------------|
|     | $10^{-7}$ erg/cm$^2$ | ph/(cm$^2$s) |   | Gpc | $10^{51}$ erg | $10^{58}$ ph/s |
| 090323 | 1000 | 12.3 | 3.79 | 31.88 | 2.66E+3 | 3.27E+1 |
| 090328 | 809 | 18.5 | 0.736 | 4.53 | 1.34E+2 | 2.62E+0 |
| 090902B | 3740 | 46.1 | 1.822 | 14.02 | 3.12E+3 | 3.84E+1 |
| 090926A | 1450 | 80.8 | 2.106 | 16.77 | 1.57E+3 | 8.76E+1 |
| 091003 | 376 | 31.8 | 0.897 | 5.79 | 7.96E+1 | 6.73E+0 |
| 100414A | 1290 | 18.22 | 1.368 | 9.81 | 6.28E+2 | 8.87E+0 |

## 9.3  RHESSI spectral lags, peak count rates, positions, redshifts

**Tab. 9.34.** Data table of the calculated RHESSI spectral lags of the GRB count light-curves at the energy channels $400-1500$ and $25-120$ keV. The errors compose of the 95 % CL statistical error and the time profile resolution. The peak count rate $F$ has been derived for the band $25-1500$ keV. The GRB sky positions and redshifts has been taken from the GRBOX[1] list. These data are explained in Section 5.

| GRB | Group | Lag | $F$ | RA | Dec | $z$ |
|-----|-------|-----|-----|-----|-----|-----|
|     |       | (ms) | (s$^{-1}$) | (deg) | (deg) |   |
| 020214 | 3 | $42.4^{+56.7}_{-35.0}$ | 8532.0 | — | — | — |
| 020218 | 3 | $607.0^{+181.9}_{-205.4}$ | 3571.3 | — | — | — |
| 020302 | 3 | — | 597.5 | — | — | — |
| 020306 | 1 | $1.2^{+15.7}_{-17.3}$ | 8875.1 | — | — | — |
| 020311 | 3 | $641.9^{+570.7}_{-519.2}$ | 1490.0 | — | — | — |
| 020313 | 3 | — | 912.6 | — | — | — |
| 020315 | 3 | — | 499.1 | — | — | — |
| 020331 | 3 | — | 336.9 | — | — | — |
| 020407 | 3 | — | 796.5 | — | — | — |
| 020409 | 3 | — | 344.4 | — | — | — |
| 020413 | 3 | — | 1051.4 | 14.91 | −7.11 | — |
| 020417 | 3 | — | 823.6 | — | — | — |
| 020418 | 3 | $108.6^{+94.2}_{-93.1}$ | 5616.8 | — | — | — |

---

[1]http://lyra.berkeley.edu/grbox/grbox.php

**Tab. 9.35.** RHESSI sample of GRB lags, peak count rates etc., continuation.

| GRB | Group | Lag (ms) | $F$ (s$^{-1}$) | RA (deg) | Dec (deg) | $z$ |
|-----|-------|----------|----------------|----------|-----------|-----|
| 020426 | 1 | — | 3568.2 | — | — | — |
| 020430 | 3 | — | 1138.6 | — | — | — |
| 020509 | 3 | — | 2600.1 | — | — | — |
| 020524 | 3 | — | 957.5 | — | — | — |
| 020525A | 3 | $452.4^{+502.2}_{-2154.8}$ | 972.1 | — | — | — |
| 020525B | 1 | — | 3291.8 | — | — | — |
| 020527 | 2 | — | 1427.8 | — | — | — |
| 020602 | 3 | — | 2630.6 | — | — | — |
| 020603 | 3 | — | 5843.1 | 236.65 | −22.25 | — |
| 020604 | 3 | — | 1151.5 | 246.91 | −29.79 | — |
| 020620 | 3 | — | 1806.9 | — | — | — |
| 020623 | 3 | — | 969.8 | — | — | — |
| 020630 | 3 | — | 1013.1 | — | — | — |
| 020702 | 3 | — | 750.4 | — | — | — |
| 020708 | 3 | — | 393.3 | 119.52 | 41.59 | — |
| 020712 | 3 | — | 658.8 | — | — | — |
| 020715A | 1 | — | 2456.5 | — | — | — |
| 020715B | 3 | $135.7^{+55.8}_{-47.9}$ | 10067.1 | 207.55 | 61.85 | — |
| 020725 | 3 | — | 3758.8 | — | — | — |
| 020801 | 3 | — | 1733.9 | — | — | — |
| 020819A | 2 | $170.6^{+127.4}_{-109.2}$ | 1814.5 | — | — | — |
| 020819B | 3 | — | 1029.7 | 351.83 | 6.27 | 0.410 |
| 020828 | 1 | $6.3^{+84.3}_{-46.7}$ | 4346.3 | — | — | — |
| 020910 | 3 | — | 1702.5 | 209.00 | −44.73 | — |
| 020914 | 3 | — | 1152.9 | — | — | — |
| 020926 | 3 | — | 616.6 | — | — | — |
| 021008A | 3 | $12.5^{+17.6}_{-16.2}$ | 51598.6 | 191.45 | 44.18 | — |
| 021008B | 3 | — | 555.7 | — | — | — |
| 021011 | 3 | — | 755.6 | — | — | — |
| 021016 | 3 | — | 647.1 | 8.43 | 46.79 | — |
| 021020 | 3 | — | 2275.4 | 322.30 | 51.92 | — |
| 021023 | 3 | — | 1786.1 | 111.23 | −16.28 | — |
| 021025 | 3 | — | 666.3 | — | — | — |
| 021102 | 3 | — | 1729.0 | — | — | — |
| 021105 | 3 | — | 815.6 | — | — | — |
| 021108 | 3 | — | 1072.1 | — | — | — |
| 021109 | 3 | — | 894.1 | — | — | — |
| 021113 | 3 | — | 390.0 | — | — | — |
| 021115 | 3 | — | 504.5 | — | — | — |

**Tab. 9.36.** RHESSI sample of GRB lags, peak count rates etc., continuation.

| GRB | Group | Lag (ms) | $F$ (s$^{-1}$) | RA (deg) | Dec (deg) | $z$ |
|---|---|---|---|---|---|---|
| 021119 | 3 | $-393.8^{+3000.4}_{-2414.5}$ | 2128.0 | — | — | — |
| 021125 | 3 | — | 1436.6 | — | — | — |
| 021201 | 1 | $10.5^{+17.9}_{-22.5}$ | 9399.0 | 121.94 | 21.24 | — |
| 021205 | 3 | — | 578.2 | — | — | — |
| 021206 | 3 | $8.9^{+4.0}_{-4.2}$ | 69508.7 | — | — | — |
| 021211 | 3 | — | 1639.9 | 122.25 | 6.74 | 1.006 |
| 021214 | 3 | — | 281.7 | — | — | — |
| 021223 | 3 | — | 1061.9 | — | — | — |
| 021226 | 1 | $16.3^{+21.1}_{-27.9}$ | 5646.1 | — | — | — |
| 030102 | 3 | — | 1064.1 | — | — | — |
| 030103 | 3 | — | 229.5 | — | — | — |
| 030105 | 2 | $6.7^{+28.4}_{-22.7}$ | 4572.0 | — | — | — |
| 030110 | 1 | — | 2736.7 | — | — | — |
| 030115A | 3 | $328.0^{+2334.8}_{-2009.9}$ | 1333.3 | 324.74 | $-80.24$ | — |
| 030115B | 3 | — | 1031.6 | 194.15 | $-48.47$ | — |
| 030127 | 3 | — | 641.4 | — | — | — |
| 030204 | 3 | $196.0^{+303.8}_{-349.4}$ | 2093.1 | 0.84 | 32.73 | — |
| 030206 | 1 | $-6.6^{+56.8}_{-16.8}$ | 3945.8 | — | — | — |
| 030212 | 3 | — | 849.5 | — | — | — |
| 030214 | 3 | — | 2163.7 | — | — | — |
| 030216 | 3 | — | 545.0 | — | — | — |
| 030217 | 3 | — | 4904.2 | — | — | — |
| 030222 | 3 | — | 801.7 | — | — | — |
| 030223 | 3 | $1108.2^{+1740.4}_{-1544.0}$ | 925.2 | — | — | — |
| 030225 | 3 | $1338.1^{+4066.8}_{-3979.4}$ | 527.7 | — | — | — |
| 030227 | 3 | $-1374.2^{+5702.5}_{-4829.1}$ | 412.0 | — | — | — |
| 030228 | 3 | — | 1278.9 | — | — | — |
| 030301 | 3 | — | 380.6 | — | — | — |
| 030306 | 3 | — | 2719.1 | 328.36 | $-10.87$ | — |
| 030307 | 3 | $248.3^{+63.8}_{-61.4}$ | 7539.4 | — | — | — |
| 030320A | 3 | $358.1^{+1341.7}_{-1042.7}$ | 1778.3 | 267.93 | $-25.31$ | — |
| 030320B | 3 | $415.8^{+1825.9}_{-1203.6}$ | 634.4 | — | — | — |
| 030326 | 3 | — | 1947.2 | 292.97 | $-11.72$ | — |
| 030328 | 3 | — | 889.1 | — | — | — |
| 030329A | 3 | $37.5^{+89.9}_{-97.7}$ | 11819.3 | 161.21 | 21.52 | 0.169 |
| 030329B | 3 | — | 563.1 | — | — | — |
| 030331 | 3 | — | 1106.0 | — | — | — |
| 030406 | 3 | $240.6^{+168.9}_{-169.7}$ | 5871.5 | — | — | — |

**Tab. 9.37.** RHESSI sample of GRB lags, peak count rates etc., continuation.

| GRB | Group | Lag (ms) | $F$ (s$^{-1}$) | RA (deg) | Dec (deg) | $z$ |
|-----|-------|----------|----------------|----------|-----------|-----|
| 030410 | 2 | $23.1^{+64.6}_{-88.5}$ | 2307.9 | — | — | — |
| 030413 | 3 | $-257.9^{+5958.8}_{-1027.3}$ | 1118.2 | 198.61 | 62.35 | — |
| 030414 | 3 | $1171.0^{+316.2}_{-258.2}$ | 2496.2 | 119.89 | $-48.59$ | — |
| 030419 | 3 | — | 7740.0 | — | — | — |
| 030421 | 3 | $69.0^{+166.4}_{-199.6}$ | 4406.1 | — | — | — |
| 030422 | 3 | — | 664.9 | — | — | — |
| 030428 | 3 | $24.9^{+22.0}_{-27.6}$ | 5049.7 | — | — | — |
| 030501A | 3 | — | 1832.0 | 283.58 | 23.88 | — |
| 030501B | 3 | — | 447.2 | 286.38 | 6.31 | — |
| 030501C | 2 | — | 3127.9 | — | — | — |
| 030505A | 3 | — | 587.2 | — | — | — |
| 030505B | 3 | $-267.8^{+749.1}_{-729.4}$ | 1449.2 | — | — | — |
| 030506 | 3 | — | 1318.9 | — | — | — |
| 030518A | 3 | $81.2^{+81.8}_{-107.4}$ | 7977.6 | — | — | — |
| 030518B | 3 | — | 553.4 | — | — | — |
| 030519A | 2 | — | 564.8 | 239.76 | $-33.49$ | — |
| 030519B | 3 | $17.0^{+17.2}_{-19.5}$ | 15276.2 | — | — | — |
| 030523 | 1 | — | 2589.7 | — | — | — |
| 030528 | 3 | — | 423.6 | 256.00 | $-22.62$ | 0.782 |
| 030601 | 3 | $465.6^{+679.7}_{-929.7}$ | 1096.6 | — | — | — |
| 030614 | 3 | — | 823.8 | — | — | — |
| 030626 | 3 | $333.6^{+714.2}_{-772.6}$ | 1346.4 | — | — | — |
| 030703 | 3 | — | 229.4 | — | — | — |
| 030706 | 3 | — | 1021.0 | — | — | — |
| 030710 | 3 | — | 1129.6 | — | — | — |
| 030714 | 3 | — | 1511.9 | — | — | — |
| 030716 | 3 | — | 377.1 | — | — | — |
| 030721 | 3 | $59.9^{+659.6}_{-318.1}$ | 11686.1 | — | — | — |
| 030725 | 3 | — | 1029.4 | 308.50 | $-50.68$ | — |
| 030726A | 3 | $57.8^{+166.1}_{-129.2}$ | 1990.1 | — | — | — |
| 030726B | 3 | — | 393.0 | — | — | — |
| 030728 | 3 | — | 655.0 | — | — | — |
| 030824 | 3 | — | 508.7 | — | — | — |
| 030827 | 3 | $18.4^{+74.2}_{-54.8}$ | 3672.7 | — | — | — |
| 030830 | 3 | — | 1818.6 | — | — | — |
| 030831 | 3 | — | 1794.9 | — | — | — |
| 030919 | 3 | — | 752.6 | — | — | — |
| 030921 | 3 | — | 2374.9 | — | — | — |
| 030922A | 3 | — | 1428.4 | — | — | — |

**Tab. 9.38.** RHESSI sample of GRB lags, peak count rates etc., continuation.

| GRB | Group | Lag (ms) | $F$ (s$^{-1}$) | RA (deg) | Dec (deg) | $z$ |
|---|---|---|---|---|---|---|
| 030922B | 3 | $38.7^{+164.4}_{-146.0}$ | 3993.2 | — | — | — |
| 030926 | 1 | — | 1990.8 | — | — | — |
| 031005 | 3 | — | 620.5 | — | — | — |
| 031019 | 3 | — | 1240.5 | — | — | — |
| 031024 | 3 | — | 689.8 | — | — | — |
| 031027 | 3 | $128.2^{+111.0}_{-125.4}$ | 4003.8 | — | — | — |
| 031107 | 3 | $-121.0^{+1243.7}_{-1523.6}$ | 1182.8 | — | — | — |
| 031108 | 3 | $88.9^{+69.3}_{-82.9}$ | 5729.7 | — | — | — |
| 031111 | 3 | $57.1^{+28.6}_{-31.6}$ | 13879.4 | 71.80 | 18.00 | — |
| 031118 | 1 | — | 4569.0 | — | — | — |
| 031120 | 3 | $1147.6^{+637.5}_{-1068.6}$ | 991.2 | — | — | — |
| 031127 | 3 | — | 634.1 | — | — | — |
| 031130 | 3 | — | 1327.9 | — | — | — |
| 031214 | 3 | — | 666.1 | — | — | — |
| 031218 | 1 | $53.4^{+137.0}_{-48.9}$ | 4640.4 | — | — | — |
| 031219 | 3 | $334.0^{+589.0}_{-350.8}$ | 3134.4 | — | — | — |
| 031226 | 3 | — | 640.2 | — | — | — |
| 031226 | 3 | — | 626.1 | — | — | — |
| 040102 | 3 | — | 1158.7 | — | — | — |
| 040108 | 3 | — | 418.2 | — | — | — |
| 040113 | 3 | — | 1132.3 | — | — | — |
| 040115 | 3 | — | 466.7 | — | — | — |
| 040125 | 3 | — | 542.5 | — | — | — |
| 040205A | 3 | — | 556.0 | — | — | — |
| 040205B | 3 | — | 323.5 | — | — | — |
| 040207 | 3 | $23.6^{+91.0}_{-84.4}$ | 3283.6 | — | — | — |
| 040211 | 3 | — | 756.3 | — | — | — |
| 040215 | 3 | — | 382.1 | — | — | — |
| 040220 | 3 | $949.6^{+1256.0}_{-1003.4}$ | 823.7 | — | — | — |
| 040225A | 3 | — | 553.7 | — | — | — |
| 040225B | 3 | — | 712.9 | — | — | — |
| 040228 | 3 | $19.2^{+39.5}_{-36.0}$ | 11235.1 | 249.00 | $-38.00$ | — |
| 040302A | 3 | — | 747.2 | — | — | — |
| 040302B | 3 | $101.8^{+47.2}_{-32.4}$ | 13903.6 | — | — | — |
| 040303 | 3 | — | 350.9 | — | — | — |
| 040312 | 1 | $2.7^{+27.4}_{-14.8}$ | 7158.5 | — | — | — |
| 040316 | 3 | $-161.3^{+316.9}_{-380.0}$ | 4343.1 | — | — | — |
| 040323 | 3 | — | 661.5 | 208.47 | $-52.34$ | — |

**Tab. 9.39.** RHESSI sample of GRB lags, peak count rates etc., continuation.

| GRB | Group | Lag (ms) | $F$ (s$^{-1}$) | RA (deg) | Dec (deg) | $z$ |
|---|---|---|---|---|---|---|
| 040324 | 1 | $2.7^{+4.5}_{-6.5}$ | 15043.5 | — | — | — |
| 040327 | 3 | — | 401.4 | — | — | — |
| 040329 | 2 | $3.6^{+9.2}_{-9.6}$ | 19948.7 | — | — | — |
| 040330 | 3 | — | 561.6 | — | — | — |
| 040404 | 3 | — | 1094.3 | — | — | — |
| 040413 | 1 | — | 5106.4 | — | — | — |
| 040414 | 3 | — | 969.4 | — | — | — |
| 040421 | 3 | $98.6^{+68.9}_{-75.1}$ | 6897.0 | — | — | — |
| 040423 | 3 | — | 715.5 | — | — | — |
| 040425 | 3 | $155.3^{+139.1}_{-126.2}$ | 7381.3 | — | — | — |
| 040427 | 3 | — | 636.3 | — | — | — |
| 040429 | 3 | — | 548.9 | — | — | — |
| 040502A | 3 | — | 3495.3 | — | — | — |
| 040502B | 3 | — | 530.6 | — | — | — |
| 040506 | 3 | — | 806.3 | — | — | — |
| 040508 | 3 | — | 241.9 | — | — | — |
| 040510 | 3 | — | 1096.5 | — | — | — |
| 040513 | 3 | — | 231.3 | — | — | — |
| 040526 | 3 | — | 525.3 | — | — | — |
| 040528 | 3 | $731.5^{+1067.9}_{-1191.3}$ | 1713.5 | — | — | — |
| 040531 | 3 | — | 1024.4 | — | — | — |
| 040601 | 3 | — | 394.1 | — | — | — |
| 040603A | 3 | — | 731.1 | — | — | — |
| 040603B | 3 | — | 149.6 | — | — | — |
| 040605A | 3 | — | 724.1 | — | — | — |
| 040605B | 1 | $26.0^{+45.0}_{-50.8}$ | 7927.1 | — | — | — |
| 040605C | 3 | — | 1208.5 | — | — | — |
| 040611 | 3 | $-2267.0^{+2983.7}_{-1345.5}$ | 739.9 | — | — | — |
| 040619 | 3 | — | 3296.3 | — | — | — |
| 040701 | 3 | $135.0^{+245.2}_{-318.9}$ | 1012.1 | — | — | — |
| 040719 | 3 | — | 1455.2 | — | — | — |
| 040723 | 3 | — | 3998.3 | — | — | — |
| 040731 | 3 | $110.2^{+244.6}_{-257.4}$ | 1686.9 | — | — | — |
| 040803 | 3 | — | 265.8 | — | — | — |
| 040810 | 3 | — | 1636.9 | — | — | — |
| 040818 | 3 | $33.2^{+97.4}_{-153.4}$ | 2445.9 | — | — | — |
| 040822 | 2 | — | 2202.9 | — | — | — |
| 040824 | 3 | — | 264.8 | — | — | — |
| 040921 | 1 | — | 2417.0 | — | — | — |

**Tab. 9.40.** RHESSI sample of GRB lags, peak count rates etc., continuation.

| GRB | Group | Lag (ms) | $F$ (s$^{-1}$) | RA (deg) | Dec (deg) | $z$ |
|---|---|---|---|---|---|---|
| 040925 | 3 | $-136.0^{+2908.9}_{-1745.9}$ | 1239.0 | — | — | — |
| 040926 | 3 | $4.1^{+92.7}_{-83.7}$ | 5646.9 | — | — | — |
| 041003 | 3 | — | 722.3 | — | — | — |
| 041006 | 3 | — | 1321.1 | 13.71 | 1.23 | 0.716 |
| 041007 | 2 | $101.0^{+119.5}_{-123.3}$ | 3087.4 | — | — | — |
| 041009 | 3 | — | 1278.5 | — | — | — |
| 041010 | 1 | $0.7^{+6.3}_{-8.0}$ | 8828.3 | — | — | — |
| 041012 | 3 | — | 278.9 | — | — | — |
| 041013A | 3 | — | 500.1 | — | — | — |
| 041013B | 1 | — | 3007.4 | — | — | — |
| 041015 | 3 | — | 1587.6 | — | — | — |
| 041016 | 3 | — | 503.5 | 26.18 | $-4.40$ | — |
| 041018 | 3 | $6774.4^{+6756.0}_{-7424.5}$ | 689.9 | — | — | — |
| 041101 | 3 | — | 931.4 | — | — | — |
| 041102 | 3 | — | 1447.2 | — | — | — |
| 041107 | 3 | — | 882.0 | — | — | — |
| 041116 | 3 | — | 265.3 | — | — | — |
| 041117 | 3 | — | 1172.5 | — | — | — |
| 041120 | 3 | — | 1111.7 | — | — | — |
| 041125 | 3 | $-7.8^{+41.0}_{-55.4}$ | 9392.0 | — | — | — |
| 041202 | 3 | $76.3^{+102.6}_{-93.7}$ | 5633.9 | — | — | — |
| 041211A | 3 | $56.3^{+657.3}_{-825.0}$ | 1077.1 | 100.80 | 20.40 | — |
| 041211B | 3 | — | 5125.7 | — | — | — |
| 041211C | 3 | $-6.0^{+12.9}_{-12.9}$ | 11805.4 | — | — | — |
| 041213 | 1 | — | 6420.2 | — | — | — |
| 041218 | 3 | — | 358.6 | 24.78 | 71.34 | — |
| 041219 | 3 | — | 546.2 | 6.12 | 62.84 | — |
| 041223 | 3 | $-1.6^{+208.1}_{-211.7}$ | 1569.9 | 100.20 | $-37.07$ | — |
| 041224 | 3 | — | 330.3 | 56.20 | $-6.66$ | — |
| 041231 | 2 | $21.6^{+187.2}_{-177.7}$ | 2047.9 | — | — | — |
| 050124 | 3 | — | 743.4 | — | — | — |
| 050126 | 3 | — | 1488.8 | — | — | — |
| 050203 | 3 | — | 2864.8 | — | — | — |
| 050213 | 3 | $44.2^{+197.8}_{-194.9}$ | 2017.9 | — | — | — |
| 050214 | 3 | — | 377.7 | — | — | — |
| 050216 | 1 | $40.2^{+73.8}_{-99.8}$ | 1900.4 | — | — | — |
| 050219 | 3 | $410.8^{+321.9}_{-186.9}$ | 4724.0 | 81.32 | $-57.76$ | — |
| 050311 | 3 | — | 451.3 | — | — | — |

**Tab. 9.41.** RHESSI sample of GRB lags, peak count rates etc., continuation.

| GRB | Group | Lag (ms) | $F$ (s$^{-1}$) | RA (deg) | Dec (deg) | $z$ |
|------|-------|----------|--------|---------|---------|-----|
| 050312 | 1 | $4.4^{+44.5}_{-31.0}$ | 4334.9 | — | — | — |
| 050314 | 3 | — | 1977.4 | — | — | — |
| 050320 | 3 | — | 851.4 | — | — | — |
| 050321 | 3 | — | 895.9 | — | — | — |
| 050326 | 3 | — | 3486.9 | — | — | — |
| 050328 | 2 | $-43.5^{+79.4}_{-84.1}$ | 12097.1 | — | — | — |
| 050404 | 3 | $29.1^{+35.6}_{-39.0}$ | 6024.9 | — | — | — |
| 050409 | 2 | $-1.7^{+16.9}_{-14.1}$ | 17432.8 | — | — | — |
| 050411 | 3 | — | 848.2 | — | — | — |
| 050412 | 3 | — | 1386.6 | — | — | — |
| 050429 | 3 | — | 1874.7 | — | — | — |
| 050430 | 3 | — | 961.4 | — | — | — |
| 050501 | 3 | — | 1567.4 | — | — | — |
| 050502 | 2 | — | 1051.7 | — | — | — |
| 050509 | 3 | $-63.6^{+397.4}_{-494.3}$ | 2216.1 | — | — | — |
| 050516 | 3 | — | 262.9 | — | — | — |
| 050525A | 3 | — | 4859.5 | 278.14 | 26.34 | 0.606 |
| 050525B | 3 | $8.5^{+50.0}_{-50.3}$ | 6795.4 | — | — | — |
| 050528 | 3 | — | 442.7 | — | — | — |
| 050530 | 2 | — | 659.1 | — | — | — |
| 050531 | 3 | $56.6^{+167.3}_{-163.6}$ | 5868.7 | — | — | — |
| 050614 | 3 | — | 374.3 | — | — | — |
| 050701 | 3 | — | 1149.8 | — | — | — |
| 050702 | 2 | — | 945.4 | — | — | — |
| 050703 | 3 | — | 2569.8 | — | — | — |
| 050706 | 3 | — | 1608.5 | — | — | — |
| 050713A | 3 | $-233.3^{+1804.9}_{-2189.4}$ | 1464.2 | 320.54 | 77.07 | — |
| 050713B | 3 | — | 443.6 | 307.81 | 60.94 | — |
| 050715 | 3 | — | 2062.1 | — | — | — |
| 050717 | 3 | $110.3^{+151.6}_{-220.0}$ | 2343.1 | 214.35 | −50.53 | — |
| 050726 | 3 | $810.5^{+1751.7}_{-1334.3}$ | 1435.6 | — | — | — |
| 050729 | 3 | — | 878.0 | — | — | — |
| 050802 | 3 | — | 503.8 | 219.27 | 27.79 | 1.710 |
| 050805 | 2 | $-6.2^{+30.0}_{-20.0}$ | 3367.3 | — | — | — |
| 050809 | 3 | — | 4613.0 | — | — | — |
| 050813 | 3 | — | 781.9 | — | — | — |
| 050814 | 1 | — | 67058.5 | — | — | — |
| 050817 | 3 | — | 874.6 | — | — | — |

**Tab. 9.42.** RHESSI sample of GRB lags, peak count rates etc., continuation.

| GRB | Group | Lag (ms) | $F$ (s$^{-1}$) | RA (deg) | Dec (deg) | $z$ |
|-----|-------|----------|----------------|----------|-----------|-----|
| 050820 | 3 | — | 680.6 | 135.60 | $-72.65$ | — |
| 050824 | 1 | $1.2^{+3.2}_{-4.0}$ | 18094.8 | — | — | — |
| 050825 | 2 | $-82.6^{+132.9}_{-71.3}$ | 4033.0 | — | — | — |
| 050902 | 3 | — | 611.7 | — | — | — |
| 050923 | 3 | — | 1951.7 | — | — | — |
| 051009 | 3 | — | 460.0 | — | — | — |
| 051012 | 3 | $462.5^{+443.7}_{-207.9}$ | 3717.8 | — | — | — |
| 051021 | 3 | $-97.8^{+716.9}_{-495.2}$ | 2125.1 | — | — | — |
| 051031 | 3 | — | 633.2 | — | — | — |
| 051101 | 3 | — | 1014.7 | — | — | — |
| 051103 | 1 | $0.6^{+2.4}_{-2.8}$ | 124293.0 | 148.14 | 68.85 | — |
| 051109 | 3 | — | 953.5 | — | — | — |
| 051111 | 3 | — | 525.4 | 348.14 | 18.37 | 1.550 |
| 051117 | 3 | $-224.8^{+2133.4}_{-3318.9}$ | 892.4 | — | — | — |
| 051119 | 3 | — | 772.8 | — | — | — |
| 051124A | 3 | — | 1029.4 | — | — | — |
| 051124B | 3 | — | 2597.9 | — | — | — |
| 051201A | 3 | — | 209.9 | — | — | — |
| 051201B | 3 | — | 988.7 | — | — | — |
| 051207 | 3 | — | 2291.3 | — | — | — |
| 051211 | 3 | $592.3^{+836.1}_{-914.9}$ | 1506.7 | — | — | — |
| 051217 | 3 | — | 415.1 | — | — | — |
| 051220A | 3 | $39.0^{+24.1}_{-26.5}$ | 18979.7 | — | — | — |
| 051220B | 3 | — | 388.4 | — | — | — |
| 051221 | 1 | $0.0^{+6.6}_{-8.7}$ | 13183.6 | 328.70 | 16.89 | 0.547 |
| 051222 | 3 | — | 341.3 | — | — | — |
| 060101 | 3 | $599.3^{+655.1}_{-778.6}$ | 1578.3 | — | — | — |
| 060110 | 3 | — | 445.3 | 72.74 | 28.43 | — |
| 060111 | 3 | $-2346.6^{+2031.1}_{-1648.5}$ | 998.6 | — | — | — |
| 060117 | 3 | — | 1293.0 | 327.90 | $-59.98$ | — |
| 060121A | 3 | — | 2918.2 | 137.47 | 45.66 | — |
| 060121B | 3 | — | 342.9 | — | — | — |
| 060123 | 3 | $18.4^{+624.7}_{-529.7}$ | 3834.0 | — | — | — |
| 060124 | 3 | — | 742.7 | 77.11 | 69.74 | 2.296 |
| 060130 | 3 | — | 2052.7 | — | — | — |
| 060203 | 2 | $-23.5^{+46.5}_{-16.1}$ | 4857.4 | — | — | — |
| 060217 | 3 | — | 2363.5 | — | — | — |
| 060224 | 3 | $-944.0^{+3150.4}_{-1486.0}$ | 803.2 | — | — | — |

**Tab. 9.43.** RHESSI sample of GRB lags, peak count rates etc., continuation.

| GRB | Group | Lag (ms) | $F$ (s$^{-1}$) | RA (deg) | Dec (deg) | $z$ |
|-----|-------|----------|----------------|----------|-----------|-----|
| 060228 | 3 | $-702.0^{+1896.9}_{-1770.9}$ | 701.8 | — | — | — |
| 060303 | 1 | $21.2^{+46.3}_{-53.5}$ | 8383.7 | 225.00 | 48.00 | — |
| 060306 | 2 | $50.3^{+12.5}_{-11.0}$ | 100390.0 | — | — | — |
| 060309 | 3 | — | 367.1 | — | — | — |
| 060312A | 1 | — | 1349.4 | — | — | — |
| 060312B | 3 | — | 266.9 | — | — | — |
| 060313 | 3 | — | 551.2 | — | — | — |
| 060323 | 3 | $-186.2^{+270.0}_{-253.4}$ | 3632.5 | — | — | — |
| 060325 | 3 | $189.3^{+235.1}_{-266.9}$ | 5780.1 | — | — | — |
| 060401 | 3 | $110.2^{+268.0}_{-207.9}$ | 2352.1 | — | — | — |
| 060408 | 3 | — | 406.1 | — | — | — |
| 060415 | 3 | — | 351.9 | — | — | — |
| 060418 | 3 | — | 712.1 | 236.43 | $-3.64$ | 1.490 |
| 060421A | 3 | — | 772.3 | — | — | — |
| 060421B | 3 | $312.3^{+208.1}_{-170.2}$ | 2663.8 | — | — | — |
| 060425 | 1 | $5.3^{+4.3}_{-4.8}$ | 8847.1 | 212.30 | $-5.07$ | — |
| 060428 | 3 | — | 263.3 | 285.22 | $-9.55$ | — |
| 060429 | 1 | $3.2^{+14.7}_{-14.5}$ | 17885.9 | 115.51 | $-24.95$ | — |
| 060505 | 3 | $277.1^{+5726.4}_{-1422.9}$ | 1240.0 | — | — | — |
| 060528 | 3 | $2441.1^{+4000.4}_{-4289.8}$ | 775.0 | — | — | — |
| 060530 | 3 | — | 974.9 | — | — | — |
| 060610 | 1 | $10.0^{+14.0}_{-21.1}$ | 9523.1 | — | — | — |
| 060614 | 3 | — | 472.3 | 320.88 | $-53.03$ | 0.125 |
| 060622 | 3 | $-1510.2^{+1932.3}_{-1524.6}$ | 980.8 | — | — | — |
| 060624 | 3 | — | 9347.1 | — | — | — |
| 060625 | 3 | — | 1057.9 | — | — | — |
| 060630 | 3 | — | 1336.5 | — | — | — |
| 060708 | 1 | $-5.6^{+21.8}_{-20.4}$ | 12889.8 | — | — | — |
| 060729 | 3 | — | 580.5 | — | — | — |
| 060805 | 3 | $18.3^{+54.8}_{-69.6}$ | 9923.4 | 272.52 | 58.16 | — |
| 060811 | 3 | — | 1702.9 | — | — | — |
| 060819 | 3 | — | 875.5 | — | — | — |
| 060823 | 2 | — | 764.2 | — | — | — |
| 060919 | 3 | — | 332.4 | — | — | — |
| 060920 | 3 | $22.4^{+61.9}_{-80.2}$ | 4492.0 | — | — | — |
| 060925 | 3 | — | 1769.0 | — | — | — |
| 060928 | 3 | $144.7^{+154.7}_{-151.5}$ | 5629.3 | 127.62 | $-42.75$ | — |
| 061005 | 3 | — | 2099.5 | — | — | — |

**Tab. 9.44.** RHESSI sample of GRB lags, peak count rates etc., continuation.

| GRB | Group | Lag (ms) | $F$ (s$^{-1}$) | RA (deg) | Dec (deg) | $z$ |
|---|---|---|---|---|---|---|
| 061006A | 1 | $9.4^{+167.7}_{-123.0}$ | 6321.4 | 111.03 | $-79.20$ | 0.438 |
| 061006B | 2 | — | 2299.5 | — | — | — |
| 061007 | 3 | $66.2^{+170.0}_{-201.1}$ | 3277.8 | 46.33 | $-50.50$ | 1.261 |
| 061012 | 3 | — | 1320.8 | — | — | — |
| 061013 | 3 | — | 780.1 | — | — | — |
| 061014 | 1 | — | 2222.3 | — | — | — |
| 061022 | 3 | — | 332.4 | — | — | — |
| 061031 | 3 | — | 923.3 | — | — | — |
| 061101 | 3 | — | 616.8 | — | — | — |
| 061108 | 3 | — | 2263.0 | — | — | — |
| 061113 | 3 | $36.4^{+67.3}_{-89.2}$ | 3195.4 | — | — | — |
| 061117 | 3 | — | 233.2 | — | — | — |
| 061121 | 3 | $-57.8^{+216.2}_{-184.8}$ | 2928.4 | 147.23 | $-13.20$ | 1.314 |
| 061123 | 3 | — | 1306.1 | — | — | — |
| 061126 | 3 | $194.3^{+166.7}_{-194.4}$ | 2760.6 | 86.60 | 64.21 | 1.159 |
| 061128 | 1 | $-3.0^{+14.5}_{-19.2}$ | 14106.4 | — | — | — |
| 061205 | 3 | — | 443.0 | — | — | — |
| 061212 | 3 | $13.0^{+19.8}_{-21.4}$ | 9789.6 | — | — | — |
| 061222 | 3 | — | 1371.4 | 358.26 | 46.53 | 2.088 |
| 061229 | 3 | — | 669.8 | — | — | — |
| 061230 | 3 | — | 547.7 | — | — | — |
| 070113 | 1 | — | 2137.1 | — | — | — |
| 070116 | 3 | — | 725.2 | — | — | — |
| 070120 | 3 | — | 393.8 | — | — | — |
| 070121 | 3 | — | 191.4 | — | — | — |
| 070125 | 3 | $-9.6^{+167.9}_{-207.8}$ | 4673.4 | 117.82 | 31.15 | — |
| 070214 | 3 | — | 500.6 | — | — | — |
| 070220 | 3 | — | 888.8 | — | — | — |
| 070221 | 3 | — | 525.8 | — | — | — |
| 070307 | 3 | — | 484.3 | — | — | — |
| 070402 | 3 | — | 848.0 | 311.18 | 27.40 | — |
| 070420 | 3 | — | 530.1 | 121.23 | $-45.56$ | — |
| 070508 | 3 | — | 1479.8 | 312.80 | $-78.39$ | — |
| 070516 | 1 | $28.9^{+59.5}_{-70.1}$ | 2095.8 | — | — | — |
| 070531 | 3 | — | 152.7 | — | — | — |
| 070614 | 1 | $2.4^{+29.2}_{-24.8}$ | 2840.7 | — | — | — |
| 070622 | 3 | $39.9^{+57.8}_{-42.9}$ | 2998.9 | — | — | — |
| 070626 | 3 | — | 1962.5 | 141.35 | $-39.87$ | — |

**Tab. 9.45.** RHESSI sample of GRB lags, peak count rates etc., continuation.

| GRB | Group | Lag (ms) | $F$ (s$^{-1}$) | RA (deg) | Dec (deg) | $z$ |
|---|---|---|---|---|---|---|
| 070710 | 3 | — | 383.6 | — | — | — |
| 070717 | 3 | — | 293.4 | — | — | — |
| 070722 | 3 | — | 315.5 | — | — | — |
| 070724 | 3 | — | 268.9 | 17.49 | 57.68 | — |
| 070802 | 2 | $15.8^{+48.0}_{-58.1}$ | 1956.4 | — | — | — |
| 070817 | 3 | — | 504.7 | — | — | — |
| 070819 | 3 | — | 880.1 | — | — | — |
| 070821 | 3 | $663.0^{+825.1}_{-764.1}$ | 1900.3 | 95.53 | −63.85 | — |
| 070824 | 2 | $-5.9^{+120.0}_{-109.8}$ | 3104.2 | 172.00 | −27.00 | — |
| 070825 | 3 | $875.1^{+1116.6}_{-1635.1}$ | 1237.9 | — | — | — |
| 070917 | 3 | — | 336.6 | — | — | — |
| 071013 | 3 | — | 927.7 | — | — | — |
| 071014 | 3 | — | 1010.1 | — | — | — |
| 071030 | 3 | — | 526.5 | — | — | — |
| 071104 | 3 | — | 657.1 | 295.61 | 14.61 | — |
| 071204 | 1 | — | 2880.2 | — | — | — |
| 071217 | 3 | — | 819.1 | — | — | — |
| 080114 | 3 | $285.7^{+233.2}_{-252.3}$ | 2182.9 | — | — | — |
| 080202 | 3 | — | 448.8 | — | — | — |
| 080204 | 3 | $60.5^{+292.9}_{-244.0}$ | 1358.5 | 285.57 | −62.90 | — |
| 080211 | 3 | $1139.6^{+2898.5}_{-1804.8}$ | 1298.7 | 43.97 | 60.04 | — |
| 080218 | 3 | — | 404.3 | — | — | — |
| 080224 | 3 | — | 1614.7 | — | — | — |
| 080318 | 3 | — | 377.8 | — | — | — |
| 080319 | 3 | $157.9^{+551.1}_{-1112.4}$ | 834.2 | 258.98 | 55.39 | 1.950 |
| 080320 | 3 | $34.0^{+381.6}_{-569.9}$ | 1639.9 | — | — | — |
| 080328 | 3 | — | 839.9 | 80.48 | 47.51 | — |
| 080330 | 3 | $26.3^{+495.1}_{-350.8}$ | 1577.3 | — | — | — |
| 080408 | 2 | $73.3^{+133.6}_{-230.4}$ | 2475.7 | — | — | — |
| 080413 | 3 | — | 507.4 | 326.14 | −19.98 | 1.100 |
| 080425 | 3 | — | 671.2 | — | — | — |

# 10    Appendix C - Uncertainties of Decimated Data

This part describes one sigma errors of the bin counts for the decimated RHESSI data mentioned in Section 5.1.

## 10.1    Full decimation

Calculation of one sigma uncertainty $\sigma_{C_{dc}}$ for the bin counts $C_{dc}$ of fully decimated data and then corrected for this decimation is following. For corrected bin counts $C_{dc}$ it holds:

$$C_{dc} = F.C_d, \tag{10.20}$$

where $F$ is the decimation factor (weight), usually equal to 4 or 6 for the RHESSI data, and $C_d$ is the number of counts in a bin of the decimated signal. If we assume that counts in a bin follow Poisson statistics then:

$$\sigma_{C_{dc}} = \frac{\partial C_{dc}}{\partial C_d}\sigma_{C_d} = F\sqrt{C_d} = \sqrt{F.C_{dc}}, \tag{10.21}$$

where $\sigma_{C_d}$ is the dispersion of the $C_d$.

## 10.2    Partial decimation

Now consider the situation that counts in a bin are only partially decimated, e.i., they consists of the non-decimated $C_1$ and the decimated signal $C_{2,d}$. This situation can happen when we sum counts over the energy band $[E_1; E_2]$, $E_1 < E_0 < E_2$, and only the counts bellow the energy $E_0$ are decimated. Then the corrected signal $C$ is equal to:

$$C = C_1 + C_{2,dc} = C_1 + F.C_{2,d}, \tag{10.22}$$

where again $F$ is the decimation factor and $C_{2,dc}$ is corrected part of the signal that was decimated. One sigma uncertainty $\sigma_C$ is then equal to:

$$\sigma_C = \sqrt{\left(\frac{\partial C}{\partial C_1}\right)^2 \sigma_{C_1}^2 + \left(\frac{\partial C}{\partial C_{2,d}}\right)^2 \sigma_{C_{2,d}}^2} =$$

$$= \sqrt{\sigma_{C_1}^2 + F^2.\sigma_{C_{2,d}}^2} = \sqrt{C_1 + F^2.C_{2,d}} = \sqrt{C_1 + F.C_{2,dc}}, \tag{10.23}$$

where $\sigma_{C_1} = \sqrt{C_1}$ is the dispersion of the non-decimated part of the bin counts and $\sigma_{C_{2,d}} = \sqrt{C_{2,d}}$ is the dispersion of the decimated part of the bin counts.

# List of Used Abbreviations

| | |
|---|---|
| AG | Afterglow |
| AGN | Active Galactic Nuclei |
| AMANDA | Antarctic Muon And Neutrino Detector Array |
| BAT | Burst Alert Telescope |
| BATSE | Burst and Transient Source Experiment |
| BH | Black Hole |
| CCF | Cross-Correlation Function |
| CGRO | Compton Gamma-Ray Observatory |
| CL | Confidence Level |
| COMPTEL | Imaging Compton Telescope |
| CPL | Cutoff Power-Law |
| EGRET | Energetic Gamma Ray Experiment Telescope |
| FWHM | Full Width at Half Maximum |
| GBM | Gamma-Ray Burst Monitor |
| GCN | GRB Coordinates Network |
| GRB | Gamma-Ray Burst |
| GRBM | Gamma-Ray Burst Monitor |
| GW | Gravitational Wave |
| HETE | High Energy Transient Explorer |
| HST | Hubble Space Telescope |
| HXD | Hard X-ray Detector |
| INTEGRAL | INTErnational Gamma-Ray Astrophysics Laboratory |
| IPN | Inter-Planetary Network |
| K-S | Kolmogorov-Smirnov |
| LAT | Large Area Telescope |
| LGRB | Long Gamma-Ray Burst |
| LIGO | Laser Interferometer Gravitational Wave Observatory |
| LISA | Laser Interferometer Space Antenna |
| ML | Maximum Likelihood |
| NFI | Narrow Field Instruments |
| NS | Neutron Star |
| OSSE | Oriented Scintillation Spectrometer Experiment |
| OT | Optical Transient |
| RHESSI | Reuven Ramaty High-Energy Solar Spectroscopic Imager |
| SFR | Star Formation Rate |
| SGR | Soft Gamma-ray Repeater |
| SGRB | Short Gamma-Ray Burst |
| SGRBEE | Short Gamma-Ray Bursts with Extended Emission |
| SN | Supernova |
| TGF | Terrestrial Gamma Flash |
| UHECRs | Ultra-High-Energy Cosmic Rays |
| UVOT | Ultra-Violet/Optical Telescope |
| VIRGO | Variability of Irradiance and Gravity Oscillations |

WAM    Wideband All-sky Monitor
WD     White Dwarf
WFC    Wide Field Cameras
XRF    X-Ray Flash
XRT    X-Ray Telescope

# List of Author's Publications

# List of Independent Citations